\documentclass{article}
\usepackage[utf8]{inputenc}
\usepackage{graphicx}
\usepackage{xcolor}
\usepackage{float}
\usepackage{hyperref}
\usepackage{natbib}
\usepackage{amsmath}
\usepackage{amssymb}
\bibpunct[,]{(}{)}{;}{a}{}{,}

\title{\textbf{Development of Bayesian methods for the characterization of galaxy star formation histories through stellar population synthesis models}}
\author{Author: Edoardo Rossi\\
Universit\`a degli Studi di Firenze - Dipartimento di Fisica e Astronomia\\
Laurea Magistrale in Scienze Fisiche e Astrofisiche \\
Supervisor: Dr. Stefano Zibetti, Ph.D. \\(INAF - Osservatorio Astrofisico di Arcetri)}

\date{Academic year 2020/2021}

\begin{document}

\maketitle
\newpage

\begin{abstract}
The goal of this project is to establish parameters and methods to characterize the Star Formation History (SFH) of galaxies beyond the mean stellar age. In particular, we use the ages at which a fixed fraction of stars is already formed to characterize the duration of star formation history. Specifically, we define the parameter $\Delta age_n\equiv(age_{10}-age{90})/age_{50}$ in order to define the SFH extension, where $age_{10},~age_{90},~age_{50}$ correspond to the age at which the 10\%, 90\% and the 50\% of formed stars is reached. Probability distribution functions of these parameters for observed galaxies are derived by means of a robust Bayesian statistical approach, which relies on the comparison between observed features in galaxy spectra and the corresponding features in models.\\
For this goal, we create vast libraries of composite stellar populations (CSP), using the existing software package SEDlibrary, which allows to implement a variety of SFH, metallicity and dust proprieties.\\
In order to compare model spectra to the observed one, we choose ten spectral features: five spectral indices (D4000n, [H$\delta_A$+H$\gamma_A$], H$\beta$, [Mg$_{2}$Fe] and [MgFe]$^\prime$) and five SDSS photometric fluxes in $ugriz$.\\

As a first result, we focus on the limiting $\Delta age_n$ above which we are able to distinguish an extended SFH from one of negligible duration, $\Delta age_{n,min}$ (time resolution). For this first result we worked on a idealized (not realistic for complex galaxies) CSP library composed by 5 million of models with no dust and no star bursts, and each million is characterized by a fixed metallicity from highly subsolar to supersolar. $\Delta age_{n,min}$ is found as the value of $\Delta age_n$, from which our set of spectral features starts depending on the duration of the SFH.\\
We find a roughly flat trend of $log_{10}(\Delta age_{n,min})$ around -0.3 dex over 4 orders of magnitude in age. $\Delta age_{n,min}$ gets lower for higher SNR, up to SNR=100, above which $\Delta age_{n,min}$ does not improve anymore.\\

In the second part of this work, we produce a mock dataset by perturbing 12\,500 models of the library with realistic errors and we study the capability of our method to retrieve the input information in terms of characteristic ages and SFH duration. We rely on a more realistic CSP library to describe a galaxy than the previous one. The 500\,000 models in this library are created considering up to 6 star bursts, dust and variable metallicity. The mock dataset is drawn from this same library.\\
We find that, we are able to constrain the SFH duration $log_{10}(\Delta age_n)$ within $\pm0.3$ dex for most of our sample. For stellar populations characterized by a strong Balmer absorption and a mean stellar age $<10^9yr$, we obtain a high uncertain ($>0.5$ dex) on $log_{10}(\Delta age_n)$, which is related to SFH degeneracies on the resulting galaxy spectra.\\
These parameters and methodology will find ample application in current and upcoming deep spectroscopic surveys of galaxies.

\end{abstract}

\newpage
\tableofcontents
\newpage

\section{Introduction}
\subsection{The realm of galaxies}
\begin{figure}
    \centering
    \includegraphics[width=\textwidth]{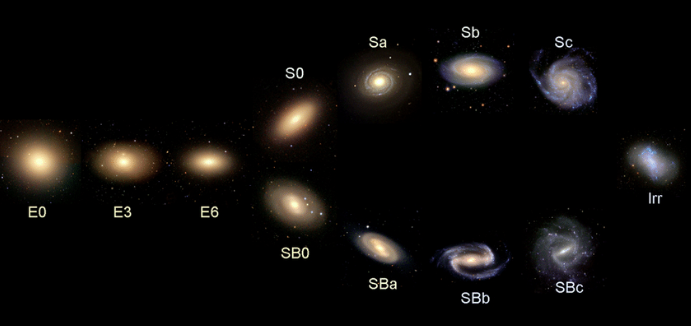}
    \caption{Edwin Hubble's classification scheme. Image credit: Department of Physics, University of Oregon.}
    \label{fig:hubble}
\end{figure}

The charming structures that we call galaxies are self gravitating objects made of stars, gas, dust and dark matter (DM). In the local Universe, galaxies are characterized by distinct structural components, which may or may not be all present and essentially belong to one of the following categories: disk, bulge and stellar halo. Based on the relative importance of these three components, on the presence and strength of the spiral arms in the disk, and on the concentration of the light distribution, galaxies are usually classified following the scheme (Fig. \ref{fig:hubble}) that was first proposed by Edwin Hubble (1929). We can classify galaxies in three basic types: spirals, ellipticals and irregulars. The first ones comprise a disk, which contains luminous spiral arms and hosts dust and young stars, a bulge and a faint stellar halo, which both host old stars. Spiral galaxies are divided into normal and barred spirals, depending on whether their spiral arms start from the nucleus or from a bar centered in the nucleus. This type of galaxies constitute more than half of the bright galaxies observed within 100 Mpc from the Sun. The elliptical galaxies have an elliptical shape and typically little dust and very few young stars. Irregular galaxies have no regular shape and are more common in the low luminosity regime. 

\begin{figure}
    \centering
    \includegraphics[width=1.2\textwidth]{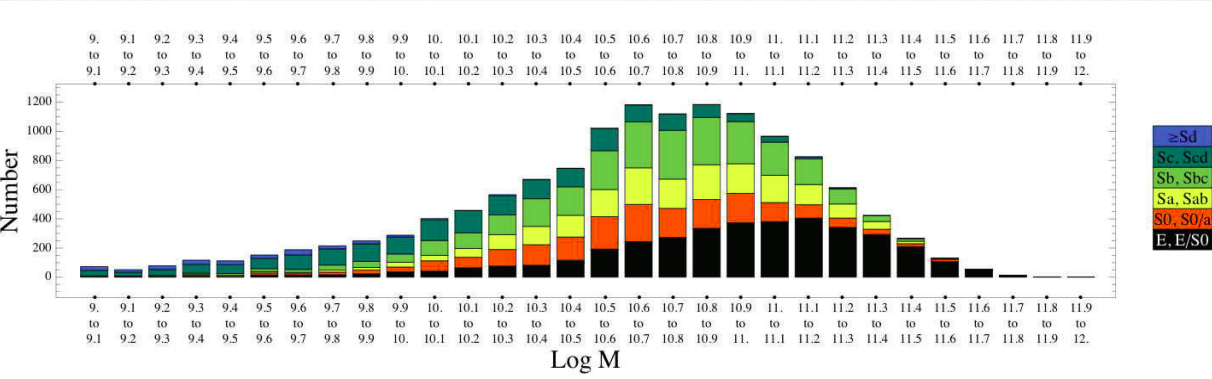}
    \caption{Morphological mix as a function of galaxies' mass. From \cite{nair2010}.}
    \label{fig:morph_mix}
\end{figure}

\begin{figure}
    \centering
    \includegraphics[width=\textwidth]{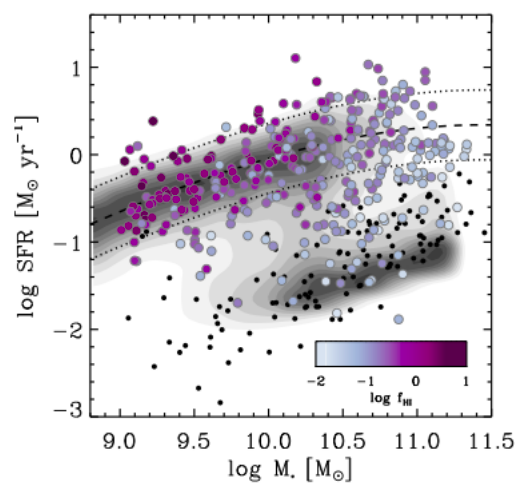}
    \caption{Distribution of xCOLD GASS sample in the SFR-$M_{*}$ plane, color-coded by atomic gas mass fraction. The small black symbols are galaxies un-detected in HI line. The grayscale contours show the overall SDSS population. The dashed and dotted lines indicate the position of the main sequence and the $\pm$0.4 dex scatter around this relation, respectively.  From Saintonge et al. 2017}
    \label{fig:sfr_mass}
\end{figure}

\begin{figure}
    \centering
    \includegraphics[width=\textwidth]{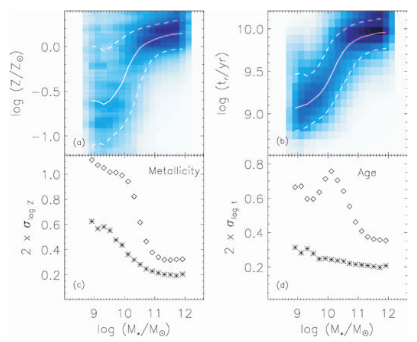}
    \caption{Distribution of stellar metallicity and age as function of stellar mass for 44 254 SDSS-DR2 galaxies with median S/N per pixel greater than 20. Top panels: solid line is the median and the dashed lines represent the 16th and 84th percentiles. Bottom panels: the diamonds are the differences between 16th and 84th percentiles, while the stars represent the mean 68 per cent confidence range in each stellar mass bin. From \cite{gallazzi2005}}
    \label{fig:role_mass}
\end{figure}

This classification scheme has proven also to be very powerful from a physical point of view, as it highlights a fundamental physical dichotomy between late-type galaxies (i.e. disk-dominated spirals and irregulars) and early-type galaxies (i.e. ellipticals and bulge-dominated galaxies). Early-type galaxies are characterized by old stars, low gas fraction and low star formation rate (SFR, i.e. mass of gas transformed into stars per unit of time). As opposed, late-type galaxies are characterized by predominance of young stars, high gas fraction and significant SFR. Furthermore, a central role in driving physical properties is played by (stellar) mass. In fact, as we can see in figure \ref{fig:morph_mix}, the morphological mix of galaxies changes as a function of stellar mass. In particular, the fraction of early-type galaxies starts growing for masses of the order of $10^{10.3}M_{\odot}$ and dominates over the fraction of the late-type galaxies at $10^{11.3}M_{\odot}$. This implies a variation of physical proprieties as a function of stellar mass. For example, if we consider mass-SFR relation (Fig. \ref{fig:sfr_mass}), we can distinguish two different sequences: the upper one (i.e. Main Sequence, MS) represents the star forming galaxies and approximately we have a mass-SFR linear relation; the lower one is populated by passive galaxies and it dominates by number at high masses. The transition to passive predominance occurs at $Log(M_{*}/M_{\odot})\sim 10.3$, a well-known characteristic mass which was identified in the early '2000 \citep[e.g.][]{kauffmann2003}. Another relevant information shown in Figure \ref{fig:sfr_mass} is the gas fraction (coded by symbol colour). Galaxies on the MS display a trend for decreasing gas fraction as mass increases. Moreover, passive galaxies (or galaxies below the MS) have a significantly lower gas fraction (not even detected in many cases) with respect to the MS galaxies of similar mass. This fact suggests that more massive systems have already transformed increasingly higher gas fraction into stars with respect to lower mass galaxies.
The importance of stellar mass for galaxy evolution can be seen also in stellar populations properties, in particular in mean stellar age and chemical composition, described in terms of metallicity (i.e. the mass of elements heavier than helium over the total mass in stars). In figure \ref{fig:role_mass} \citep[from][]{gallazzi2005} we see that both stellar metallicity and the mean stellar age increase with increasing stellar mass. These observations suggest that the more massive galaxies have formed and evolved more rapidly and have been more efficient to convert gas into stars. On the contrary, less massive galaxies, in particular with mass lower than $10^{10.3}M_{\odot}$, are still forming stars by conversion of gas into stars.
This scenario of galaxy evolution is what we refer to as ``anti-hierarchical''.

\subsection{Cosmological context}
If we look at the universe on Mpc scales, we find it dominated by individual galaxies and therefore very clumpy. However, if we go on larger scales ($\sim$10 Mpc) a web of cosmic structures emerges, in which groups and clusters of galaxies are connected via filaments. On even larger scales the Universe begins to display a much more homogeneous and isotropic picture. This entire cosmic web is known to follow a cosmological expansion, as first pointed out by Hubble (\cite{hubble1929}). According to the most widely accepted cosmological theory, the so-called Hot Big Bang Theory, our Universe started out of a state of high density and temperature. After $\sim$300\,000 years, radiation and matter decoupled, leaving a black body radiation, the so-called cosmic microwave background (CMB), as a fossil record of this epoch. Measurements of the CMB spectra have shown that the Universe at that epoch was in almost perfect thermodynamic equilibrium and very homogeneous, with temperature fluctuations $\delta T/T\sim10^{-5}$. The tiny amplitude of these fluctuations implies that they would have not be able to grow to form the present-day structures if the Universe was done by ordinary baryonic matter alone. Hence, some form of ``dark matter'' (i.e. interacting only via gravity but not via electromagnetic forces) is required to have started growing fluctuations well before the decoupling between baryons and photons. Furthermore, from the large scale distribution of galaxies and their clustering, we can say that the best assumption about the DM is that it is ``cold'' (CDM), i.e. DM is composed by not-relativistic particles at the decoupling time. As demonstrated by theoretical works \citep[e.g.][]{press_schechter1974} and N-body simulations \citep[e.g.][]{NFW1996}, the introduction of the CDM defines a ``bottom-up'' or ``hierarchical'' scenario: smaller systems are formed first and the bigger ones are formed subsequently by merging of the smaller ones. 

\subsection{Galaxy formation and evolution: the key role of baryons}
In a scenario where galaxy evolution would mirror the hierarchical assembly of DM halos, we would expect the most massive galaxies to be still in the process of formation, actively star forming, harboring young stars and with an incomplete degree of chemical enrichment. Vice versa,  we would expect low-mass galaxies to have already completed their evolution and be essentially passive. This is actually the opposite of what we observe.

\begin{figure}
    \centering
    \includegraphics[width=\textwidth]{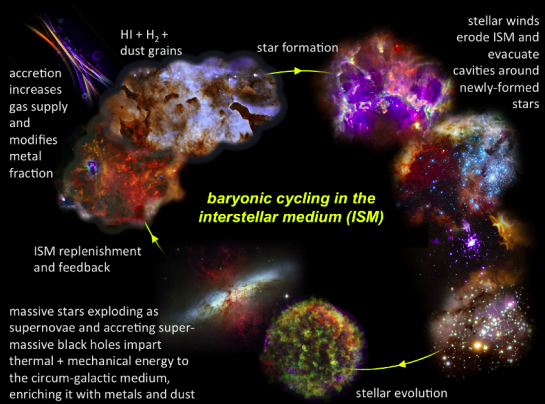}
    \caption{Baryonic cycle. Courtesy of L. K. Hunt \textcopyright 2015}
    \label{fig:bar_circ}
\end{figure}

The key to understand the opposite evolution of galaxies and DM halos has to be found in the physics that regulates how the different baryonic components relate to the each other and, in particular, how gas is transformed into stars and eventually recycled into the interstellar and circumgalactic medium.
As originally proposed by \cite{white_rees_1978}, the cycle is initiated by the shock heating of the primordial gas that falls into the potential of a DM halo. Subsequently, this hot gas cools and sinks towards the centre of the DM halo. When low enough temperatures and high enough density are reached, giant molecular clouds are formed. These clouds eventually undergo fragmentation and further collapse, until in the cores high densities can be reached  that allow the onset of thermonuclear reactions and the actual birth of the stars.
The efficiency of these processes is very sensitive to the physical condition of the gas itself (e.g. its dynamical state, turbulence, chemical composition, magnetic fields) and its environment (e.g. external radiation field, pressure by surrounding gas, gravitational perturbations). Some of the ``external'' conditions that regulate the efficiency of star formation can be seen as a direct consequence of the galaxy formation process itself and the processes they implement are often referred to as ``feedback''. In particular, studies have focused onto two main categories of feedback: the stellar feedback and the AGN feedback.
Stellar feedback mainly consists in the ionization of the surrounding gas by the UV radiation of young stars and in injection of energy and momentum into the ISM by powerful winds and Supernovae explosions (the death of massive stars). In the second type, instead, $feedback$ comes from the Active Galactic Nucleus (AGN) of a galaxy. An AGN is characterized by emission that is powered by the accretion onto a supermassive ($\sim 10^8 M\odot$) black hole that lurks in the centre of the galaxy and grows thanks to the infall of the surrounding gas (and stars). The powerful jets and winds that originate from an AGN can have a great impact on the host galaxy by injecting energy and momentum into the ISM and into the circumgalactic medium. These feedback mechanisms can limit the ability of gas to cool and collapse and form new stars: the same mechanisms that feed the star formation in a galaxy, are eventually responsible for triggering the feedback mechanisms that hamper the star formation itself. An illustration of these processes and of their circular dependence can be seen in the ``baryonic cycle'' shown in Fig. \ref{fig:bar_circ} (courtesy of L. K. Hunt).

In our current understanding of galaxy formation and evolution \citep[e.g.][]{croton_2006,delucia_2006}, star formation efficiency and feedback are the key mass-dependent parameters to reconcile the hierarchical growth of DM structure and the anti-hierarchical properties of galaxies. Today's massive galaxies must have experienced a very efficient early phase of star formation, followed by an intense feedback that has prevented further gas accretion and/or star formation. For lower masses, the hypothesis is that they had both lower initial efficiency and less intense subsequent feedback, which have allowed them to have a more prolonged and slower evolution.

\subsection{Understanding galaxy formation and evolution through observations, theoretical models and simulations}
While the \emph{qualitative} picture described at the end of the previous section has emerged in the last decades, connecting and reconciling the theory of structure formation with the observations of galaxy evolution in a \emph{quantitative} way remains the fundamental problem in extragalactic astrophysics. On the one hand we have to refine our observational knowledge of galaxy evolution with these two approaches to gain more detail and more secure constraints. On the other hand we have to develop models and simulations with realistic physical recipes that are able to produce galaxies starting from the well known physics of DM halos. These simulations in astrophysics have the same role of experiments in physics, that is testing hypothesis in a controlled framework.

From the observational side, the research has been following two complementary approaches: the so-called ``galaxy archaeology'' approach and  ``look-back'' approach.

In the galaxy archaeology approach the goal is to reconstruct the history of a galaxy through its ``relics''. Since the star formation process is irreversible (stars cannot go back to the gas from which they formed), stars of all generations accumulate in a galaxy over time. Other fossils of a galaxy's evolution are the products of stellar evolution (metals) which are locked in stars and in the ISM, and the stellar kinematics. Unfortunately these ``fossil remnants'' degrade over time. For instance, distinguishing between stellar population of different ages become more and more difficult as age increases, also because they are weaker in luminosity and are easily outshined by younger generations of stars. Also from the stellar dynamical point of view, dynamical relaxation washes out the signature of the early dynamical state of a galaxy as time goes by. As a result, the archaeological approach suffers from significant limitations as we push it to early cosmic times.

The ``look-back'' approach is the investigation at different cosmological eras. Because of finite speed of light, $c$, the farther a light source is, the more we observe it in the past. So, we observe galaxies of different cosmological epochs on which we make a statistical analysis. The goal of that approach is to reconstruct the ``cosmic evolution movie''. The ``look-back'' approach has been very successful in the last two decades in providing us with a demographic description of the evolution of galaxies as populations through cosmic time, e.g. by identifying a peak in overall star formation rate and black hole accretion rate at $z\sim2$ (the so-called ``cosmic noon''). The problem is that we can not observe the same galaxies across the cosmic time (in other words, we do not know which are the ``progenitors'' of galaxies at different epochs), and this severely hampers our capability of reconstructing the evolutionary trajectories of individual galaxies. 

As we mentioned, each of these approaches has important limitations and in order to make a good investigation on galaxy evolution, we need to combine information from all of them.

Given the key importance of the SFH to characterize a galaxy's evolution, in this thesis we want to investigate the possibility of retrieving basic information about the SFH of a galaxy from its spectrum and spectral energy distribution (SED). We build this work on the Bayesian statistical method introduced by \cite{gallazzi2005} and lately developed by \cite{zibetti2017}, that takes also models degeneracy into account, as we describe in the next section. In particular, we will not focus on the reconstruction of the full SFH, rather on a \emph{robust} characterization of its overall duration, in addition to its characteristic time-scale (i.e. stellar age).

\newpage
\section{Reconstructing the star formation histories of galaxies}\label{chap:SFH}
Every galaxy is characterized by its own star formation history (SFH), which describes the build up of the present stellar mass across time. The SFH is defined by the star formation rate (SFR) function: SFR(t)=$\frac{dM_{*}}{dt}(t)$, i.e. the stellar mass produced per unit time as function of time. By ``reconstructing the SFH of a galaxy'' we mean characterising this function. In an ideal case, we would like to be able to find the intensity of SFR(t) for any time. However, in reality we can only approximate this function either by reverting to some analytic fitting function or by sampling it with a finite time resolution, for instance by discretizing the SFR(t) as $\frac{\Delta M_{*}}{\Delta t}(t)$ in bins of finite $\Delta t$ (see Sec. \ref{sec:SPS}). 

To constrain the shape of a galaxy SFH we compare its observed starlight spectrum with model spectra of ``stellar populations'' (i.e. physically motivated ensambles of stars that are built using stellar population synthesis models, see Sec. \ref{sec:SPS}). Unfortunately, this approach is hampered by serious problems, most notably noisy data and degeneracy between different physical parameters, as we will see in detail in Sec. \ref{sec:inversion_problem}. So, it is reasonable to try to characterize the SFHs of galaxies by determining some parameters through a robust statistical analysis. In this project, in particular, we focus on a description of the SFH by means of a limited number of characteristic \emph{ages} (e.g. light-weighted age, mass-weighted age, ages corresponding to the time at which a given fraction of the stellar mass was in place). These are estimated based on a robust Bayesian statistical approach, which will be discussed in detail in Sec. \ref{sec:SFHparams}.

\subsection{Fundamentals of stellar population synthesis}\label{sec:SPS}
To get information about galaxy SFH we have to compare observed starlight to models of starlight. To do this we need techniques that are able to reproduce photometric and spectral global features of a group of stars by superposing spectral emission of different stars and different stellar generations. These are the so-called stellar population synthesis (SPS) techniques \citep[e.g.][]{tinsley1978,bruzual_charlot1993,Bressan_chiosi1994,fioc1997,maraston1998,vazdekis1999}.\\

The simplest group of stars that we can define is the so-called ``simple stellar population'' (SSP). A SSP consists of stars born at the same time in a burst of star formation activity of negligible duration, all with the same initial chemical composition. The stars belonging to the same SSP are drawn from an initial distribution in mass, dubbed ``initial mass function'' (IMF). The mass distribution and the light emitted by the stars evolves with time according to the physics of stellar evolution. Once a given IMF and an initial chemical composition is assumed, the spectral evolution of a SSP as a function of time is fully determined by the (complex and still partly unknown) physics of stellar evolution.

However, galaxies are much more complex than a single SSP, since they are usually composed by several stellar populations of different ages and chemical composition. The overall stellar population of a galaxy can be thought as a ``composite stellar population'' (CSP) given by the superposition of different SSPs, which work as the ``building blocks''.

In real galaxies not all the light that is emitted by stars reaches the observer because interstellar dust partly absorbs and scatters it off the line of sight, producing the phenomenon of dust attenuation, which is parameterized via the effective optical depth $\tau_d$. In order to properly compare spectral models with observations, dust attenuation must be taken into account. 
A typical spectral energy distribution (SED, i.e. specific luminosity as a function of frequency or wavelength) of a CSP, $f_{CSP}(\lambda)$, observed at a given time $t$ since the beginning of the SFH, can be expressed as a linear superposition of SSP spectra of different age $t'$ and metallicity $Z$, $f_{SSP}(\lambda, t', Z)$:
\begin{equation}\label{eq:CSP_cont}
    f_{CSP}(\lambda)=\int_{0}^{t}\int_{0}^{Z_{max}}(SFR(t-t')P(Z,t-t')f_{SSP}(\lambda | t', Z)e^{-\tau_{d}(t',Z)})dt'dZ 
\end{equation}
The double integral extends over the parameter space of all possible ages $t'$ and metallicities $Z$, whose distribution is provided as a function of cosmic time $t-t'$ via the $SFR(t-t')$ (the star formation history) and $P(Z, t-t')$ (the metallicity probability distribution), respectively. 
Note the dust attenuation factor $e^{-\tau_{d}(t',Z)}$, which, in principle, can be different for each SSP. \\
For any practical application, we are forced to discretize Eq. \ref{eq:CSP_cont}, by replacing the integrals with sums, as follows:
\begin{equation}\label{eq:CSP_double_sum}
    f_{CSP}=\sum_{j}\sum_{i}w_{i,j}f_{SSP}(t_{i},Z_{j})
\end{equation}
where the index $i$ runs over the age, while $j$ over the metallicities, so that the weights $w_{i,j}$ account for the SFH $SFR(t-t')$ and the metallicity distribution $P(Z, t-t')$.\\

\subsection{The inversion problem}\label{sec:inversion_problem}
From Eq. \ref{eq:CSP_double_sum} we see that the star-formation and chemical history of a galaxy can be reconstructed by solving for the weights $w_{i,j}$.
This is what is called the ``inversion problem''.
By rearranging the array of SSPs, Eq. \ref{eq:CSP_double_sum} can be simplified as:
\begin{equation}\label{eq:CSP_simple_sum}
    f_{CSP}=\sum_{i}w_{k}\cdot f_{SSP}(t_k, Z_k)=\sum_{k}w_{k}\cdot f_{SSP,k}
\end{equation}
In this equation we can see each SSP spectrum as the line of a matrix, with each column corresponding to a different wavelength. Using this SSP matrix, $\textbf{f}_{SSP}$, we can write Eq. \ref{eq:CSP_simple_sum} in a ``matricial'' form:
\begin{equation}\label{eq:inversion_matrix}
    f_{CSP}=\textbf{f}_{SSP}\cdot w
\end{equation}
where $f_{CSP}$ is the galaxy spectra and $w$ is the vector of weights. The goal is to obtain $w$ by inverting $\textbf{f}_{SSP}$. \\
This approach is referred to as ``full inversion'', and the vector of weights can be written by inverting the equation \ref{eq:inversion_matrix}:
\begin{equation}
    w=(\textbf{f}_{SSP}^T\cdot \textbf{f}_{SSP})^{-1}\cdot \textbf{f}_{SSP}^T \cdot f_{CSP}
\end{equation}
The solution is obtained by minimizing the $\chi^2$ of the observed spectrum (point by point) relative to the models, taking into account the observational errors. However, we have to consider that real data, $y$, are affected by noise. In fact, once Singular Decomposition Value (SDV) of $\textbf{f}_{SSP}$ is introduced, our solution is given by the sum of an unperturbed component and a component related to the noise. By comparing these two terms, we see that to constrain the weight for a decent number ($\geq$ 4) of eigenvectors we need high S/N spectra \citep[see][]{ocvirk2006}.\\

\begin{figure}
    \centering
    \includegraphics[width=\textwidth]{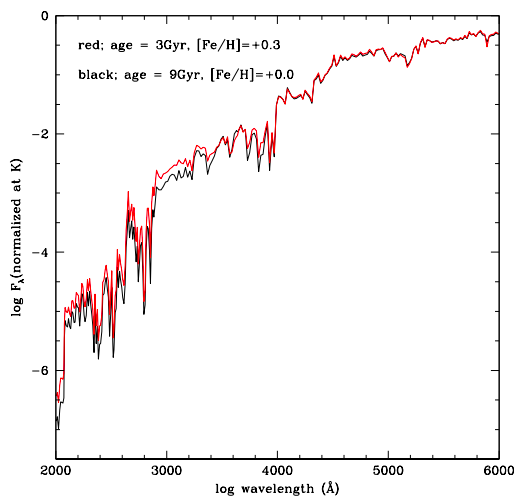}
    \caption{Age-metallicity degeneracy. From \cite{chavez2011}}
    \label{fig:age_Z_degeneracy}
\end{figure}

The problem is further complicated by the fact that the SSP spectra that define the $\textbf{f}_{SSP}$ matrix are affected by significant degeneracy with respect to their age and metallicity, i.e. very similar stellar spectra result from different combinations of age and metallicity. Fig. \ref{fig:age_Z_degeneracy} illustrates the ``age-metallicity degeneracy'' effect by showing that for a given SSP spectrum corresponding to an age of 3 Gyr and a metallicity of twice solar, a very similar spectrum can be obtained by increasing the age by a factor 3 and simultaneously reducing the metallicity by a factor 2.\\
In such a degenerate parameter space, the maximum likelihood solution is hardly robust from a statistical point of view, as it critically depends on small measurement fluctuations. Moreover, the maximum likelihood approach is not suited to properly quantify the parameter uncertainties connected with model degeneracies.

So, it is reasonable trying to give a characterization of galaxy SFH through the determination of a more restricted parameter set and considering SFR(t) as a distribution (in the mathematical sense). In this work we consider in particular two different ``mean'' stellar age: the ``light-weighted'' age (equation 6) and the ``mass-weighted'' age (equation 5).
\begin{equation}\label{eq:light_age}
    Age_{l-w}=\frac{\int_{t=0}^{t_{form}}dt(t_{form}-t)SFR(t)L(t)}{\int_{t=0}^{t_{form}}dtSFR(t)L(t)}
\end{equation}
\begin{equation}\label{eq:mass_age}
    Age_{m-w}=\frac{\int_{t=0}^{t_{form}}dt(t_{form}-t)SFR(t)}{\int_{t=0}^{t_{form}}dtSFR(t)}
\end{equation}
where $SFR(t)$ is the star formation rate and $L(t)$ is the luminosity arising per unit of stellar mass from an SSP of age $t$, and $t_{form}$ the time of observation since the beginning of the SFH.
It is worth noting that, relatively to the mass-weighted age, the light-weighted age is more directly related to the starlight and less impacted by the uncertain mass contribution of the oldest stellar populations, which have very high mass-to-light ratio. Notably, the mass-weighted age is the first moment of SFR distribution.\\

Following \cite{pacifici2016} we define the ages corresponding to a given fraction of total mass formed. From the integration of the $SFR(t)$, the mass fraction formed within a given time $\bar{t}$ can be obtained as:
\begin{equation}
    \frac{M(\bar{t})}{M_{tot}}=A\cdot\int_{0}^{\bar{t}}SFR(t')dt'
\end{equation}
where $A$ is a normalization coefficient, so that $\frac{M(\bar{t})}{M_{tot}}$ is unity at the time  $t_{form}$ after the beginning of the SFH:
\begin{equation}
    A=\frac{1}{\int_{t=0}^{t_{form}}SFR(t')dt'}
\end{equation}
Now, we can define the time $t_f$ at which a given formed mass fraction $f$ is reached (i.e. percentiles of the distribution $SFR(t)$):
\begin{equation}
   \frac{M(t_{f})}{M(t_{form})}=f
\end{equation}
\begin{equation}
    f=A\cdot\int_{0}^{t_{f}}SFR(t')dt'
\end{equation}
$t_{f}$ is obtained by solving this integral equation. We can then compute the age $age_{f}$ corresponding to $t_{f}$ as:
\begin{equation}\label{eq:f_age}
    age_{f}=t_{form}-t_{f}
\end{equation}
For our analysis, we will use $age_{10}, age_{50}$ and $age_{90}$ (i.e. look-back times at which the 10\%, 50\% and 90\% of the stellar mass is produced) and we also define the characteristic SFH duration $\Delta t_{10,90}$, as the difference between the look-back time at which 10\% and 90\% of the total formed mass is reached:
\begin{equation}
    \Delta t_{10-90}=age_{10}-age_{90}
\end{equation}

Since this parameter scales with the order of magnitude of $age_{50}$ and spans a wide dynamic range, it is convenient express the SFH duration in relative terms. So, we define the following quantity:

\begin{equation}
    \frac{\Delta age_{10-90}}{age_{50}}=\frac{age_{10}-age_{90}}{age_{50}}\equiv \Delta age_{n}
\end{equation}
$\Delta age_{n}$ thus describes the duration of star formation activity relative to the median stellar age $age_{50}$. 

\subsection{Derivation of SFH parameters}\label{sec:SFHparams}
In this section we describe the basis of the Bayesian statistical approach used in this work (Sec. \ref{subsec:bayesian_stats}) and
the library of CSP models that are needed to infer the physical (SFH) parameters (Sec. \ref{subsec:speclibrary}). In the last subsection (Sec. \ref{subsec:BaStA}) we describe in detail the algorithm used to get the information about the parameters of interest.

\subsubsection{Bayesian statistics}\label{subsec:bayesian_stats}
In this section we describe the statistical method used to produce a statistical probability function for the physical parameters introduced in Sec. \ref{sec:SFHparams}, based on the comparison between observations and models \citep[see][]{kauffmann2003}.
Start supposing that we have a library of CSP models, for which we know all relevant physical parameters and observational quantities. For each model we can define a physical parameters vector, \textbf{P}. From an astronomical observation (e.g. a spectrum) we obtain the vector of observed spectral features, \textbf{O}, which can be directly compared to the corresponding features in the models. From Bayes' theorem, the ``posterior'' probability distribution function (PDF) of the model property \textbf{P} given the data is obtained as:
\begin{equation}
    f(\textbf{P}|\textbf{O})d\textbf{P}=Af_{p}(\textbf{P})Pr\{\textbf{O}|\textbf{P}\}d\textbf{p}
\end{equation}
Here $f_{p}(\textbf{P})$ is the ``prior'' probability distribution function, and $Pr\{\textbf{O}|\textbf{P}\}$ is the likelihood function. $A$ is a constant that is adjusted so that $f(\textbf{P}|\textbf{O})$ normalizes correctly to unity. As we adopt a fixed model library, the probability density distribution of the models in the space of physical parameters defines the prior $f_{p}(\textbf{P})$.\\
The likelihood function is typically assumed $\propto exp(-\chi^2/2)$, where $\chi^2$ is defined on the set of spectral features as follows:
\begin{equation}\label{eq:chi2}
    \chi^2=\sum_{i}\bigg(\frac{obs_{i}-\overline{obs_{i}}}{\sigma_{i}}\bigg)^2
\end{equation}
where $obs_{i}$ and $\overline{obs_{i}}$ are spectral features from data and from models respectively, and $\sigma_{i}$ is the corresponding observational error. This expression for the likelihood function is rigorously justified only in case of gaussian and uncorrelated errors on the observational quantities. In most real cases, this assumption is considered a good approximation. \\
The posterior PDF of a derived parameter $Y$ can be obtained from the full posterior PDF $f(\textbf{P}|\textbf{O})d\textbf{P}$ through the so-called ``marginalization'':
\begin{equation}\label{eq:pdf}
    f(Y|\textbf{O})dY=\int_{-dY/2}^{+dY/2}f(\textbf{P}|\textbf{O})d\textbf{P}
\end{equation}
where the integral extends over all $\textbf{P}$ for which $Y$ lies in a specified bin $Y\pm dY/2$. 

This approach is very robust against degeneracy, as the PDF takes into account all models in the library. The width of the posterior probability distribution can be used to quantify the uncertainties of the considered parameter. 

\subsubsection{The spectral library}\label{subsec:speclibrary}
As the first step to implement the Bayesian statistical analysis described above we have to create a model library. We rely on the SEDlibrary package, a set of programs written in C language by Stefano Zibetti, that allows to create spectral libraries starting from three ingredients: SFHs (SFR and metallicity parameterization), dust proprieties and an SSP grid for different ages and metallicities \citep{zibetti2017}. \\

The SSPs are formed by combination of stellar spectra (MILES stellar library, \cite{sanchez2006}, \cite{Falcon2011}) produced by the stellar population synthesis code BC03 of \cite{bruzual_charlot2003} in its 2016 version \citep[][]{cheva2016}. As initial mass function is concerned, we assume the IMF parameterization by \cite{chabrier2003}.

\begin{figure}
    \centering
    \includegraphics[scale=0.7]{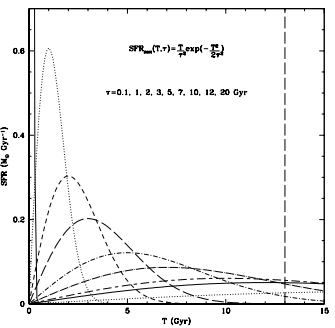}
    \caption{Sandage parameterization. From \cite{gavazzi2002}}
    \label{fig:sandageSFH}
\end{figure}

For the star formation history we assume a continuum component with at most six superposed random bursts.\\ 
For the continuous component of the SFH we adopt the parameterization introduced by \cite{sandage1986}: $SFR\propto \frac{t}{\tau^2}exp(t^2/2\tau^2)$). This parametrization has the advantage of modeling a broad range of shapes and timescales with two parameters ($\tau$ and $t_{form}$). SFHs of this form can be peaked in the past with a rapid decline, present slow raising phases followed by a decline, or even display a continued raising trend.\\
As shown in Fig. \ref{fig:sandageSFH}, $\tau$ determines the position of the SFR peak and is related to the slope of $SFR(t)$ before the peak. $t_{form}$ sets the origin of star formation activity, thus allowing for a span final ages. 

During a galaxy's SFH random events can occur (e.g. galaxy mergers, stochastic gas accretion) that increase the SFR over a short time interval (i.e. star bursts). This stochastic component is included in our parameterization, supposing that in a single SFH at most 6 star bursts can occur, each one contributing to the stellar mass and chemical composition. Bursts are parameterized via their age, metallicity and fraction of stellar mass contributed relative to the continuum component.\\
The evolution of the stellar metallicity for the continuous component of SFH is parameterized as a function of formed mass fraction $M(t)/M_{final}$ as follows:
\begin{equation}
    Z(t)=Z_{final}-(Z_{final}-Z_{0})(1-M(t)/M_{final})^{\alpha}, \alpha \geq 0
\end{equation}
Here $M(t)$ and $M_{final}$ are the stellar mass formed up to time $t$ and the stellar mass formed at the end of SFH respectively. $Z_{final}$ and $Z_{0}$ are the stellar metallicity at the end and at the beginning of SFH. The parameter $\alpha$, which is randomly generated between 0 and 1, defines how rapidly the final metallicity is reached. In particular, for small values we have delayed transitions to $Z_{final}$, while for large values, metallicity starts growing earlier towards the final value. The initial and final metallicity are chosen from a logarithmically uniform distribution in the interval 1/50-2.5$Z_{\odot}$.

We know that stellar spectra can be affected by dust attenuation, which can be related to the presence of the interstellar medium (ISM) or to the presence of dense clouds in which stars form (i.e. birth cloud). The recipe adopted to take into account this effect is the same contained in the work of \cite{CharlotFall2000}: all stellar populations are affected by the contribution of the ISM attenuation, which is $\propto \lambda^{-0.7}$; stellar populations younger than $10^7$ yr are affected also by the birth cloud attenuation, which is $\propto \lambda^{-1.3}$.

\subsubsection{Estimating stellar population parameters from galaxy spectra: the BaStA algorithm}\label{subsec:BaStA}
In order to get information from an observed galaxy spectrum via the Bayesian statistics described in Sec. \ref{subsec:bayesian_stats}, we have to compare it with our library spectra based on a selected set of spectral features, using the \textbf{Ba}yesian \textbf{St}ellar population \textbf{A}nalysis (BaStA) algorithm developed by A. Gallazzi and S. Zibetti. We thus select a set of five spectral absorption indices, following the prescriptions of \cite{gallazzi2005}, complemented by five broadband photometric fluxes as in \cite{zibetti2017}. The five spectral indices are: D4000n, [H$\delta_A$+H$\gamma_A$], H$\beta$, [Mg$_{2}$Fe] and [MgFe]$^\prime$ (see appendix B for detailed definition of the passbands). The five photometric bands in which fluxes are measured are the standard $ugriz$ bands of the SDSS \citep[][]{york2000, gunn1998}.\\
D4000n is defined as the ratio between the average flux densities in the narrow bands 4000\AA-4100\AA and 3850\AA-3950\AA. The H$\beta$ index measures the relative intensity of the Balmer line H$\beta$(4861\AA). H$\delta_A$+H$\gamma_A$ represents the sum of the two Balmer indices for the H$\delta_A$(4102\AA) and H$\gamma_A$(4341\AA). The use of the composite index H$\delta_A$+H$\gamma_A$ is preferred to the individual indices because it was found to be better reproduced by the models \citep[][]{gallazzi2005}. These Balmer line indices are quantified via their ``equivalent width'' ($w$), which is defined by the following equation:
\begin{equation}\label{eq:equivalent_width}
    w=\int_{line}\frac{I_{cont}-I_{\lambda}}{I_{cont}}d\lambda
\end{equation}
where $I_{cont}$ is the intensity of the (pseudo)continuum and $I_{\lambda}$ represents the intensity across the entire wavelength of interest. To get the quantity $w$ we have to define three regions: a central band, in which the spectral feature is located, and two side band from which $I_{cont}$ is obtained.\\
These three spectral indices are (mostly) age-sensitive. However, because of the age-metallicity degeneracy the best constraints on the SFH can be obtained only if we simultaneously constrain the metallicity. For this reason we include two more composite indices, which are (mostly) metal-sensitive and show a small dependence on $\alpha$-element abundance relative to iron-peak elements \citep{gallazzi2005, thomas2003}: [Mg$_{2}$Fe] and [MgFe]$^\prime$. They are defined as a combination of individual indices:
\begin{equation}\label{eq:mg2fe}
    [Mg_{2}Fe]=0.6Mg_{2}+0.4log(Fe4531+Fe5015)
\end{equation}
\begin{equation}\label{eq:mgfep}
    [MgFe]^\prime=\sqrt{Mgb(0.72Fe5270+0.28Fe5335)}   
\end{equation}
$Mg_{2}, Fe4531, Fe5015, Mgb, Fe5270$ and $Fe5335$ are defined through the equivalent width of the absorption features caused by (blends of) line transitions mainly due to Mg and Fe.\\
The five SDSS photometric fluxes in $ugriz$ are measured using the AB magnitude system. The monochromatic AB magnitude ($m_{AB}$) is defined as follows:
\begin{equation}
    m_{AB}=-2.5log_{10}\bigg(\frac{f_{\nu}}{3631 Jy}\bigg)
\end{equation}
where 1$Jy=10^{-23}erg\cdot s^{-1}\cdot Hz^{-1}\cdot cm^{-2}$ and the quantity $f_{\nu}$ represents the flux density.
In order to compare the models to the observations, we compute broad-band AB magnitudes following the prescriptions of \cite{fukugita1996}:
\begin{equation}
    m_{AB,bb}=-2.5log\frac{\int d(log\nu)f_{\nu}S_{\nu}}{\int d(log\nu)S_{\nu}}
\end{equation}
where $S_{\nu}$ corresponds to the system quantum efficiency, that includes atmospheric absorption.

The intrinsic spectrum of a galaxy is redshifted as a consequence of cosmological expansion and convolved with the line-of-sight velocity distribution of its stars, due to classical Doppler shift.
The redshift $z$ impacts on the broadband fluxes that are taken in shifted spectral windows with respect to the nominal restframe filter passbands.
The line-of-sight velocity dispersion ($\sigma_{v}$) causes spectral features to be broadened.
Therefore, in order to make a proper comparison between the observed spectral properties and the models, one needs process the model spectra to bring them in the same observational conditions as the observed ones. Specifically, we have to adjust the model spectra with the measured value of $z$ and of $\sigma_{v}$. \\

The CSP models are computed for $1\,M_\odot$ of present-day stellar mass. Therefore, in order to compare their luminosity density with the luminosity density of real galaxy spectra we need to rescale the models by a factor $M_*$, which corresponds to the present-day stellar mass of the galaxy, under the assumption that its stellar population is represented by the model CSP.
For any given models we thus evaluate the best estimate of $M_{*}$ by minimizing the $\chi^2$ in this equation:
\begin{equation}
    \chi^2=\sum_{j}\bigg(\frac{f_{obs,j}-M_{*}f_{mod,j}}{\sigma_{j}}\bigg)^2
\end{equation}
Here $j$ is the index running over all the photometric bands, $\sigma_{j}$ is a typical error from an astronomical observation, while $f_{obs,j}$ and $f_{mod,j}$ are the observed luminosity densities and the model luminosity densities, respectively. When $M_{star}$ is found for each model, Bayesian statistical analysis (section 2.3.1) can be made finding the ``posterior'' probability distribution of one (or more) parameter of interest. In order to obtain the PDF by following equation \ref{eq:pdf}, we have to divide the considered parameter $Y$ into different bin, $\pm dY/2$, and to sum the likelihood of the models in each bin. Once the sum in each bin is done, we normalize this distribution obtaining the PDF of the parameter $Y$. One could think of PDF as the sum of the weighted models for its likelihood-function.\\
In order to give a good estimate of a parameter $Y$, we calculate the median value of the posterior distribution. Regarding the error associated to $Y$, we calculate the width of its PDF, i.e. the difference between the 16$^{th}$ and 84$^{th}$ percentile.

\clearpage
\section{Theoretical limits to the estimate of SFH duration}\label{chap:time_resolution}
We focused the previous section on the approach used to characterize SFH of galaxies and on the method by which we intend to determine the duration of star formation activity (i.e. $\Delta age_{n}$).\\
In Sec. 3.1, we describe typical errors in spectroscopic and photometric measurements, and in Sec. 3.2, we aim at finding the theoretical limit for the minimum $\Delta age_{n}$ for which an extended SFH can be distinguished from one of negligible duration. For a given set of observational constraints at given signal-to-noise ratio, this limit represents the ``time resolution'', $\Delta age_{n,min}$, that we can achieve. \\

\subsection{Errors in spectroscopic and photometric measurements}\label{errors}
The goal of this section is to simulate an astronomical observation. Given a CSP model, we aim at perturbing the spectral features, used in this work: $D4000n$, $H\delta_A+H\gamma_A$, $H\beta$, $Mg_2Fe$, $[MgFe]^\prime$ and the five photometric fluxes in $ugriz$. First of all we have to investigate the different error sources for a galaxy spectrum, and then we describe more in detail the errors assumed. \\

In an astronomical observation, photons from the galaxy and from the background (mainly atmosphere glow and scattered light) reach the CCD at the same time. The CCD electronics also contribute to the noise mainly due to the so-called read-out noise. In order to measure the galaxy spectrum, we have to remove the offset of the background signal.
There are two main contributions to the noise in a spectrum: the background noise (due to the fluctuations in the background signal) and the Poisson noise from the source itself. In typical deep spectroscopic observations, the noise is dominated by the background component. For our simulations we assume that this is the case and, furthermore, that the noise level is constant as a function of wavelength. We neglect systematic effects connected to calibration processing. Therefore the noise spectrum associated to a source is simply given by $\sigma(\lambda)=constant$.\\

The way we simulate the set of observations we consider in our analysis is as follows. We consider a model spectrum, which we assume is observed with a given signal-to-noise ratio (SNR) per restframe Angstrom in a given reference wavelength range. From the SNR we compute the noise spectrum, i.e. $err$, which we assume constant across the full wavelength range:
\begin{equation}
    err=\frac{\int_{\lambda_{0}}^{\lambda_{1}}f_{\lambda}d\lambda}{<SNR>\Delta\lambda}
\end{equation} 
Here $\Delta\lambda$ is the considered wavelength interval, and $f_\lambda$ is the flux density per unit of wavelength.
The spectral indices are measured on the model spectrum and the corresponding errors are computed via standard error propagation, assuming that the error in each spectral pixel is uncorrelated to the others. The "observed" index is obtained by perturbing the model index by a random quantity drawn from a Gaussian distribution with sigma given by the error.
Concerning the photometric fluxes, we assume an error equal to 0.03 mag for $g,~r,~i$, and 0.05 mag for $u,~z$, which are the typical accuracies that can be reached in the photometric calibration in these bands. Also in this case, we obtain the ``observed'' flux by perturbing the model flux by a random quantity drawn from a Gaussian distribution with sigma given by the error.\\
Contrary to the other indices, which are defined with two side bands and hence are very little sensitive to flux calibration errors, for the D4000n index (defined as the flux ration in two bands) the accuracy in flux calibration is the main limiting factor. Therefore we correct the error on $D4000n$ derived from the noise spectrum with an extra term added in quadrature, as follows:
\begin{equation}
    err_{D4000n}=\sqrt{err^2+FCE^2}
\end{equation}
where FCE (Flux Calibration Error) is a fixed relative flux uncertainty between the blue and red bands of the index, which we choose as 0.03 based on our previous experience with spectroscopic surveys.

\subsection{Computing time resolution on SFH duration}
In this experiment we want to see how changing $\Delta age_n$ affects galaxy spectra. In particular, we aim at finding the theoretical limits, within which the SFH duration can be changed without affecting the set of spectral features used to obtain PDF of physical parameters.
To do that we fix all the other parameters than $\Delta age_n$ (metallicity, dust, mean stellar age) and consider only simple star formation history a la Sandage (without star bursts).
In fact, we use a library of idealized CSP models, in which we assume: fixed metallicity; no random starburst but just the continuum Sandage component; null dust attenuation. We consider five subsets of models with fixed metallicity ($2\cdot 10^{-2}Z_{\odot}, 2\cdot 10^{-1}Z_{\odot}, 4\cdot 10^{-1}Z_{\odot}, 1 Z_{\odot}, 2.5 Z_{\odot} $) and for each of them we realize 1 million models with SFH a la Sandage with randomly generated $t_{\rm{form}}$ and $\tau$.
This library is not appropriate for representing the complexity of real galaxies, but for the aim of this section we want to find the theoretical limit of time resolution that can be achieved in the simplest scenario, leaving aside all possible degeneracies linked to (variable) metallicity, random bursts and dust. We will test the performance of our method on more realistic models in the next chapter.
\\

\begin{figure}
    \centering
    \includegraphics[width=\textwidth]{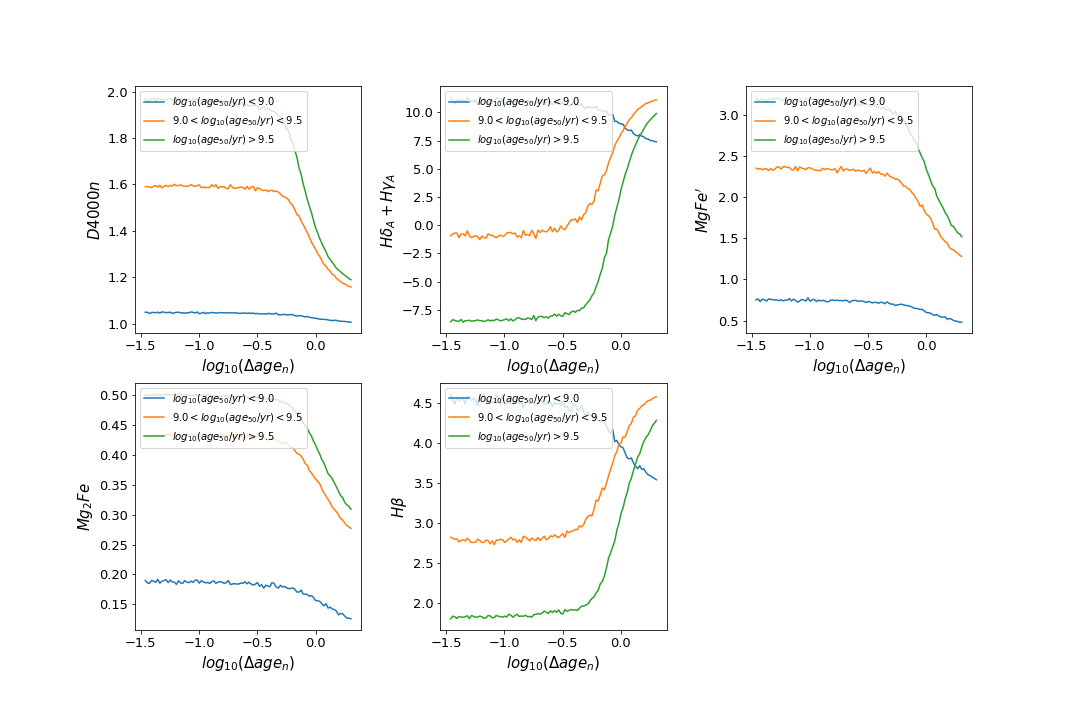}
    \caption{Relation between perturbed (SNR=20) spectral indices and $log_{10}(\Delta age_{n})$ for solar metallicity. The three different painted lines represent the indices median as function of $log_{10}(\Delta age_{n})$ for different range of $age_{50}$. The green line, the orange one and the blue one are characterized by $log_{10}(age_{50})$ lower than 9.0, between 9.0 and 9.5, and higher than 9.5 respectively.}
    \label{fig:idx_tres}
\end{figure}
\begin{figure}
    \centering
    \includegraphics[width=\textwidth]{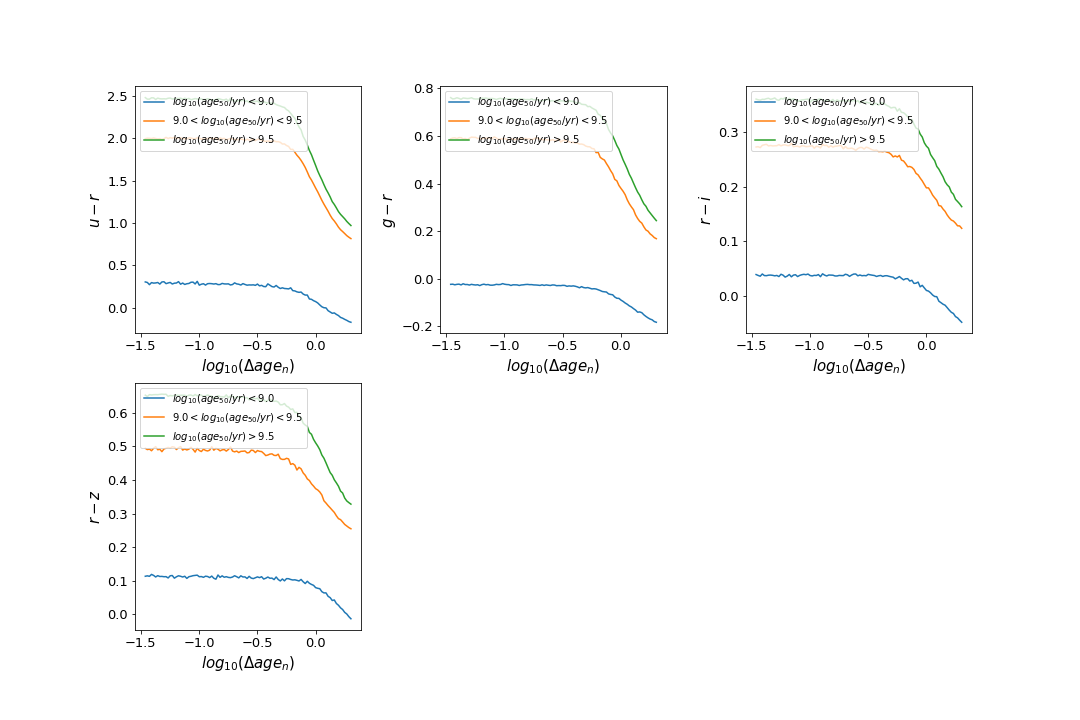}
    \caption{Relation between perturbed colours and $log_{10}(\Delta age_{n})$ for solar metallicity. The three different painted lines represent the indices median as function of $log_{10}(\Delta age_{n})$ for different range of $age_{50}$. The green line, the orange one and the blue one are characterized by $log_{10}(age_{50})$ lower than 9.0, between 9.0 and 9.5, and higher than 9.5 respectively.}
    \label{fig:col_tres}
\end{figure}
As in the previous chapter, we consider the five spectral indices ($D4000n$, $H\delta_A+H\gamma_A$, $H\beta$, $[Mg_{2}Fe]$, $[MgFe]^\prime$) and the photometric information as well. As far as photometry is concerned, we must consider that the overall normalization is a free parameter ($M_*$), hence, out of the five magnitudes, only four are independent. For this reason we opt to use the four independent colours that can be derived from the five bands: $u-r,~g-r,~r-i,~r-z$.\\

The library covers a wide range of $\Delta age_n$, from $\sim$0.04 up to $\sim$1.6. For the lowest values of $\Delta age_n<10^{-1}$, models are characterized by a star-formation-history very similar to an instantaneous star burst, so that the duration of SFH is not resolved. We aim at observing how model spectra (and spectral features) change increasing $\Delta age_n$, and how they differ from those with a SFH of negligible duration.\\
Our set of spectral features depends on $age_{50}$ to the first order, and secondarily on $\Delta age_n$. If we consider the full $age_{50}$ range, the scatter of the spectral features is too broad to appreciate their dependence on $\Delta age_n$.
So, in order to underline the dependence of the spectral features on $\Delta age_n$, we sample the models in three $log_{10}(age_{50})$ bins: $log_{10}(age_{50})<$9.0; 9.0$<log_{10}(age_{50})<$9.5; $log_{10}(age_{50})>$9.5. Then we plot spectral features (indices and colors) as function of $\Delta age_n$. We observe that the points of each spectral feature show an initially flat trend, and from a critical value on, we can see a deviation.\\
In order to appreciate better the trend of each spectral feature, we compute the running median of the spectral features in the three $age_{50}$ bins, and in figures \ref{fig:idx_tres} and \ref{fig:col_tres} we show the median value of each index and color as function of $log_{10}(\Delta age_n)$.
In each panel (one for each individual spectral  feature) we can see three different colored lines, representing the running-median of the considered feature for the three different $log_{10}(age_{50})$ range. We observe that below $log_{10}(\Delta age_n)\sim$-0.5 (i.e. $\Delta age_{n} \sim$0.3) all indices and colours display no dependence on the duration if the SFH. At larger values we start to see significant deviations. This means that we can not appreciate differences in spectra of model with a SFH duration (i.e. $\Delta age_{10,90}$) lower than $\sim$ 1/3 of $age_{50}$.\\
Figures \ref{fig:idx_tres} and \ref{fig:col_tres} indicate that it is possible to define a range of $\Delta age_{n}$, common to all spectral features, where the SFH duration is by no means distinguishable from an instantaneous burst. We define $log_{10}(\Delta age_n)<-1.0$ as the reference range against which we compare our models and establish whether or not they are significantly different from an instantaneous burst.\\
In order to be quantitative and include all information carried by the different spectral features\footnote{This is analogous to the 
definition of $\chi^2$ (Equation \ref{eq:chi2}) for the likelihood.} we introduce the following parameter:
\begin{equation}\label{eq:delta}
    \delta^j=\sum_{i}\bigg(\frac{obs_{i}^{j}-obs_{i,\rm{ref}}}{\sigma_{i}}\bigg)^2
\end{equation}
where the index $i$ runs over our observable set, while $j$ indicates the model number. Each spectral feature is associated with a fixed reference value, $obs_{i,ref}$, from which the difference has to be quantified. As $obs_{i,ref}$ we simply choose the mean of $obs_{i}^{j}$ in the reference range. Furthermore, each difference has to be ``weighted'' on the scatter,  $\sigma_i$, which is specific to each spectral and photometric feature. This allows us to quantify the significance of the deviations from the reference values. $\sigma_i$ has to take into account both the intrinsic scatter of the feature in the reference range due to the different SFHs and the observational scatter due to the noise. For this analysis, we perturbed spectral features of our models with the expected errors of astronomical observations at different SNR (Sec. \ref{errors}). For each feature $i$, $\sigma_i$ is simply given by the standard deviation of $obs_{i}^{j}$ in the reference range.\\

In order to minimize the contribution to the intrinsic scatter from sources other than the \emph{shape} of the SFH, in addition to having dust and metallicity fixed, we analyze models in small $log_{10}(age_{50})$ bins of 0.05 dex, in a range from 6.0 to 10.15.\\

We consider that $\delta$ start depending on $\Delta age_{n}$, when the running median is away two times the standard deviation from the mean value of the reference level. So, in order to find $\Delta age_{n,min}$, we want to find the intersection point between $\delta$ running 16$^{th}$ percentile and $\delta=\delta_{mean,ref}+\delta_{stdev,ref}$, where $\delta_{mean,ref}$ is the mean value of $\delta$ in the reference level and $\delta_{stdev,ref}$ is the standard deviation of the reference level. From then on, we will define time resolution as $\Delta age_n$ corresponding to the intersection point.
\begin{figure}
    \centering
    \includegraphics[width=\textwidth]{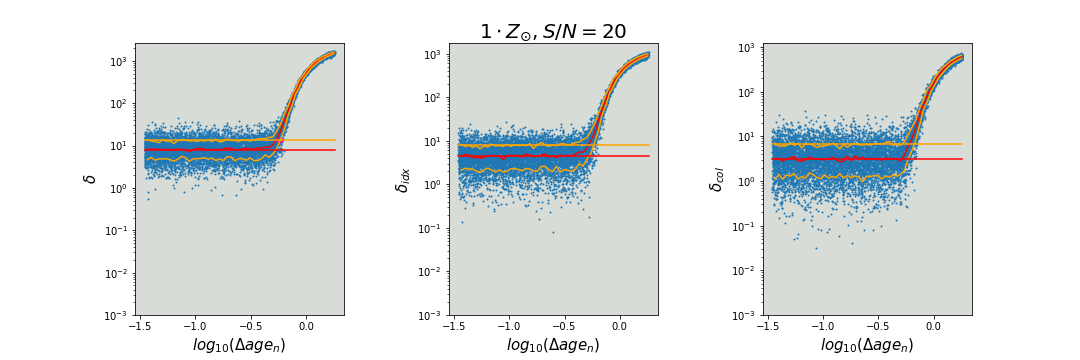}
    \caption{$\delta$ as function of $log_{10}(\Delta age_{n})$ considering the sample of models characterized by SNR=20, $log_{10}(age_{50})$ range between 9.7 and 9.75 and metallicity $Z=Z_\odot$. The orange and the red horizontal lines represent the standard deviation and the mean value of $\delta$ for the models below the reference level (i.e. $log_{10}(\Delta age_{n})<$-1.0). The two noisy orange lines are the ``running'' percentile 16 and 84, while the noisy red one is the ``running'' median.}
    \label{fig:97975delta}
\end{figure}

\begin{figure}
    \centering
    \hspace*{-3cm}
    \includegraphics[width=1.5\textwidth]{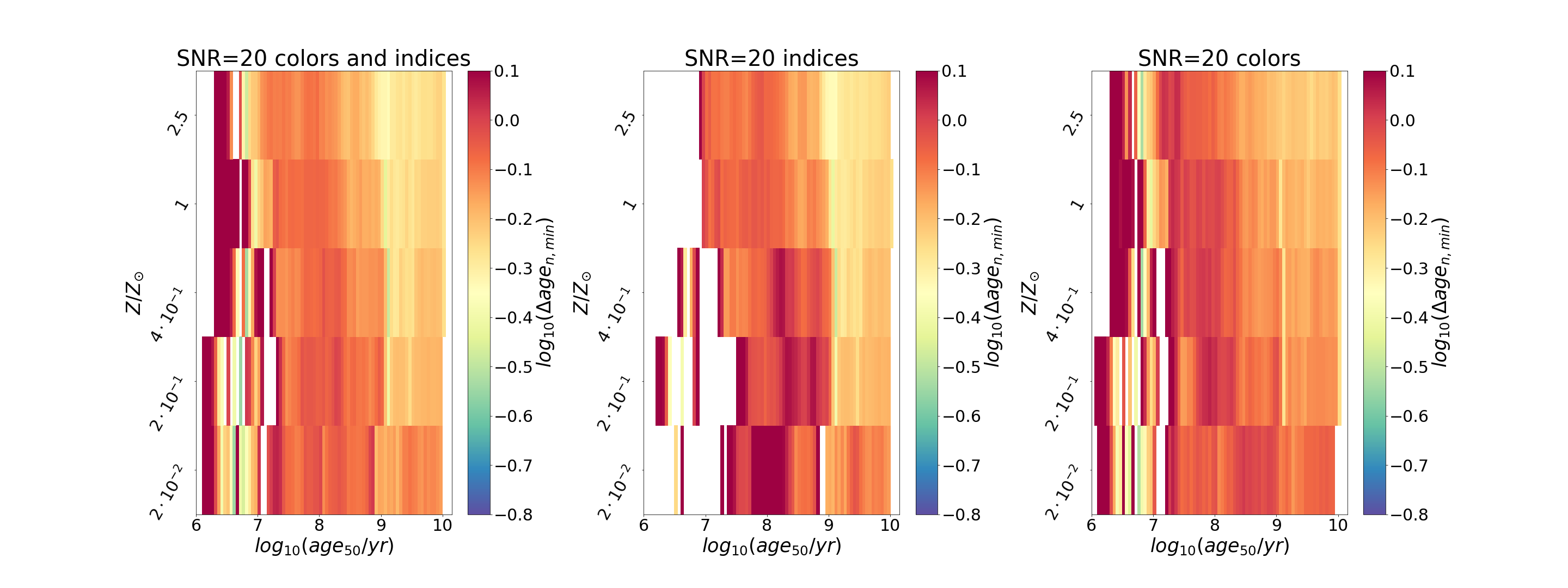}
    \caption{Time resolution map at fixed SNR=20. Along the horizontal axis $log_{10}(age_{50})$ is represented, while along the vertical axis stellar metallicity, i.e. $Z$, increases from bottom to top. The five horizontal bands represent the five fixed $Z$: from bottom to top we have metallicity 32, 42, 52, 62, 72 (i.e. $2\cdot 10^{-2}Z_{\odot}, 2\cdot 10^{-1}Z_{\odot}, 4\cdot 10^{-1Z_{\odot}}, 1 Z_{\odot}, 2.5 Z_{\odot} $). The bin-size on the horizontal axis is 0.05 dex}
    \label{fig:tres_map}
\end{figure}

The three panels in Figure \ref{fig:97975delta} display the typical behaviour of $\delta$ as function of $log_{10}(\Delta age_{n})$, specifically for the sample of models with metallicity $Z=Z_{\odot}$, perturbed with a signal to noise ratio SNR=20 and characterized by 9.7$<log_{10}(age_{50})<$9.75. From right to left we have $\delta$ computed by considering only colours, considering only spectral indices and taking into account indices and colors. In each panel we can observe two horizontal lines, which represent the mean and the standard deviation of the reference level, in red and orange respectively, and three lines displaying the running 16$^{th}$ percentile and the running 84$^{th}$ percentile (in orange), and the running median of $\delta$ (in red). In the left panel we can observe the intersection point, given by the mean plus the standard deviation of the reference level and the running 16$^{th}$ percentile. In this case, we have a sample of models, whose spectral features (indices and color) start depending on SFH duration, when $log_{10}(\Delta age_n)$ is $\sim$-0.25 dex as the first panel from left shows. For this metallicity ($Z=Z_{\odot}$) and this median stellar age (9.7$<log_{10}(age_{50}<$9.75), at SNR=20, we obtain a very similar resolution, by considering only colours (first panel from right) or only indices (middle panel). However, there are age bins in which colours and indices individually provide different time resolution.\\

\begin{figure}
    \centering
    \includegraphics[width=\textwidth]{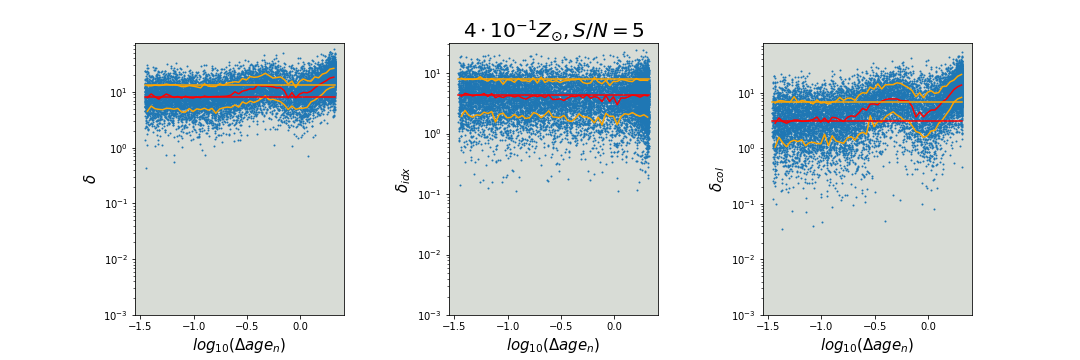}
    \caption{$\delta$ as function of $log_{10}(\Delta age_{n})$ considering the sample of models characterized by SNR=5, $log_{10}(age_{50})$ range between 7.0 and 7.05 and metallicity $4\cdot 10^{-1}Z_\odot$. The orange and the red horizontal lines represent the standard deviation and the mean value of $\delta$ for the models below the reference level (i.e. $log_{10}(\Delta age_{n})<$-1.0). The two noisy orange lines are the ``running'' percentile 16 and 84, while the noisy red one is the ``running'' median.}
    \label{fig:7705delta}
\end{figure}

\begin{figure}
    \centering
    \hspace*{-3cm}
    \includegraphics[width=1.5\textwidth]{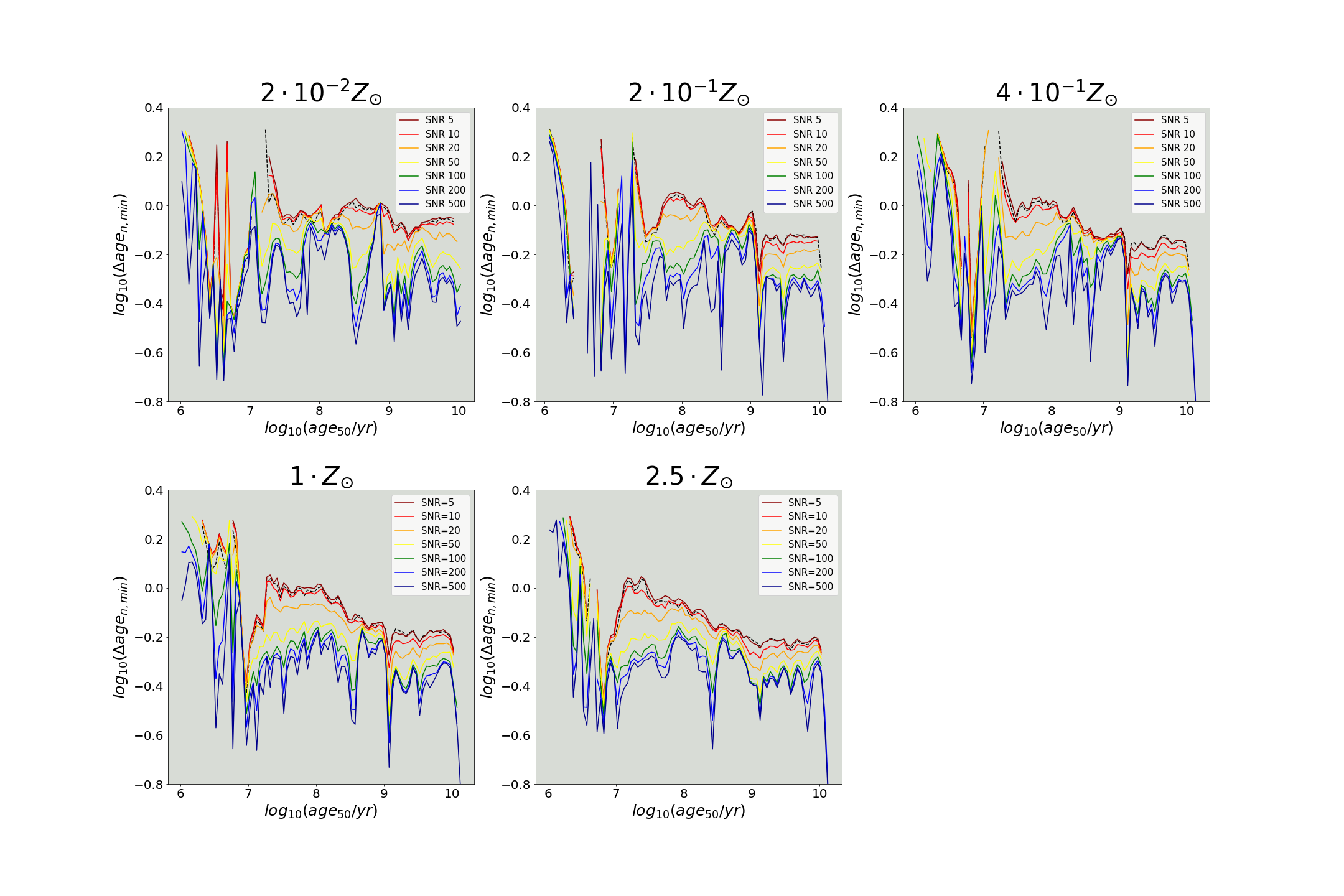}
    \caption{Time resolution as function of $log_{10}(age_{50})$ for different signal to noise ratio at fixed stellar metallicity. Different painted lines represent time resolution at different SNR values. The dashed black line is the time resolution given only by considering the colours without spectral indices.}
    \label{fig:tres}
\end{figure}
Once we find the limiting time resolution for each sample of models, we can study its dependence on stellar metallicity and median stellar age $age_{50}$. The results are shown in figure \ref{fig:tres_map}: the three panels (from left to right) represent $log_{10}(\Delta age_{n,min})$, at fixed SNR=20 considering spectral indices and colours together, spectral indices only and colours only. For $log_{10}(age_{50})<$7.5, the $\Delta age_{n,min}$ is largely determined by the photometry, while for $log_{10}(age_{50})>$9.0 $\Delta age_{n,min}$ is driven by spectral indices. \\
In the map some null ``pixels'' are present. These represent a few rare $age_{50}$ bins, in which the dependence of $\delta$ on $\Delta age_n$ is flat or not well behaved, so that it is impossible to extract information on $\Delta age_n$ from the spectra. Figure \ref{fig:7705delta} illustrates this problem for the models with $log_{10}(age_{50})$ between 7.0 and 7.05. There is no intersection point between the running 16$^{th}$ percentile and the mean plus standard deviation of the reference level in the first panel from left. This means that for these values of $age_{50}$ and for this metallicity, we are not able to get information about the duration of SFH. \\

By increasing SNR, we expect to reach lower values of limiting time resolution. The maps analogous to figure \ref{fig:tres_map} for SNR=5 and SNR=100 are reported in appendix A. In figure \ref{fig:tres} each panel corresponds to a fixed metallicity, and each colored line displays $log_{10}(\Delta age_{n,min})$, as function of $log_{10}(age_{50})$ for a given signal-to-noise ratio. Furthermore, in each panel, a black dashed line is represented indicating the resolution given only by the colours (whose uncertainties are fixed). We can observe that for a low SNR (5 or 10) the resolution is completely driven by photometry, so, at those SNRs spectral indices provide negligible information as far as SFH duration is concerned. \\
We can observe also that $\Delta age_{n,min}$ reaches the lowest values for high signal-to-noise ratio, until this becomes equal to 100. From this value on, the time resolution for SFH duration remains more or less the same.\\
Another relevant fact is that $log_{10}(\Delta age_{n,min})$ is characterized by a remarkably small range, considering that $age_{50}$ covers a broad range of more of 4 orders of magnitude (from $10^6yr$ to $10^{10.15}yr$). So, we can say that the relative time resolution is roughly constant, and it covers a range between $\sim-0.6$ dex to $\sim0.2$ dex. Looking at the distribution in more detail, systematic trends emerge, and they can be better appreciated in figure \ref{fig:tres_map}. In particular, considering colors and indices (the first panel from left), we can note three different regions:  for $age_{50}<10^7yr$ we note an irregular shape of $\Delta age_n$; for $10^7yr<age_{50}<10^9yr$ we have $log_{10}(\Delta age_{n,min})\sim-0.1$; and for $age_{50}>10^9yr$ we reach $log_{10}(\Delta age_n)<-0.2$. 
\begin{figure}
    \centering
    \includegraphics[width=\textwidth]{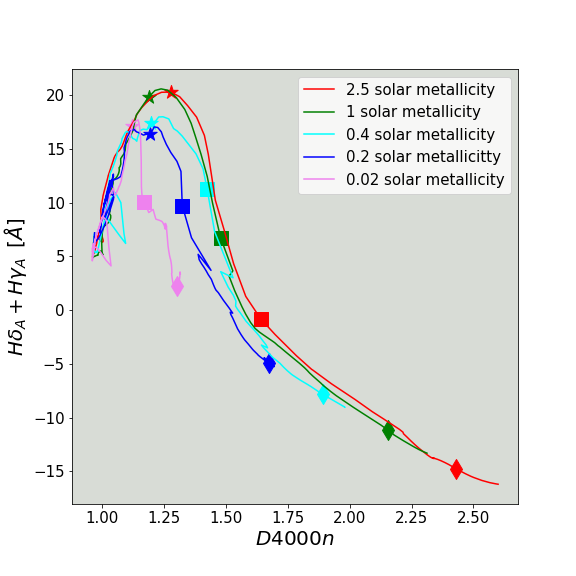}
    \caption{SSP tracks from BC03 (2016 version) for different stellar metallicity. The stars correspond to stellar age$\sim2.6\cdot 10^8yr$, the squares to stellar age$\sim10^9yr$, and the diamonds to stellar age$\sim10^{10}yr$.}
    \label{fig:ssp_traces}
\end{figure}
These different regions are related to different stellar evolutionary phases, and can be better understood by looking at the different SSP evolutionary tracks shown in figure \ref{fig:ssp_traces} for different fixed metallicity. Over the vertical axis, the $H\delta_A$+$H\gamma_A$ is shown, while over the horizontal one we have the 4000\AA-break, which increases as stellar age increases. We can note that each track is characterized by an initial and noisy evolution, typically for mean stellar age below $\sim10^7yr$. Then, the peak of the Balmer lines is reached for $age\sim2.6\cdot 10^8yr$. At this moment the stellar population is dominated by A-type stars, and spectra are characterized by the strong presence of Balmer lines. The maximum rate of variation in the strength of hydrogen lines is reached for stellar age $\sim10^9yr$. When the stellar age reaches higher values of $10^9yr$, Balmer lines become weak and evolve slowly, while other metal indices appear.\\
The three regions in the first panel from left of figure \ref{fig:tres_map} represent this: the ``irregular'' region for $age_{50}<10^7yr$; the Balmer-dominated region for $10^7yr<age_{50}<10^9yr$; the minimum value of time resolution for $age_{50}$ related to the maximum rate of variations of Balmer lines; and the region of $age_{50}>10^9yr$ when stars become colder and metal indices appear. Thanks to the presence of other metal indices we have additional constrains and low values of $\Delta age_{n,min}$ can be reached.\\

\begin{figure}
    \centering
    \title{9.5$<log_{10}(age_{50})<$9.55, Z=Z$_{\odot}$}
    \includegraphics[width=\textwidth]{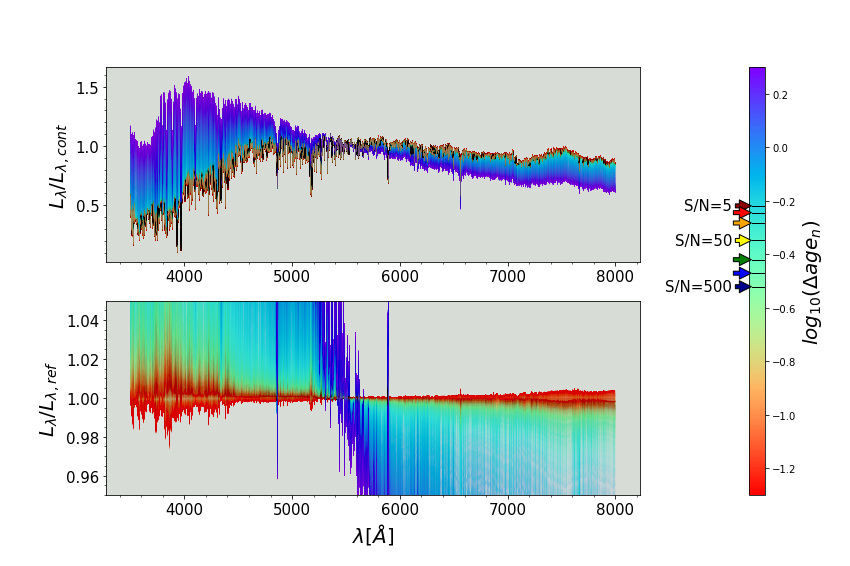}
    \caption{
    \emph{Top panel}: spectra distribution for the models characterized by $log_{10}(age_{50})$ between 9.5 and 9.55. The spectra are normalized between 5450\AA\, and 5550\AA and color-coded according to their $\Delta age_n$ as shown in the colorbar. \emph{Bottom panel}: ratio between each spectrum and the ``reference'' spectrum, defined as the spectrum of the model with the lowest $\delta$-value computed by perturbing spectral feature with SNR=500.
    The painted arrow next to the colorbar represent $log_{10}(\Delta age_{n,min})$ for different signal-to-noise ratio.  Dark red: SNR=5; red: SNR=10; orange: SNR=20; yellow: SNR=50; green: SNR=100; blue: SNR=200; dark blue: SNR=500}
    \label{fig:specm62_95}
\end{figure}
\begin{figure}
    \centering
    \title{9.5$<log_{10}(age_{50})<$9.55, Z=Z$_{\odot}$}
    \includegraphics[width=\textwidth]{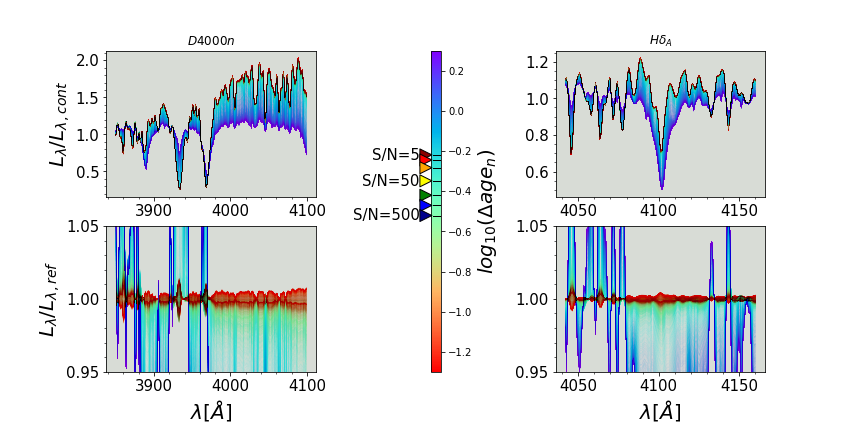}

    \caption{\emph{Top panels}: distributions of $D4000n$ and of $H\delta_A$ for the models characterized by $log_{10}(age_{50})$ between 9.5 and 9.55. For the $D4000n$, spectra are normalized on the range 3850\AA-3950\AA and for $H\delta_A$ the two side bands are used for normalization. Each spectrum is color-coded according to its $\Delta age_n$, as shown in the colorbar. \emph{Bottom panel}: ratio between each spectrum and the ``reference'' spectrum, defined as the spectrum of the model with the lowest $\delta$-value computed by perturbing spectral feature with SNR=500.
    The painted arrow next to the colorbar represent $log_{10}(\Delta age_{n,min})$ for different signal-to-noise ratio.  Dark red: SNR=5; red: SNR=10; orange: SNR=20; yellow: SNR=50; green: SNR=100; blue: SNR=200; dark blue: SNR=500}
    \label{fig:d4000nhdm62_95}
\end{figure}
\begin{figure}
    \centering
    \title{9.5$<log_{10}(age_{50})<$9.55, Z=Z$_{\odot}$}
    \includegraphics[width=\textwidth]{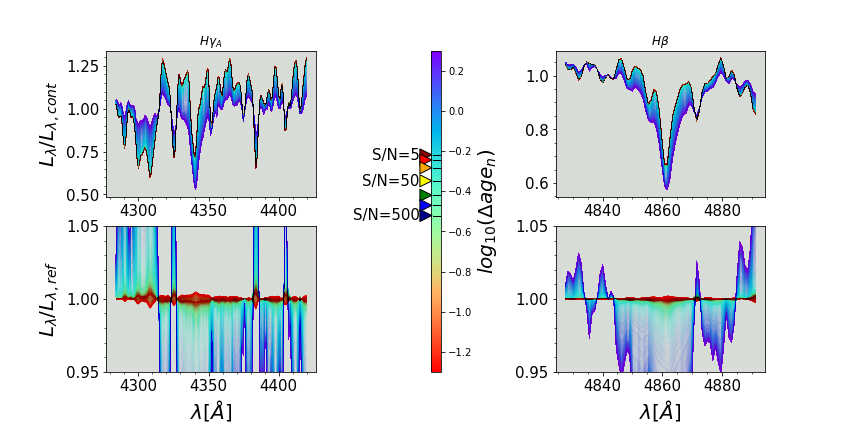}

    \caption{Same as figure \ref{fig:d4000nhdm62_95} but for $H\gamma_A$ and $H\beta$. In the top panels, spectra are normalized on the index side bands.}
    \label{fig:hghbm62_95}
\end{figure}
\begin{figure}
    \centering
    \title{9.5$<log_{10}(age_{50})<$9.55, Z=Z$_{\odot}$}
    \includegraphics[width=\textwidth]{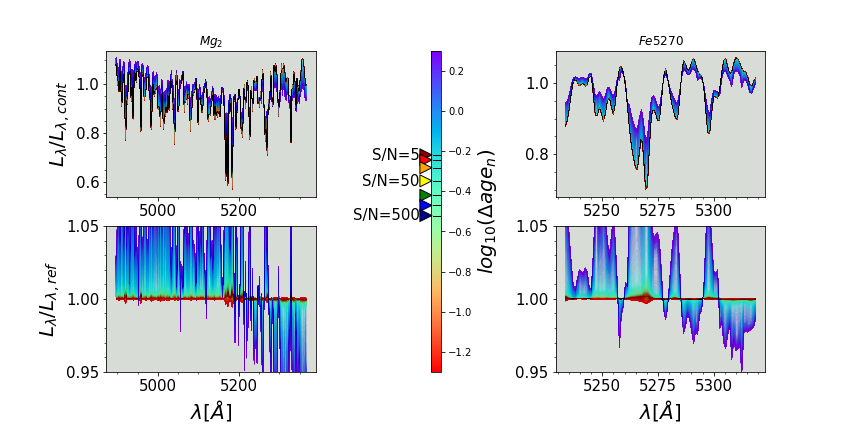}

    \caption{Same as figure \ref{fig:d4000nhdm62_95} but for $Mg_2$ and $Fe5270$. In the top panels, spectra are normalized on the index side bands.}
    \label{fig:mg2fe5270m62_95}
\end{figure}

\subsection{Effects of SFH duration on spectral shape and absorption features}
In the last section of this chapter we take a closer look at the effect of increasing $\Delta age_n$ on the spectra, both in terms as of overall spectral shape and in terms of strength of absorption features.\\
In the top panel of figure \ref{fig:specm62_95} we show how spectra of increasing $\Delta age_n$ differ from a reference spectrum of negligible duration. We focus on models characterized by solar metallicity and 9.5$<log_{10}(age_{50})<$9.55 representative of relatively old stellar populations. Each spectrum is normalized by the mean luminosity density in 5450\AA$<\lambda<$5550\AA, $L_{\lambda,cont}$, and plotted in a colour that represents its $\Delta age_n$, from small in red, to large in blue/violet.\\
We further define a reference spectrum as the spectrum of the model characterized by the lowest $\delta$-value computed for a SNR=500. The ratio between each spectrum and the reference spectrum is shown in the bottom panel. As for the top panel, each spectrum in the plot is color-coded by its $log_{10}(\Delta age_n)$, as indicated by the colorbar on the right side. The arrows beside the colorbar indicate the time resolution $log_{10}(\Delta age_{n,min})$ for different signal-to-noise-ratio. Arrow colors for different signal-to-noise ratio follow the same coding as in figure \ref{fig:tres}.\\
In the top panel we can observe that spectra with a higher $\Delta age_n$ are characterized by an increasing luminosity at shorter wavelengths. In fact, when the median stellar age is fixed, a longer SFH is obtained by adding both younger stars and older stars. The younger stars, however, are hotter and more luminous per unit mass, so the net effect on the composite spectrum is to make it bluer.\\
The bottom panel represents the ratio between the spectra and the reference spectrum, $L_{\lambda}/L_{\lambda,ref}$. We can observe that models with $log_{10}(\Delta age_n)$ lower than -1.0 dex show variations relative to the reference spectrum of $\sim2\%$ at most for wavelength lower than 4000\AA. For wavelength greater than 4000\AA, differences are $<$1\%. We reach fluctuations higher than $5\%$ for $log_{10}(\Delta age_n)>$-0.3 dex. Below the limiting time resolution, each spectrum is comparable to the reference spectrum, while for $\Delta age_n>\Delta age_{n,min}$ the systemic differences increases rapidly, as shown by the color code.\\
We further investigate the effect of $\Delta age_n$ on index strength by zooming in on the relevant absorption features (Fig. \ref{fig:d4000nhdm62_95}, \ref{fig:hghbm62_95}, and \ref{fig:mg2fe5270m62_95}). In figure \ref{fig:d4000nhdm62_95} we can observe the distribution of the spectra in the regions corresponding to $D4000n$ (left top panel) and of $H\delta_A$ (right top panel), while the corresponding differences with the reference spectrum are shown in the bottom panels. For $D4000n$, the flux is normalized between $3850\AA$ and $3950\AA$, i.e. the blue band of $D4000n$ index. We can see that increasing the duration of SFH, the intensity in the interval between $4000\AA$ and $4100\AA$ decreases, since young stellar populations cause a decrease of $D4000n$ value. Regarding the Balmer index $H\delta_A$, the normalization is made on the two side bands. In this case, by increasing $\Delta age_n$ the presence of the Balmer absorption line becomes more evident, as we can see also in the top panels of figure \ref{fig:hghbm62_95} regarding $H\gamma_A$ and $H\beta$. In order to better appreciate the variations between the spectra, we focus on the bottom panels of these two figures. In particular, we notice that fluctuations of $D4000n,~H\delta_A,~H\gamma_A~$and $H\beta$ for the models with $log_{10}(\Delta age_n)$ lower than -1.0 dex are within 1\% of the reference spectrum.\\
In order to get a good constrain on age and metallicity, in this project we included also two composite indices, $[Mg_{2}Fe]$ and $[MgFe]^\prime$, which are (mostly) metal-sensitive. These are defined in equations \ref{eq:mg2fe} and \ref{eq:mgfep}, and include three different $Fe$ and $Mg$ each. For illustration, we show the response to $\Delta age_n$ for $Mg_2$ and $Fe5270$. In figure \ref{fig:mg2fe5270m62_95}, we can observe the distributions of the spectra around these two indices (top panels), normalized on the two side bands, and and the ratio with the reference spectrum. With respect to the Balmer indices, the red line (i.e. models distribution with $log_{10}(\Delta age_n)<-1.0$ dex) is thinner and better centered around 1, because they are (mostly) metal-sensitive and the models of this library are characterized by a fixed metallicity. Hence their sensitivity to $\Delta age_n$ is more limited.\\
\begin{figure}
    \centering
    \title{8.5$<log_{10}(age_{50})<$8.55, Z=Z$_{\odot}$}
    \includegraphics[width=\textwidth]{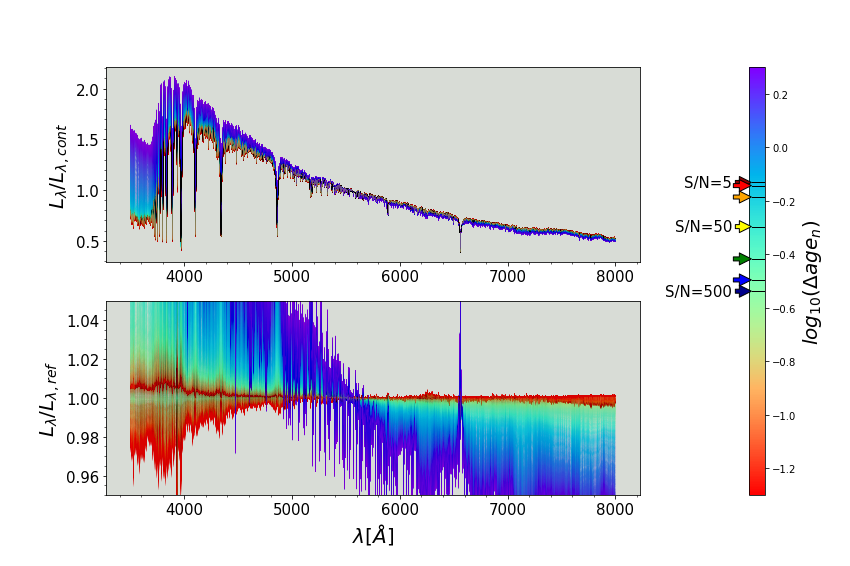}
    \caption{Same of figure \ref{fig:specm62_95}, but for 8.5$<age_{50}<$8.55 and $Z=Z_\odot$}
    \label{fig:specm62_85}
\end{figure}
\begin{figure}
    \centering
    \title{8.5$<log_{10}(age_{50})<$8.55, Z=Z$_{\odot}$}
    \includegraphics[width=\textwidth]{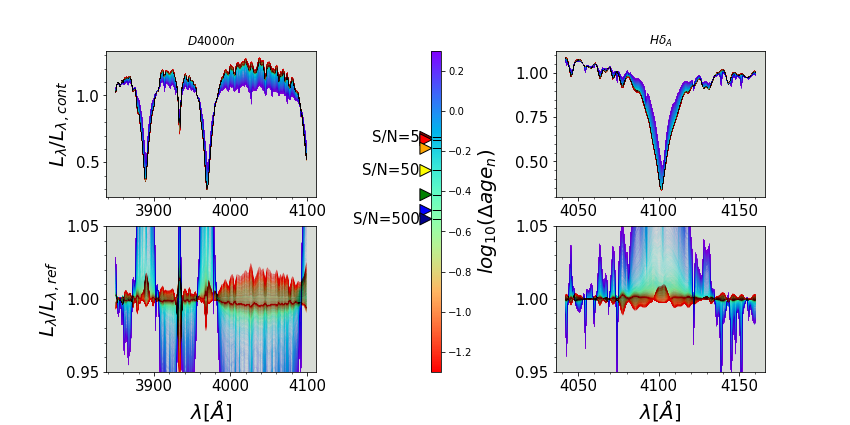}

    \caption{Same of figure \ref{fig:d4000nhdm62_95}, but for 8.5$<age_{50}<$8.55 and $Z=Z_\odot$}
    \label{fig:d4000nhdm62_85}
\end{figure}
\begin{figure}
    \centering
    \title{8.5$<log_{10}(age_{50})<$8.55, Z=Z$_{\odot}$}
    \includegraphics[width=\textwidth]{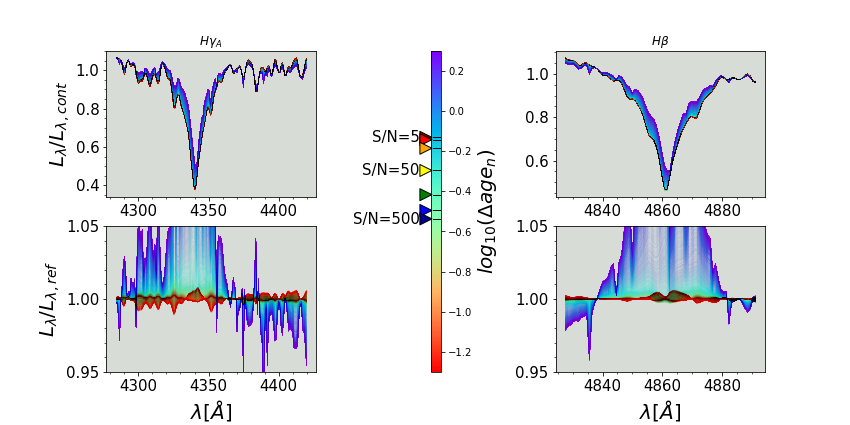}

    \caption{Same of figure \ref{fig:hghbm62_95}, but for 8.5$<age_{50}<$8.55 and $Z=Z_\odot$}
    \label{fig:hghbm62_85}
\end{figure}
\begin{figure}
    \centering
    \title{8.5$<log_{10}(age_{50})<$8.55, Z=Z$_{\odot}$}
    \includegraphics[width=\textwidth]{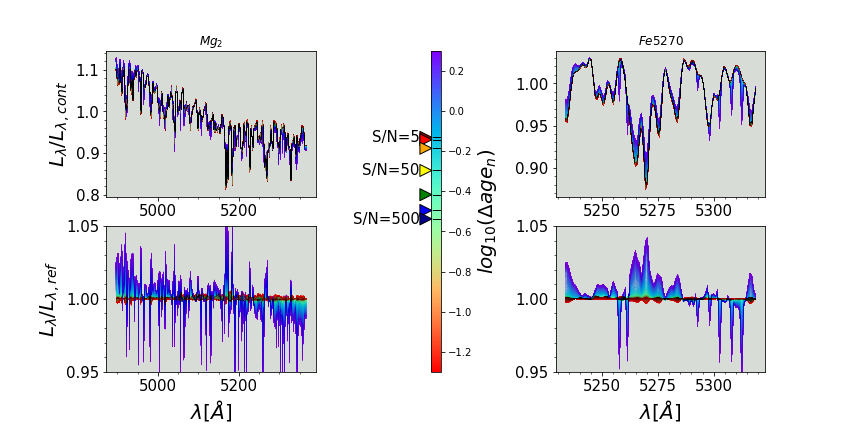}

    \caption{Same of figure \ref{fig:mg2fe5270m62_95}, but for 8.5$<age_{50}<$8.55 and $Z=Z_\odot$}
    \label{fig:mg2fe5270m62_85}
\end{figure}

As further example of the spectral response to $\Delta age_n$ in different $age_{50}$ regime, we consider the models with $log_{10}(age_{50})$ between 8.5 and 8.55 and solar metallicity, which is the young range where we observe a minimum of the time resolution. In this case time resolution got worse for SNR$<$50 than in the previous case. The top panel of figure \ref{fig:specm62_85} shows that the spectra distribution is narrower, and spectra are more similar to each other than in the previous case, as we can observe more in detail by comparing the bottom panels.\\
We note that the fluctuations from the reference spectrum are within 4\% for models characterized by $log_{10}(\Delta age_n)<$-1.0 dex in the blue region. For wavelength $>$ 5000\AA\, fluctuations in the reference $\Delta age_n$ regime become lower than 0.5\%. In fact, for $age_{50}$ between $10^{8.5}$ and $10^{8.55}$, spectra are dominated by Balmer lines, thanks to the strong presence of A-type stars. Looking at figures \ref{fig:d4000nhdm62_85} and \ref{fig:hghbm62_85}, we can see that Balmer lines are more intense than in the previous case, and we can note that fluctuations of $D4000n$, from the reference spectrum, are within 2\% for models with $log_{10}(\Delta age_n)<$-1.0 dex opposite to the 1\% of the previous case. For $~H\delta_A,~H\gamma_A,~H\beta$ fluctuations in the reference $\Delta age_n$ regime are within 1\% similar to the previous case.\\
Considering figure \ref{fig:mg2fe5270m62_85}, we can observe that, for this $age_{50}$ value, metal indices, as $Mg_2$ and $Fe5270$, are difficult to be noted. As for the fluctuations from the reference spectrum, they are within 5\% over the all range of $\Delta age_n$, because they do not vary when  $\Delta age_n$ changes.\\

\begin{figure}
    \centering
    \title{8.0$<log_{10}(age_{50})<$8.05, Z=Z$_{\odot}$}
    \includegraphics[width=\textwidth]{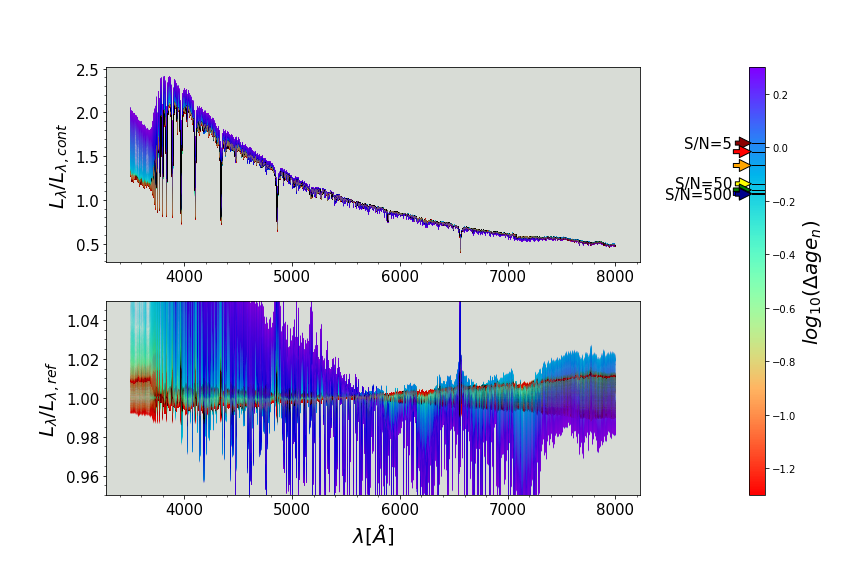}
    \caption{Same of figure \ref{fig:specm62_95}, but for 8.0$<age_{50}<$8.05 and $Z=Z_\odot$}
    \label{fig:specm62_8}
\end{figure}
\begin{figure}
    \centering
    \title{8.0$<log_{10}(age_{50})<$8.05, Z=Z$_{\odot}$}
    \includegraphics[width=\textwidth]{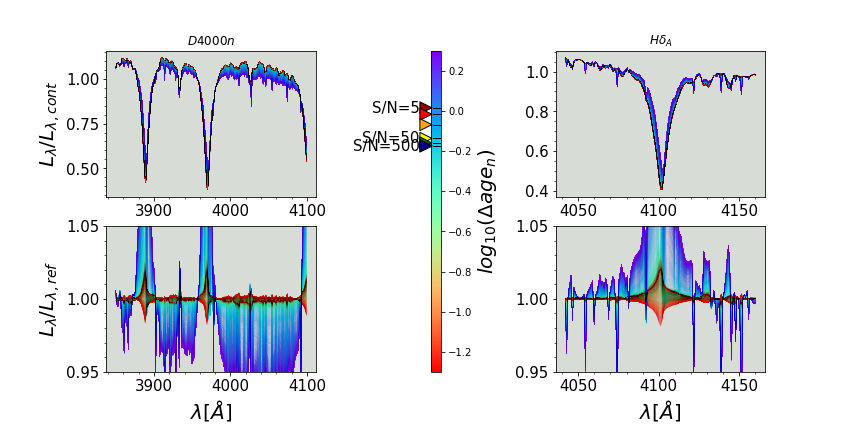}

    \caption{Same of figure \ref{fig:d4000nhdm62_95}, but for 8.0$<age_{50}<$8.05 and $Z=Z_\odot$}
    \label{fig:d4000nhdm62_8}
\end{figure}
\begin{figure}
    \centering
    \title{8.0$<log_{10}(age_{50})<$8.05, Z=Z$_{\odot}$}
    \includegraphics[width=\textwidth]{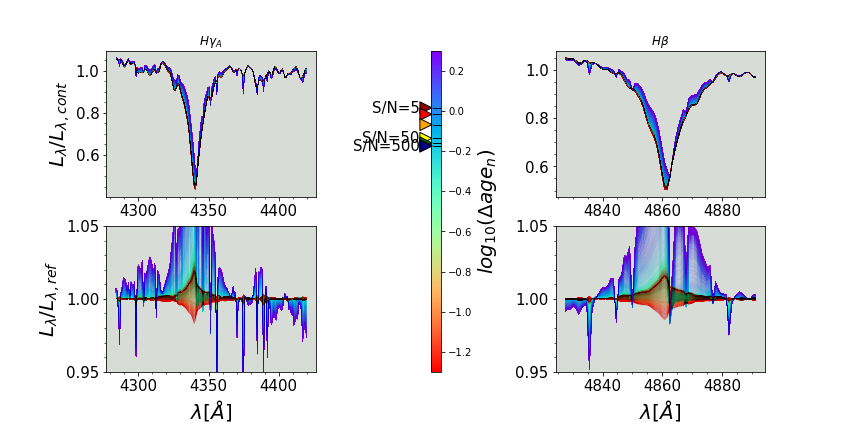}

    \caption{Same of figure \ref{fig:hghbm62_95}, but for 8.0$<age_{50}<$8.05 and $Z=Z_\odot$}
    \label{fig:hghbm62_8}
\end{figure}
\begin{figure}
    \centering
    \title{8.0$<log_{10}(age_{50})<$8.05, Z=Z$_{\odot}$}
    \includegraphics[width=\textwidth]{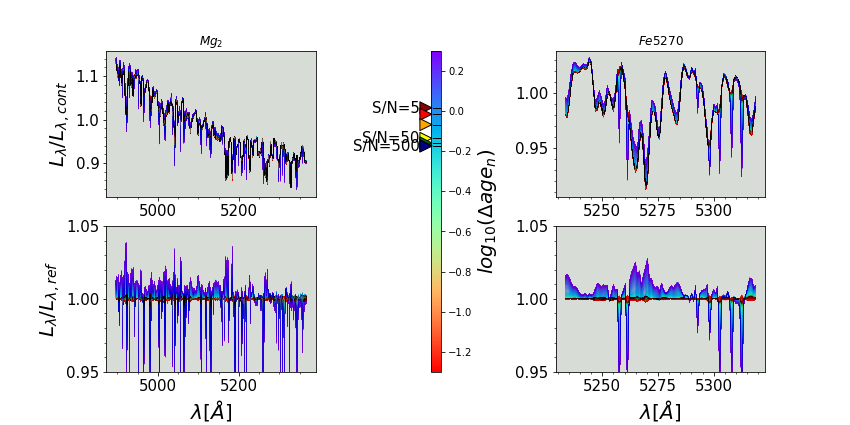}

    \caption{Same of figure \ref{fig:mg2fe5270m62_95}, but for 8.0$<age_{50}<$8.05 and $Z=Z_\odot$}
    \label{fig:mg2fe5270m62_8}
\end{figure}

The last case considered is characterized by solar metallicity and $log_{10}(age_{50})$ between 8.0 and 8.05, which is representative of relatively young stellar populations. Even in this case, we are in the region of $age_{50}$, in which stellar populations are dominated by hot stars. So, we expect strong Balmer absorption and very weak metal absorption.\\
First of all, we have to notice that for this metallicity and $age_{50}$ range, time resolution is the worst of these considered cases, as we can see from the ``arrows'' of the colorbar of figure \ref{fig:specm62_8}. In particular $log_{10}(\Delta age_{n,min})$ is greater than -0.2 dex for all signal-to-noise ratios. \\
We note that spectra distribution is narrower than in the previous cases. In fact, fluctuations from the reference spectrum, for wavelength lower than 4000\AA, are within 1.5\% for models with $log_{10}(\Delta age_n)<$-0.1 dex. At larger wavelengths, the fluctuations in the reference $\Delta age_n$ regime become smaller. In particular, for wavelength higher than 5000\AA, all the models display at most fluctuations $\sim$5\%.\\
Looking at $D4000n$ and Balmer lines distributions, we note that they vary in a narrow interval. We observe that only models with $log_{10}(\Delta age_n)>$0.2 dex show $D4000n$ fluctuations greater than 5\%.\\ 
Regarding the three Balmer lines (i.e. $H\delta_A,~H\gamma_A,~H\beta$), we note that they are very strong. However, at the centre of each Balmer lines, there are fluctuations 2\% for all the models with $log_{10}(\Delta age_n)<$-1.0 dex. This is due to the intrinsic scatter of the models, related to the chaotic evolution of the Balmer lines in this age range, as we can observe in figure \ref{fig:ssp_traces}, before the peak of $H\delta_A+H\gamma_A$.\\
For $age_{50}$ between $10^8yr$ and $10^{8.05}yr$, metal indices are very weak, as the spectrum is dominated by hot stars. In fact, as figure \ref{fig:mg2fe5270m62_8} displays, the metal absorption, corresponding to $Mg_2$ and $Fe5270$ respectively, is hardly detectable in the upper panels. Furthermore, as seen in the lower panels, all the fluctuations, with respect to the reference spectrum, are smaller than 5 \%.\\
So, the low resolution values of this case are related to the intrinsic scatter of the models, which can be better appreciated in this young stellar age regime ($ age_{50}\sim10^8yr$), because of the chaotic evolution of the Balmer lines and the spectra are dominated by Balmer absorption.\\

Even if we analysed an ideal case (fixed metallicity, without dust and with no star bursts), we find that time resolution depends on metallicity and on median stellar age. In particular, as figure \ref{fig:tres_map} shows, we can get better time resolution for stellar population with a median stellar age greater than $10^9yr$, when other indices (metal indices) besides Balmer indices appear in the spectra. The worse time resolution for young stellar populations ($age_{50}\sim10^8$) is due to the instrinsic scatter of the models, related to the chaotic evolution of the Balmer lines (figure \ref{fig:ssp_traces}).\\
Furthermore, we find that we are not able to get time resolution values lower than $\sim -0.5$ dex (i.e. $age_{10,90}\sim 1/3\cdot age_{50}$) in any case, as figure \ref{fig:tres} displays. \\
This is a fundamental result, in order to define a theoretical limit for SFH reconstruction. Furthermore, thanks to this result, we can create an optimal library of CSP models for the aim of this work.\\
Looking at the figures \ref{fig:specm62_95}, \ref{fig:specm62_85} and \ref{fig:specm62_8}, we note that photometry could in principle be sensitive to lower $\Delta age_n$ than the actual time resolution we have estimated, but this would require photometry accuracy better than 0.01 mag. This is not realistic in most of astrophysical applications. Furthermore, we should consider that in real galaxies, the degenerate effect of dust and metallicity can greatly increase the intrinsic scatter of the models at fixed $age_{50}$, so that in practice the formal sensitivity that we see in figures \ref{fig:specm62_95}, \ref{fig:specm62_85} and \ref{fig:specm62_8}, is not available. So, in order to lift the degeneracy and approach the time resolution that we have compute in this chapter, spectroscopy of SNR$>$10 is mandatory.

\clearpage
\section{Characterization of complex star formation histories}\label{sec:charact_csfh}
In the previous chapter we determined to which extension it is possible to distinguish an extended SFH from one of negligible duration.\\
A crucial step for this work is to test how accurately and reliably we can measure the parameters that characterize a complex SFH. This is the scope of this chapter. Given an observed galaxy spectrum, we would like to obtain information on its SFH by using the statistical approach described in chapter \ref{chap:SFH} to estimate parameter such as mean age, $age_{50}$, $\Delta age_n$ and mean metallicity.\\
From a CSP library of 500\,000 models, we will perturb spectral indices and absolute magnitudes of a sample of 12\,500 models with astronomical errors (Sec. \ref{errors}), simulating astronomical observations in order to see how well we are able to measure the parameters of interest. In section  4.2, we show and analyze the results, compared at different signal-to-noise ratio, in particular for SNR equal to 5, 20 and 100.

\subsection{Testing our SFH characterization method with mock observations}\label{sec:mock_test}
We create a spectral library as described in section \ref{subsec:speclibrary}, and we consider it as the base of our Bayesian fitting algorithm BaStA. As we mentioned, the continuous component of the SFH is determined by two parameters: $\tau$ that determines the position of SFH peak, and $t_{form}$ that sets the origin of star formation activity. Our spectral library is defined by 500\,000 models with $tau/t_{form}$ between 1/20 and 2, considering dust, variable metallicity and at most six star bursts.\\
From the 500\,000 models, we perturb 12\,500 models as described in Sec. \ref{errors}. These perturbed (``mock'') spectra will be analyzed as if they were real observations. Then we use the Bayesian statistical method, described in chapter \ref{subsec:BaStA}, in order to measure the parameters of interest (i.e. light-weighted stellar age, light-weighted stellar metallicity, mass-weighted stellar age, median stellar age and  $\Delta age_n$). Finally, we compare the results to the true input values to estimate biases and uncertainties in different physical and observational regimes. We perform this analysis at different signal-to-noise ratio, namely at SNR=5, SNR=20 and SNR=100. 

\subsubsection{Light-weighted stellar age}\label{sec:lw_age}

\begin{figure}
    \centering
    \centerline{\includegraphics[width=1.6\textwidth]{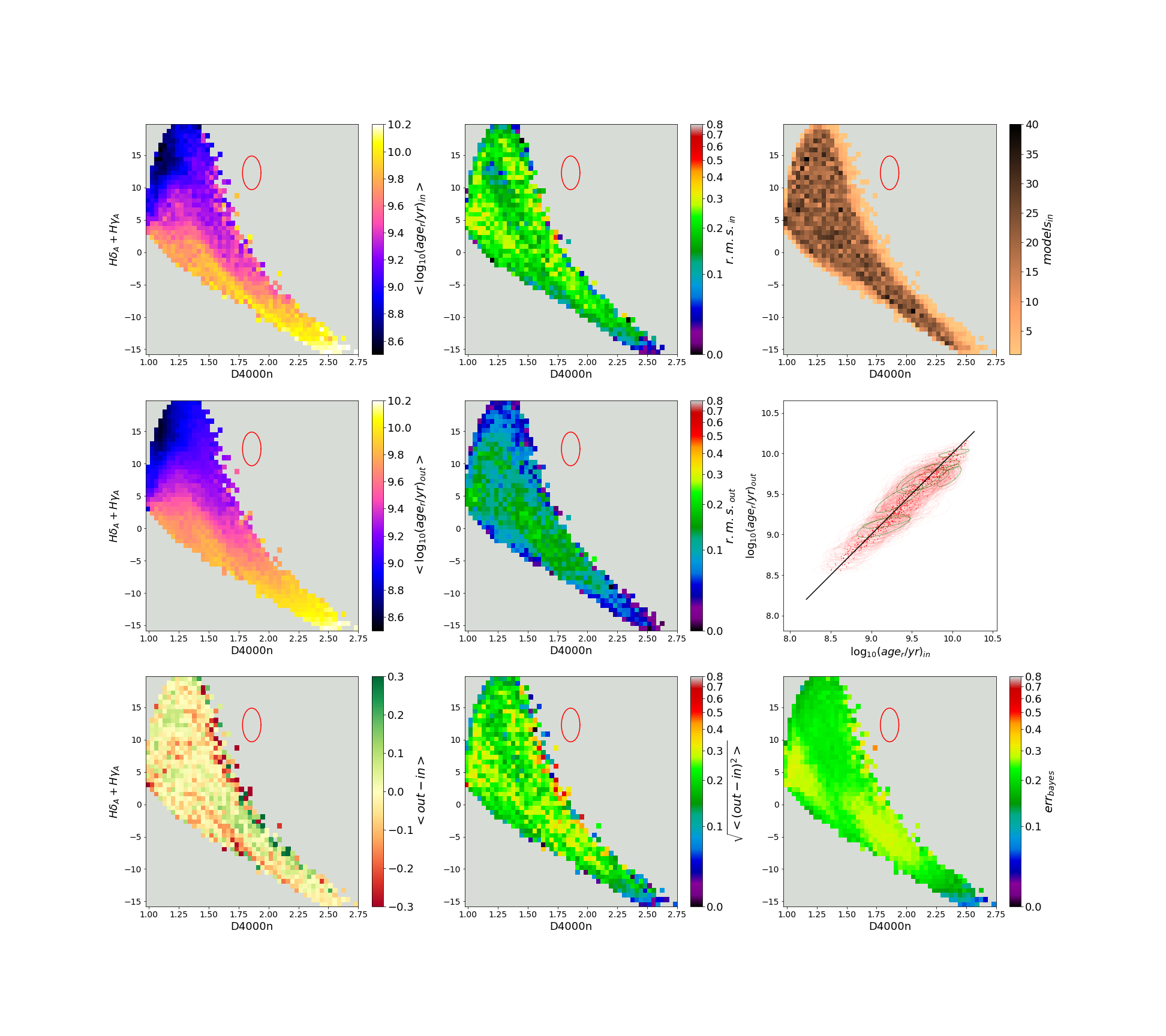}}
    \caption{Measurements of light-weighted age at SNR5. \emph{Top row} (from left to right): mean input distribution, rms of the input values, density of the input data. \emph{Middle row} (from left to right):mean output distribution, rms of the output values, correlation between input and outpu values. \emph{Bottom row} (from left to right): mean of difference between output and input values, standard deviation, Bayesian error. See text for details.}
    \label{fig:lwage_snr5}
\end{figure}
\begin{figure}
    \centering
    \centerline{\includegraphics[width=1.6\textwidth]{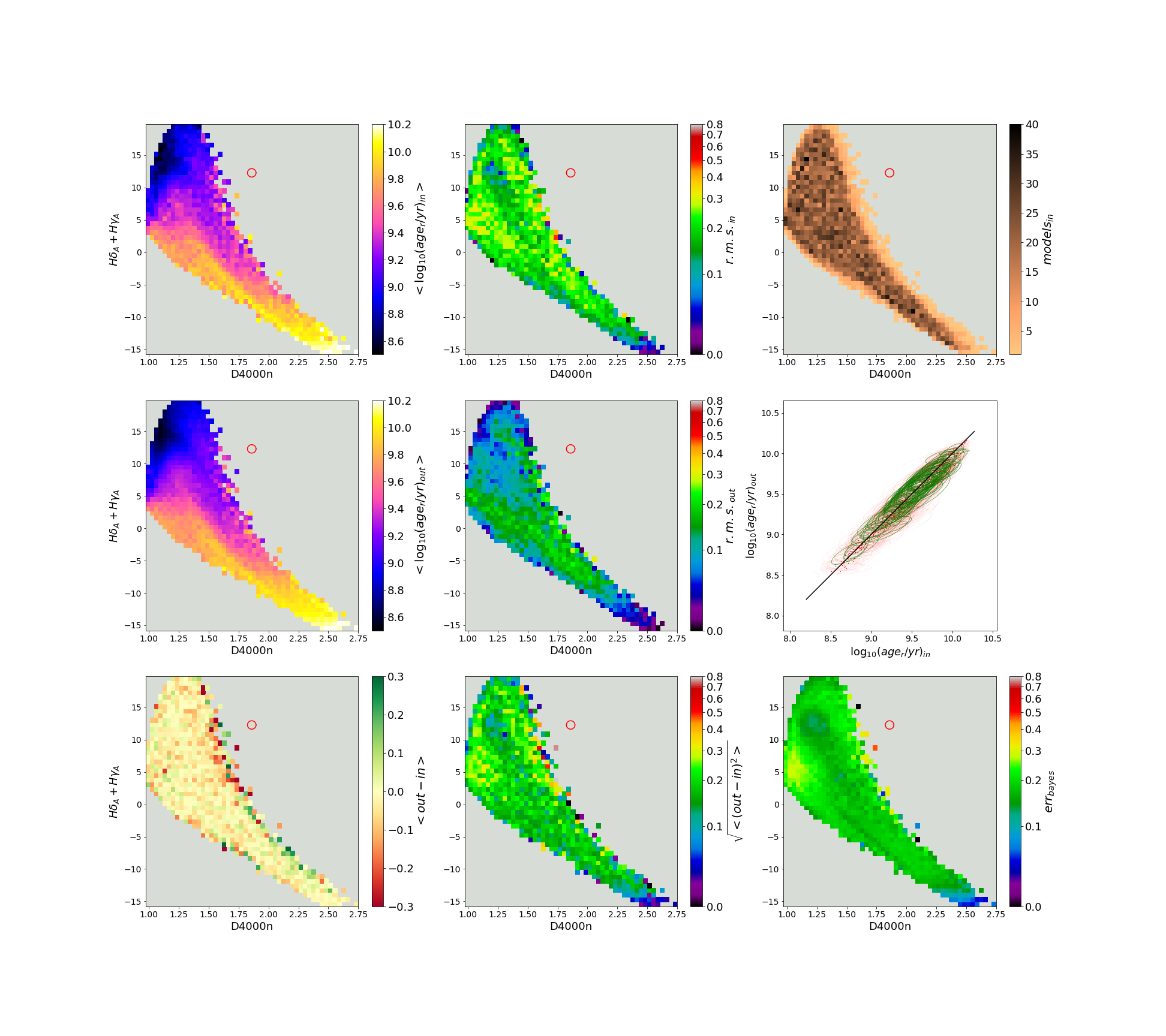}}
    \caption{Same of figure \ref{fig:lwage_snr5}, but at SNR=20.}
    \label{fig:lwage_snr20}
\end{figure}
\begin{figure}
    \centerline{\includegraphics[width=1.6\textwidth]{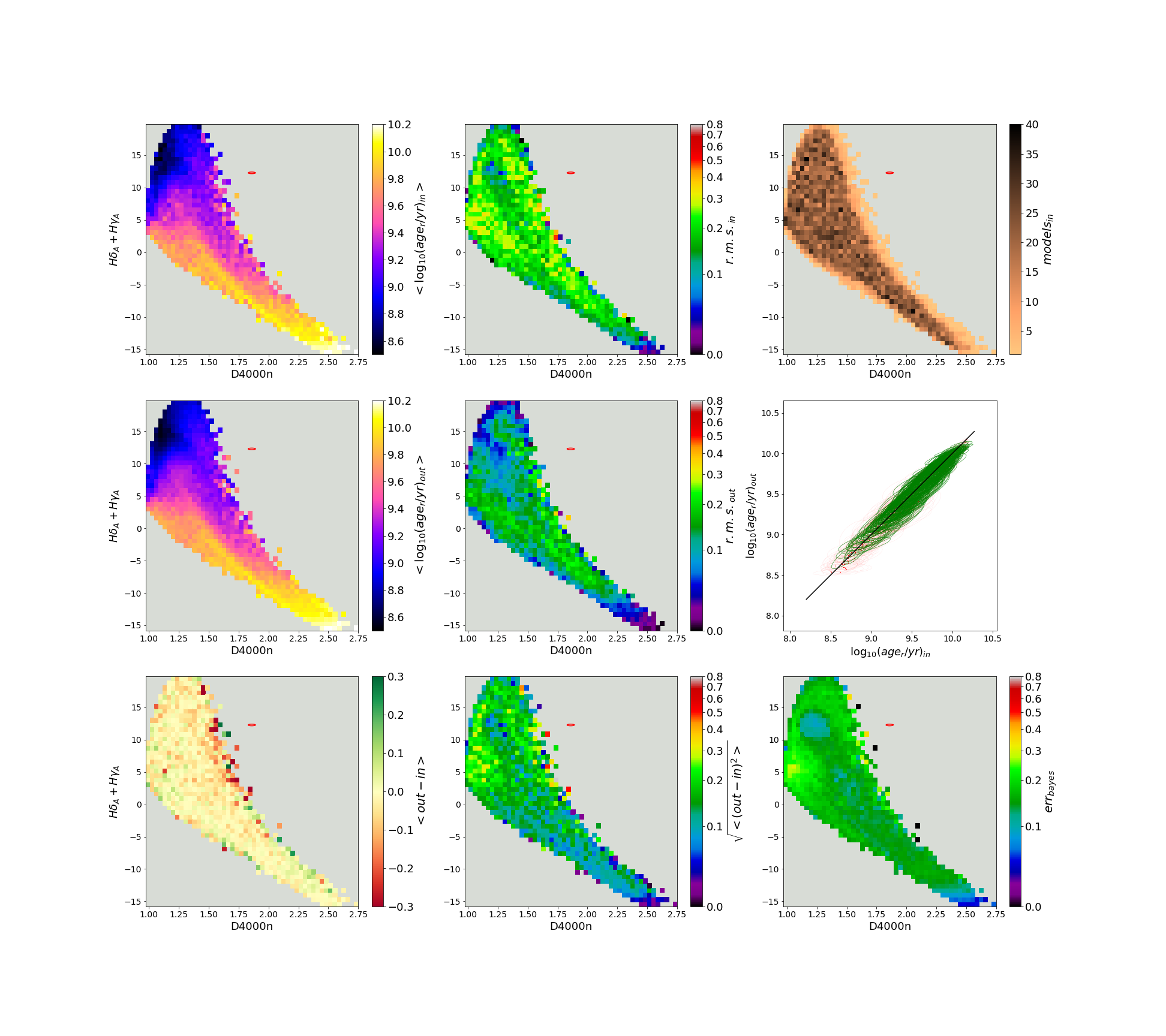}}
    \caption{Same of figure \ref{fig:lwage_snr5}, but at SNR=100}
    \label{fig:lwage_snr100}
\end{figure}

Considering light-weighted age in $r$-band, $age_{r}$, figure \ref{fig:lwage_snr5} summarizes the results of the analysis conducted at SNR=5. It useful to represent these results on a plane called Balmer plane, whose axes are $H\delta_A+H\gamma_A$ (the vertical axis) and $D4000n$ (the horizontal axis). This was first introduced by \cite{kauffmann2003} as SFH diagnostic, and it is very useful because $D4000n$ and Balmer absorption are (mostly) age sensitive. In particular, \cite{kauffmann2003} showed the power of the Balmer plane as a diagnostic for bursty vs continuous SFHs.\\
There are 8 maps created by a grid 50x50 bins in the Balmer plane and 1 panel which shows the relation between the input and the output parameter.\\
From left to right, the three top panels display the input distribution of the mean value in each bin, $<log_{10}(age_{r})>$, the width (i.e. the half difference between the percentile 84 and 16) of the input distribution and the model density number respectively. The first panel from left gives important information about the distribution of the parameter on the Balmer plane. As we can observe, stellar population with a mean stellar age $<10^9 yr$ are characterized by spectra dominated by Balmer lines, and $D4000n$ is at most equal to 1.25. $H\delta_A+H\gamma_A$ reaches its maximum value (Balmer peak) for stellar populations with $age_r\sim3\cdot10^8yr$, when luminosity is dominated by A-type stars. Then, for $age_r>10^9yr$ the light becomes dominated by cooler stars, Balmer lines become weak, and $D4000n$ increases. In particular, stellar populations with $age_r\sim3\cdot 10^9yr$ are characterized by a 4000\AA-break at most 2.25, and $D4000n>$2.25 can be reached only for $age_r>6\cdot 10^9yr$. \\
In the second row, we have two maps and a linear relation. The maps represent the mean value of the output distribution and the associated rms deviation in each bin respectively. The third panel from left shows the relation between $<log_{10}(age_{r})_{in}>$ and $<log_{10}(age_{r})_{out}>$ in each bin on the \emph{Balmer plane}. In order to illustrate the relation between the input data and the output data in each bin, we use the covariance matrix\footnote{$\sigma_{i,j}$, where $\sigma_{i,j}$ is equal to the variance of $x_i$ if $i=j$, while for $i\neq j$, $\sigma_{i,j}$ is given by the covariance between $x_i$ and $x_j$}. From this matrix, we can compute eigenvectors that give us the direction of the major axis of the points distribution, and the eigenvalues, whose square roots represent the amplitude of the axis. Hence, given a covariance matrix, we can define the associated covariance ellipse, with axes along the direction of the eigenvectors, and axis amplitude equal to the square root of the eigenvalues.\\
In this panel there is also a black line, which represents 1:1 relation. We can note that the ellipses are painted with two different colors, for different values of Pearson correlation coefficient, defined as follows:
\begin{equation}
    \rho_{in,out}=\frac{\sigma_{in,out}}{\sigma_{in}\sigma_{out}}
\end{equation}
where $\sigma_{in,out}, \sigma_{in}, \sigma_{out}$ are the covariance between the input and output values, the standard deviation of input values and standard deviation of output values respectively. In particular, the ellipses are green or red depending on whether the input and output data within a bin are strongly related with $\rho_{in,out}>0.7$, or not. In the three bottom panels, from left to right we have the mean of the difference between output data and input data ($<out-in>$), i.e. the bias, the square root of the mean squared deviation ($\sqrt{<(out-in)^2>}$) and the Bayesian error obtained from the difference between 84$^{th}$ percentile and 16$^{th}$ percentile of the posterior probability distribution of $age_{r}$. We can also note a red ellipse in each panel, that represents the error ellipse on the Balmer plane, in particular its horizontal axis is given by the error on $D4000n$, while the vertical one is the $H\delta_A+H\gamma_A$ error.\\
From these nine panels we can see that the distribution of the mean value is reproduced except for a mean bias of the order of 0.15 dex in absolute value. Furthermore, the bias may occasionally reach absolute values as high as 0.3 dex, most notably in the upper border at intermediate $D4000n$. At this SNR value we are not able to reproduce the distribution of the data inside the single bin, as we can see by comparing the rms of the age distributions in individual bins in input and in output (central panels of the first and second row). By comparing the input and output mean distribution (right panel of second row), we observe clearly a linear relation, which means that the distribution of the mean values are well reproduced. Yet most the ellipses are red, meaning that there is no ``hidden'' information inside the bins, i.e. input and output inside each bin are not significantly correlated. Finally, the error on light-weighted age in r-band obtained from Bayesian statistics is of the same order of the mean squared deviation, $\sim$0.2 dex.\\

It is interesting to observe the same results at different SNR. Figure \ref{fig:lwage_snr20} displays the results at signal-to-noise ratio equal to 20. Comparing to the previous case, the output rms deviation for $10^{9.4}<age_r<10^{9.6}$ reproduces the input one better than for SNR=5, as confirmed by the predominance of green ellipses in the correlation plot.\\ 
Looking at the bias, we can affirm that general trends of the input mean distribution are reproduced, except for small fluctuations of the order of $\sim\pm$0.1 dex. In fact, we can note a residual bias $\lesssim$0.1 dex in absolute value in most of the Balmer plane. However, we observe a region corresponding to ages of $10^9yr$ up to $3\cdot 10^9yr$ ($D4000n\sim1.6$ and $0<H\delta_A+H\gamma_A<15$), in which the bias reaches values of the order of -0.3 dex. This is due to dust and star bursts.\\
The Bayesian error is $\sim$0.2 dex in most of the plane, and it is well represented by $\sqrt{<(out-in)^2>}$. Close to the border of the distribution around [$D4000n\sim1.25$, $10<H\delta_A+H\gamma_A<15$] the error reaches a minimum value of the order of 0.15 dex. This region is characterized by a small input rms deviation ($\sim0.1$ dex). This means that this region is characterized by a small range of $age_r$, and it implies a smaller uncertain than in most of Balmer plane. \\ 

Fig. \ref{fig:lwage_snr100} shows the diagnostic plot for $age_{r}$ with a SNR=100. As expected, the Bayesian error value is lower than in the two previous cases. In particular, Bayesian error for SNR=100 is of the order of 0.15 dex in most of Balmer plane, and it reaches a minimum value ($<0.1$ dex) for old stellar populations ($age_r\sim10^{10.2}$). We note another minimum of $err_{bayes}\sim0.1$ dex, for young stellar populations with $H\delta_A+H\gamma_A\sim 12.0$ and $D4000n\sim 1.25$. Young stellar populations, characterized by an higher value of Balmer lines, have a Bayesian error $\sim$ 0.2 dex. Furthermore, the SNR=100 significantly improves the accuracy of the rms deviation, in particular for $age_r\sim 10^{9.5}yr$. This means that we have an optimal data relation in each single bin, as can also be seen from the strong presence of green ellipses in the correlation plot.\\
The distribution of the mean value (first panel of the middle line) is better reproduced than for SNR=20. The bias distribution has smaller amplitude and is better centered around zero. However, we can note a border region ($2<H\delta_A+H\gamma_A<15$ and $1.50<D4000n<2$), where the bias is $\sim$-0.3 dex. This is related to the degeneracy effect due to dust, metallicity and star bursts.\\

\subsubsection{Mass formed-weighted stellar age}
Beyond the light-weighted age described in sec. \ref{sec:lw_age}, we consider also the mass formed-weighted age (i.e. $age_{wmf}$), defined in sec. \ref{sec:inversion_problem}. This stellar age is computed on the formed stellar mass. This means that we take into account all the stars formed from the SFH beginning. In the previous case (light-weighted age), old stars have a negligible impact because they are overshined by the young ones. So, we expect older stellar age than in the previous case, especially in extended SFH.\\
\begin{figure}
    \centerline{\includegraphics[width=1.6\textwidth]{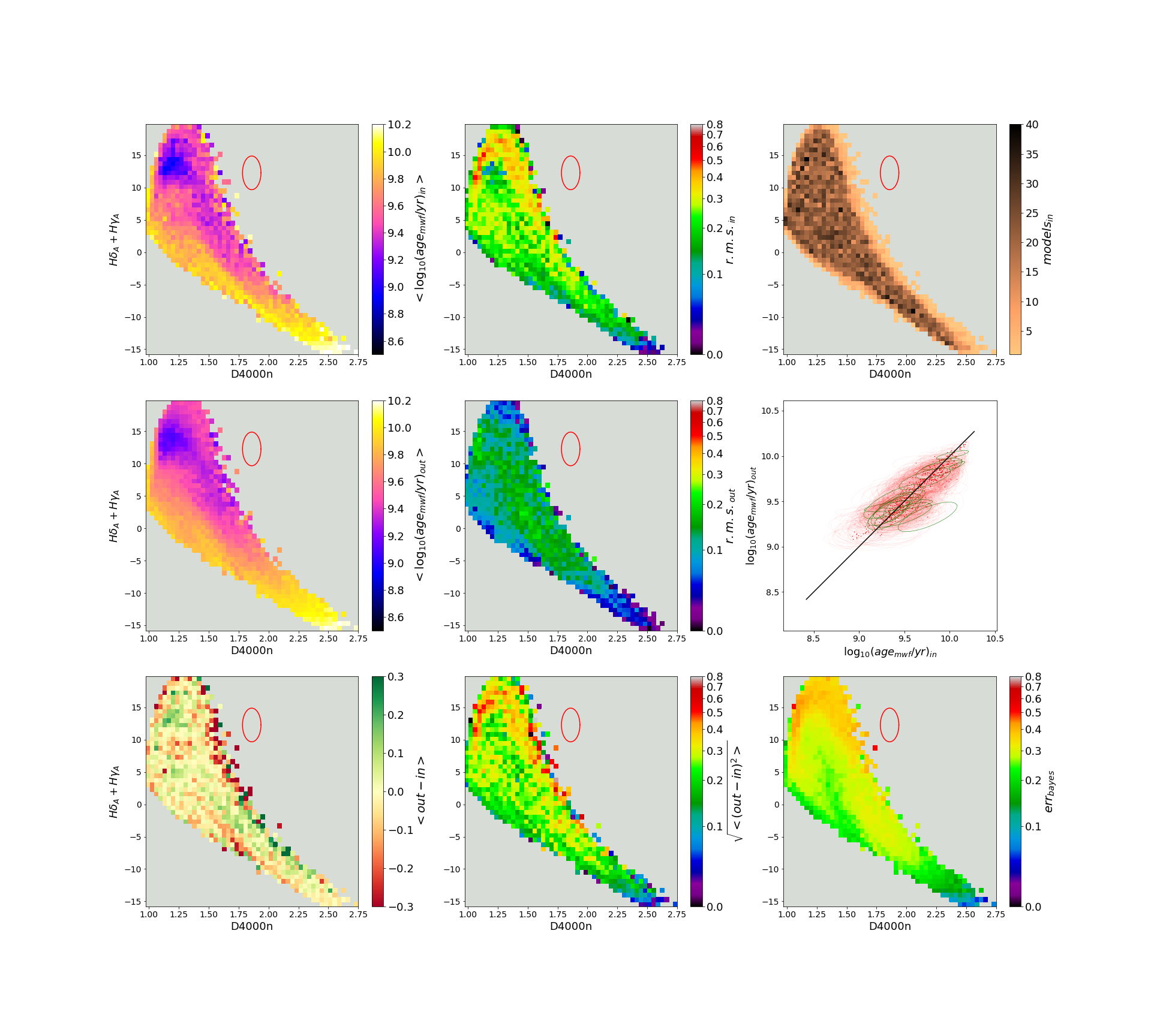}}
    \caption{Same of figure \ref{fig:lwage_snr5}, but for mass formed-weighted age at SNR=5.}
    \label{fig:wmf_age_snr5}
\end{figure}
\begin{figure}
    \centerline{\includegraphics[width=1.6\textwidth]{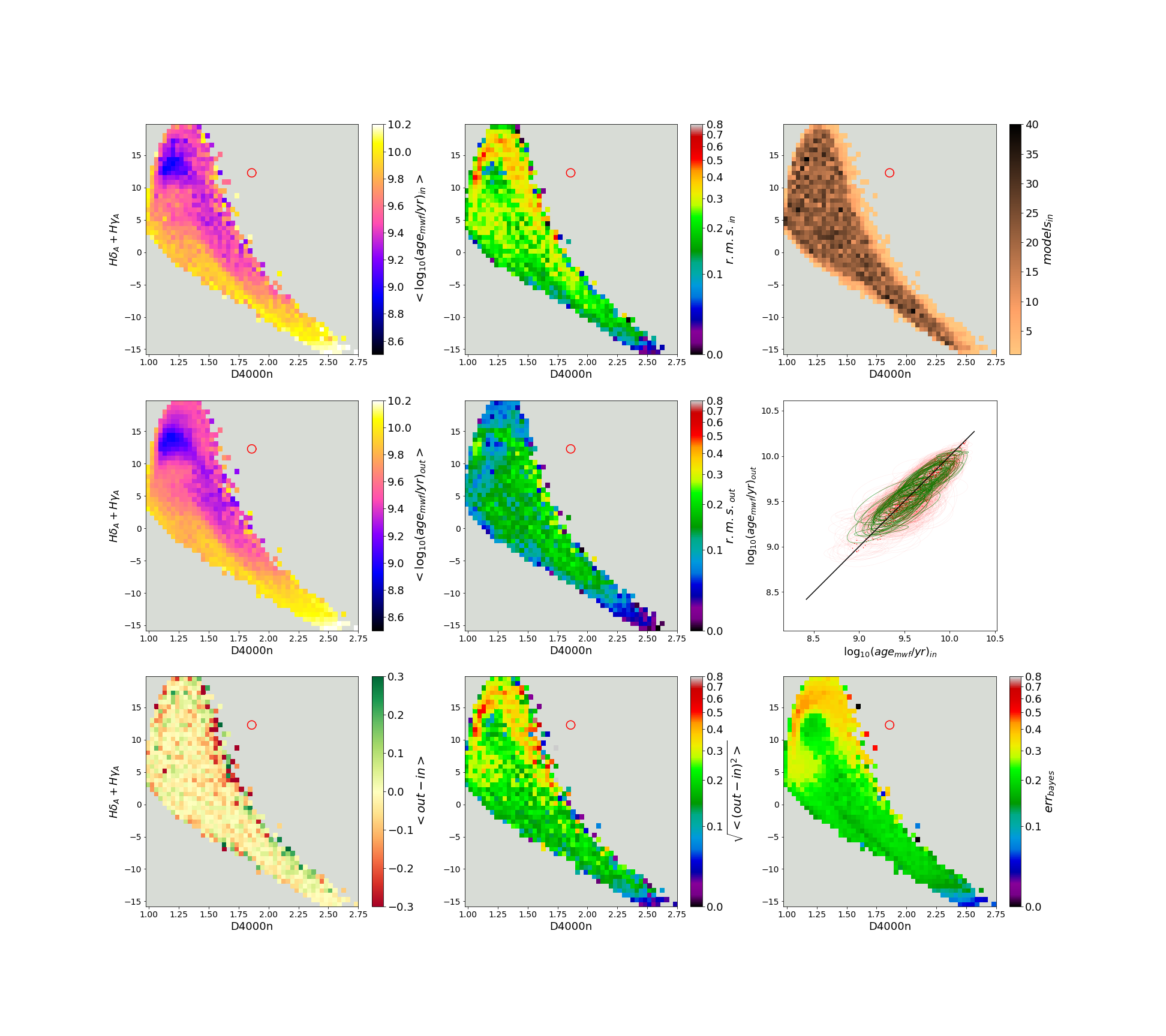}}
    \caption{Same of figure \ref{fig:lwage_snr5}, but for mass formed-weighted age at SNR=20.}
    \label{fig:wmf_age_snr20}
\end{figure}
\begin{figure}
    \centerline{\includegraphics[width=1.6\textwidth]{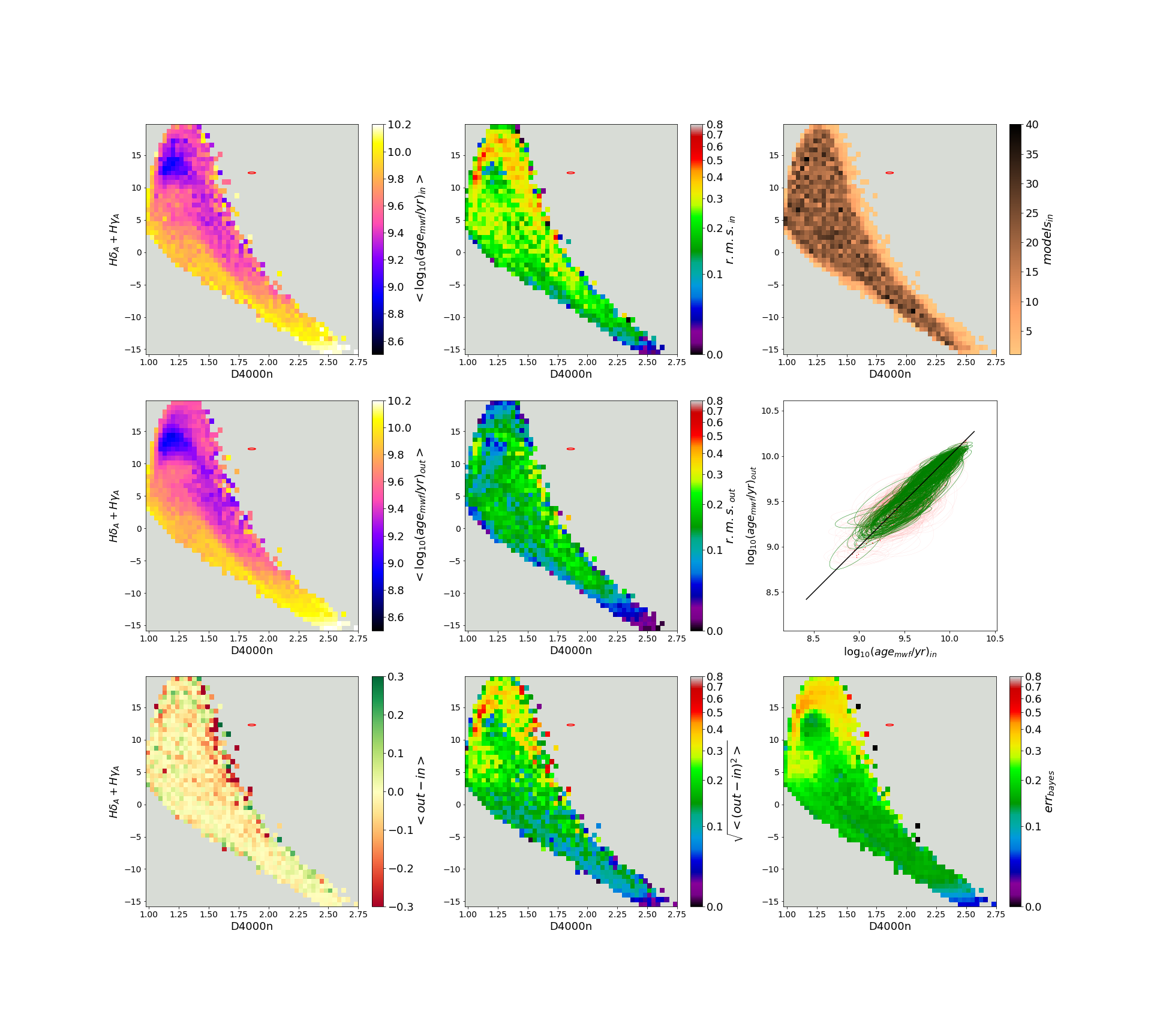}}
    \caption{Same of figure \ref{fig:lwage_snr5}, but for mass formed-weighted age at SNR=100.}
    \label{fig:wmf_age_snr100}
\end{figure}
Figure \ref{fig:wmf_age_snr5} shows us the results of mock test regarding $age_{wmf}$, considering spectra characterized by signal-to-noise ratio equal to 5. The first top panel from left and the first from left of the middle line represent the input and output mean distribution of $log_{10}(age_{wmf})$.\\
Looking at the input mean distribution, we can note that young stellar populations ($age_{mwf}<10^9yr$) are characterized by $H\delta_A+H\gamma_A \sim 12$ and a Balmer break lower than 1.25. For $age_{mwf}\sim10^9yr$, we reach the top values of the Balmer lines, thanks to the presence of A-type stars. We note a border on the top of the Balmer plane, which is characterized by a $age_{mwf}\sim 10^{9.7}yr$. These models have a corresponding mean stellar age ($age_r$) of the order of $10^{8.8}yr$. This difference implies that there is a non-negligible old component in stellar mass, and a young component that overshines the old one. We can now affirm that these stellar populations are characterized by an extended SFH, which is peaked in the past, and by a remarkable young star burst, which is evident in light but not in mass contribution. \\
Above $age_{wmf}\sim10^9yr$, as $age_{mwf}$ increases, $D4000n$ increases and Balmer lines decrease. In particular, for stellar populations with $age_{mwf}>10^{10}yr$, $D4000n$ values as large as 2.25 are reached.\\
At this low signal-to-noise ratio of 5, small fluctuations can not be appreciated, and the output distribution tends to be more uniform than the input one. This can be seen in the first bottom panel from left, which represents the mean value of the difference between output and input value of $log_{10}(age_{wmf})$. In particular, there is a bias $\sim-0.3$ dex in the border on the top of the Balmer plane that we described above, and for $age_{mwf}\sim10^{9.7}yr $ there is a bias $\sim$0.2 dex. \\
The rms deviation is represented in the second bottom panel, which is well reproduced by the error computed through the difference between the 84$^{th}$ percentile and 16$^{th}$ percentile of the PDF (the third bottom panel from left). This means that the error obtained from our analysis is in agreement with the uncertainties, with which we estimate  $log_{10}(age_{wmf})$. Looking at the Bayesian error distribution, we note that for stellar populations dominated by A-type stars (peak of $H\delta_A+H\gamma_A$), the Bayesian error on $log_{10}(age_{mwf})$ is $\sim$0.4 dex. In most of the Balmer plane the Bayesian error is $\sim$0.2 dex, and for $age_{mwf}>10^{10}yr$ becomes of the order of 0.1 dex. \\
For this value of SNR, we are not able to reliably characterize the $log_{10}(age_{wmf})$ distribution in a single bin, as we can observe by comparing the rms deviation of the input distribution (second panel from left of the top row) and the rms deviation of the output distribution (second panel from left of the middle row).\\
The strong presence of red ellipses, in the first panel from right of the middle row, underlines the lack of correlation between the input and the output data in each bin, even if a relation between the \emph{mean} input and output distribution is observed.\\

In figure \ref{fig:wmf_age_snr20}, we analyze the same results at SNR=20. We note that the bias is $\sim\pm$0.1 dex in most of the Balmer plane, except in a region ($D4000n\sim1.6$ and $0<H\delta_A+H\gamma_A<15$) in which the bias reaches $\sim\pm$0.3 dex.\\
The output rms deviation reaches values between 0.2 dex and 0.3 dex for stellar populations characterized by $9.4<log_{10}(age_{mwf})<9.8$, in agreement with the input rms deviation. This agreement can be noted also from the remarkable presence of green ellipses in the first panel from right of the middle row.\\
Concerning the uncertainties, the Bayesian error is $\sim$0.2 dex in most of the plane, and it is well represented by the standard deviation. For $H\delta_A+H\gamma_A>12$, the error is $\sim$0.4 dex. This is related to the SFH degeneracy, which produces similar galaxy spectra for different SFH. In particular, in this region SFHs are characterized by a remarkable young star burst, which leaves more freedom to the old stellar component. The minimum of the Bayesian error ($\sim$0.1 dex) is reached for $D4000n>2.25$, which corresponds to $age_{wmf}>10^{10}yr$.\\

Next, we analyze the same results considering a  SNR value equal to 100 (fig. \ref{fig:wmf_age_snr100}). In this case the mean input and output distribution are more similar to each other than in the previous cases. In fact, as we can observe also in the first bottom panel from left, $<out-in>$ is well centered on zero-value with fluctuations within $\pm0.1$ dex in most of the Balmer plane. There is only a region on the Balmer plane ($1.50<D4000n<1.75$ and $0<H\delta_A+H\gamma_A<15$), in which the bias is $\sim$-0.3 dex. This is due to dust and star bursts, as mentioned in sec. \ref{sec:lw_age}.\\ 
Furthermore, the distribution of $log_{10}(age_{wmf})$ is better reproduced, as we can see by comparing the second panel from left of the top and of the middle row. In particular, rms of the output values is $\sim0.2$ dex in most of the plane, and it is comparable to the rms of the input values. Another relevant result is given from the predominance of green ellipses in the first panel from right of the middle line, which indicates a significant correlation between input and output distribution of $log_{10}(age_{wmf})$ in each single bin.\\
The Bayesian error is $\sim$0.15 dex in most of the Balmer plane, and for stellar populations with $age_{mwf}>10^{10}yr$ the error reaches values below 1.0 dex. However, we can note that in the region dominated by A-type stars, the Bayesian error is $\sim$0.4 dex as in the case SNR=5. This is due to the SFH degeneracy already mentioned above. In fact, the top of the Balmer plane is dominated by stellar populations characterized by extended SFH, in which the old stellar component is overshined by the young one. This produces a high uncertainty on the old stellar component, which is reflected on the Bayesian error of $age_{wmf}$.\\

\subsubsection{Median stellar age}
\begin{figure}
    \centerline{\includegraphics[width=1.6\textwidth]{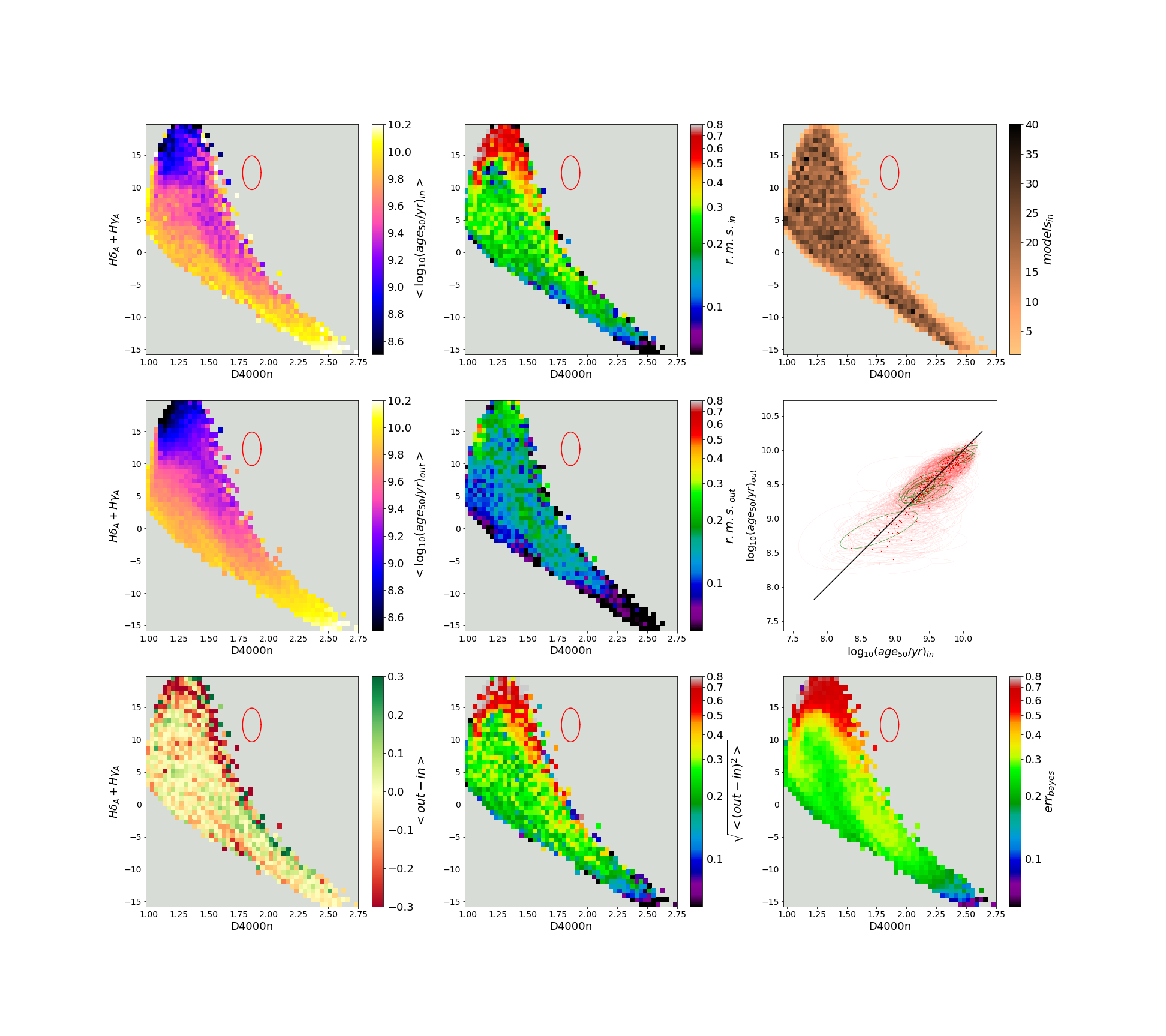}}
    \caption{Same of figure \ref{fig:lwage_snr5}, but for $age_{50}$ at SNR=5.}
    \label{fig:age50_snr5}
\end{figure}
\begin{figure}
    \centerline{\includegraphics[width=1.6\textwidth]{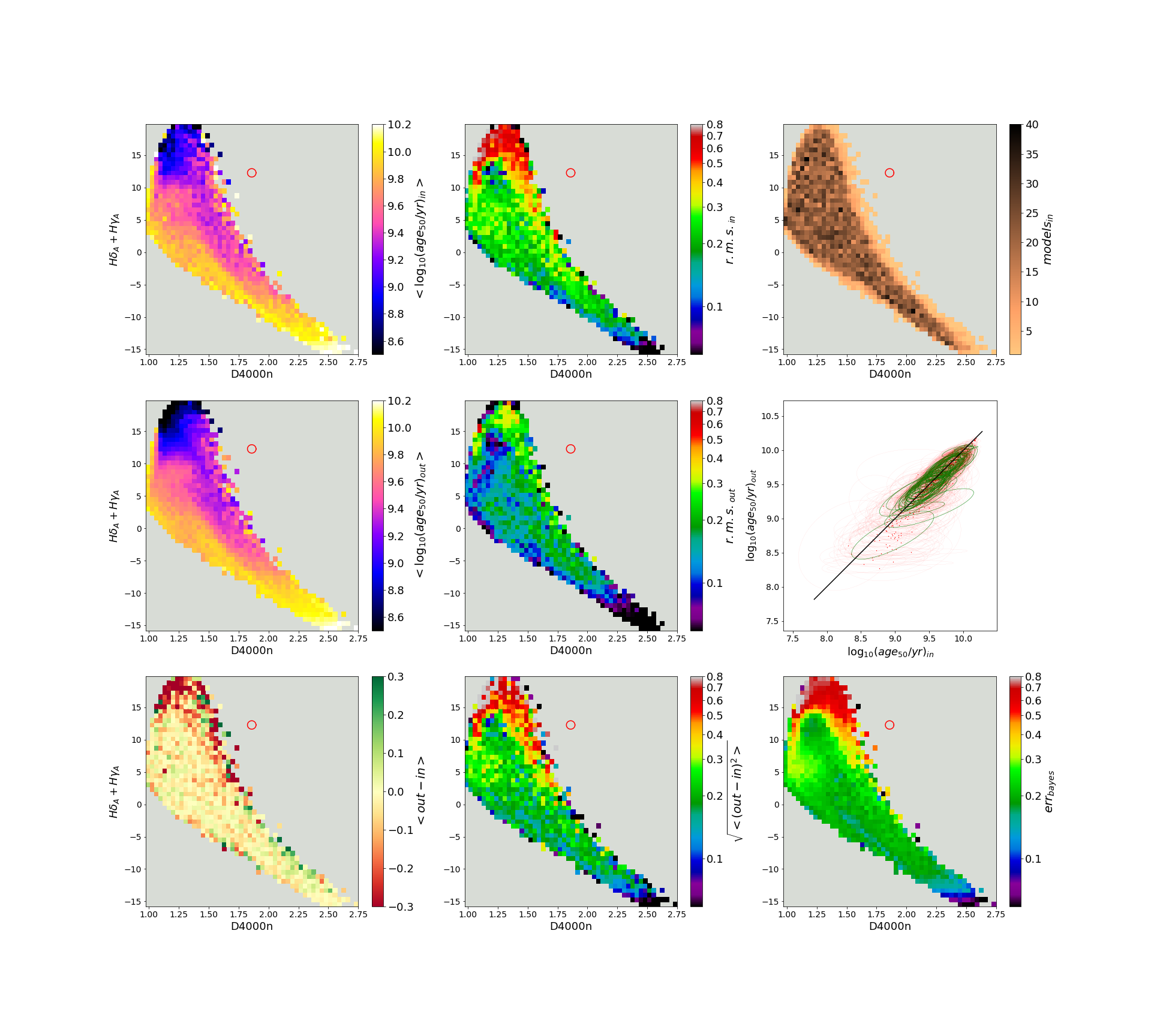}}
    \caption{Same of figure \ref{fig:lwage_snr5}, but for $age_{50}$ at SNR=20.}
    \label{fig:age50_snr20}
\end{figure}
\begin{figure}
    \centerline{\includegraphics[width=1.6\textwidth]{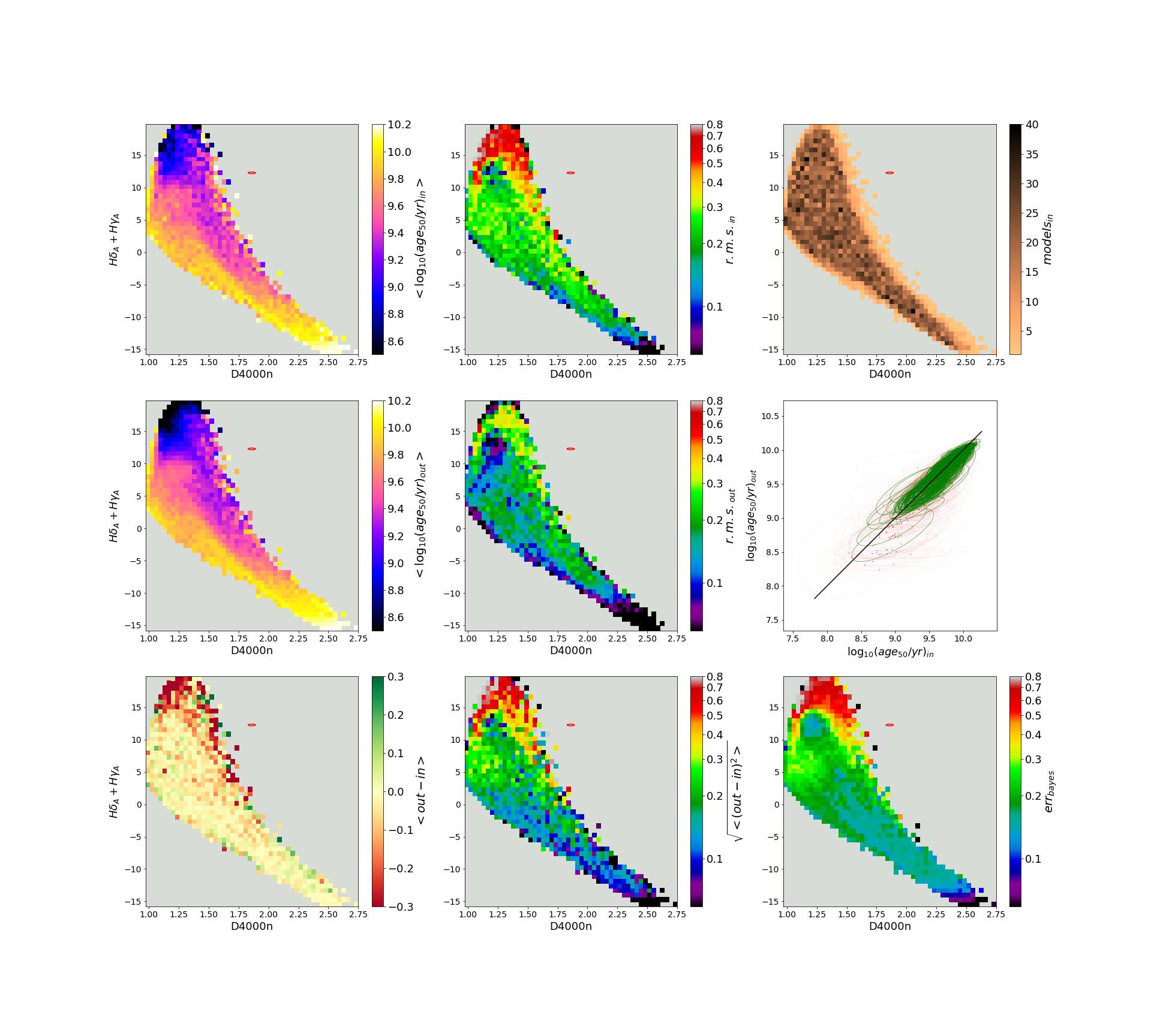}}
    \caption{Same of figure \ref{fig:lwage_snr5}, but for $age_{50}$ at SNR=100.}
    \label{fig:age50_snr100}
\end{figure}
In section \ref{sec:inversion_problem}, we defined the median stellar age $age_{50}$ as the look-back time at which the 50\% of the stellar mass is produced. In order to obtain a good characterization of star formation history of galaxies, it is fundamental to obtain this parameter. Figure \ref{fig:age50_snr5} displays the results of the mock test for $age_{50}$ with SNR=5. By comparing the first column of panels from left, we see that the general trends of the mean input distribution (top panel) are reproduced, but its fine structure is lost in output. This gives rise to local biases of up to 0.3 dex in absolute value, as we can observe in the first bottom panel from left.\\
Looking at the input distribution (first top panel from left), we can note that the largest values of the Balmer lines are typical of young stellar populations ($age_{50}<10^9yr$), and they are characterized by a 4000\AA-break lower than 1.5. For $H\delta_A+H\gamma_A$ lower than 10, we find $age_{50}\sim10^{9.5}yr$, and $D4000n$ assumes values till 2.0. Then, old stellar populations ($age_{50}\sim10^{10}yr$) are characterized by weak Balmer lines, and by a notable 4000\AA-break ($D4000n>$2.0).\\
The distribution of the data inside the individual bins are not well reproduced, as we can see by comparing the second panels from left of the top, i.e. width of input data distribution in each bin, and of the middle line, i.e. width of output data distribution in each bin. The width of the output distribution is smaller than the width of the input distribution. In fact, for all mocks in a given bin we obtain estimates of $age_{50}$ that are very close to the median value of the prior in that bin. This lack of relation between input and output data inside the single bin is reflected in the first panel from right of the middle line. As one can see, red ellipses are dominant, implying that there is very little additional information beyond what is provided by the location in the Balmer plane, but the red dots are around the black line, meaning that mean input distribution is broadly re-constructed. \\
Even in this case, the standard deviation (middle panel of the bottom row) is well reproduced by the Bayesian error (third bottom panel from left), which is of the order of 0.6 dex for $log(age_{50})<$9.0. For 9.0$<log(age_{50})<$10.0 we have a Bayesian error of the order of 0.25 dex, while for $log(age_{50})$ higher than 10.0 dex the error is $\sim$0.1 dex.\\

Next we analyze the same results for SNR=20, displayed in figure \ref{fig:age50_snr20}. Looking at the first panel from left of the bottom row, we note a region characterized by a bias $\sim$-0.3 dex, in the upper part of the Balmer plane for $D4000n<1.75$. However, in most of the plane general trends and fluctuations are reproduced with a bias $\sim\pm0.1$ dex.\\
In the same region, characterized by a bias $\sim$-0.3 dex, we have a Bayesian error $>0.5$ dex, corresponding to a factor $>$3 on $age_{50}$. In most of the plane we have an error $\sim$0.18 dex, while for $D4000n>2.25$ it is $\sim$0.1 dex.\\
The rms deviation is better reproduced than in the case of SNR=5, in particular for $10^{9.25}yr<age_{50}<10^{9.50}yr$.\\

Figure \ref{fig:age50_snr100} shows the same parameter considering a SNR=100. As expected, the agreement between the input and output distribution clearly improves. By comparing these, we observe that small fluctuations are better reproduced for $age_{50}>10^{9.30}yr$. In fact, looking at the first panel from right of the bottom row, for $age_{50}>10^{9.30}yr$ there is a bias at most 0.1 dex in absolute value. Instead, in the Balmer plane regions dominated by A-type stars we obtain a bias of the order of 0.3 dex in absolute value.\\
Furthermore, the strong presence of green ellipses means that inside each bin data are well related, as we can observe by comparing the second panels from left of the top and middle line. These two plots, in fact, describe the width of the input and output data distribution in each individual bin respectively. In this case, the rms of the output values reaches $\sim$0.35 dex in regions dominated by A-type stars, where rms of the input values is greater than 0.5 dex, because of SFH degeneracy.\\
Also in this case, the Bayesian error (first bottom panel from left) reproduces well the mean square deviation (middle bottom panel). We can see that models characterized by $<log(age_{50})><$9.0 are affected by an error greater than 0.5 dex. In general, for larger ages we have an error between 0.2 dex and 0.15 dex. For models with $log(age_{50})$ greater than 10.0 dex, we have an error of the order of 0.12 dex, and this decreases with the increasing of median stellar age.

\subsubsection{Comparing different stellar ages}
As we have seen in the previous sections, different stellar ages can be defined: light-weighted age (eq. \ref{eq:light_age}), mass formed-weighted age (eq. \ref{eq:mass_age}) and median stellar age (eq. \ref{eq:f_age}).\\
Comparing the mass formed-weighted and the light-weighted age (first panels from left of the top row of figures \ref{fig:lwage_snr100}, \ref{fig:wmf_age_snr100}), we observe that the general trends are similar: Balmer lines decrease as stellar age increases, and we reach high values of 4000\AA-break ($D4000n>$2.20) only for stellar ages $\sim10^{10}yr$. The main difference regards the Balmer peak\footnote{The region on the Balmer plane characterized by the high value of $H\delta_A+H\gamma_A$ ($H\delta_A+H\gamma_A>15$).}, where A-type stars dominate. Considering light weighted-age, this region ($1<D4000n<1.5$ $10<H\delta_A+H\gamma_A<20$) is dominated by stellar populations characterized by stellar age$<10^9yr$. If we compute the mean stellar age by weighting on mass formed, we note that the lowest value the mean stellar age is equal to $\sim10^9yr$ in the region $12<H\delta_A+H\gamma_A<16$, $1.10<D4000n<1.30$. This offset is due to the greater weight in brightness that the young populations have with respect to the old ones. In other words, the old stellar populations in the Balmer peak, which are overshined in light by the young ones, are highlighted by weighting on mass and their contribution results as an increase of the mean mass-weighted stellar age.\\
Looking at the first panels from left of the top row of figure \ref{fig:wmf_age_snr100}, we observe a border at the top of Balmer peak of old $age_{mwf}\sim10^{9.5}yr$. This is not observed in the light weighted age, because they are stellar populations characterized by an old peak of star formation activity, and by a young star burst, which overshines the old stars in the case of light-weighted age. \\
Also for the median stellar age ($age_{50}$) the general trends can be compared to the other stellar ages. As far as the region $age_{50}<10^9yr$ is concerned, it is more similar to the light-weighted age. In fact, the old border at the top of the Balmer peak observed in the mass formed-weighted age, for $age_{50}$ it is not observed. This confirms that these stellar populations are characterized by a remarkable young star burst, in which more than 50\% of the observed stellar mass is produced and overshines the old stars.\\
Another relevant difference involves the uncertainties of these stellar ages computed from the width of the PDF. Considering the results for SNR=20, we can observe that for most of the Balmer plane $err_{bayes}$ on light-weighted age is $\sim$0.15 dex, even in the peak of the Balmer lines. For the mass formed-weighted age and for the median age, we note that the Bayesian error is between $\sim$0.2 dex and $\sim$0.15 dex for most of the bins characterized by $H\delta_A+H\gamma_A<15$. Concerning the Balmer peak ($H\delta_A+H\gamma_A>15$), we have $err_{bayes}\sim0.4$ dex for the mass formed-weighted age, and $err_{bayes}>0.5$ dex for the median stellar age. These high uncertainties are being related to the old stars overshined by the young ones. In fact, in the Balmer peak young stars dominate in brightness, giving more freedom to the old component of stars, which implies more uncertain on the mass-weighted quantities.\\ 
The difference between mass and light weighted stellar ages is due to the different weights given to the young and to the old stellar component. In particular, these differences emerge for extended SFH, where both old and young stars are present, so that old stars are overshined by the young ones.\\
Since we get photons from astronomical observations and old stellar populations are overshined by the young ones, light weighted quantities are better constrained and in closer agreement with the true values.

\subsubsection{Duration of star formation history}

\begin{figure}
    \centerline{\includegraphics[width=1.6\textwidth]{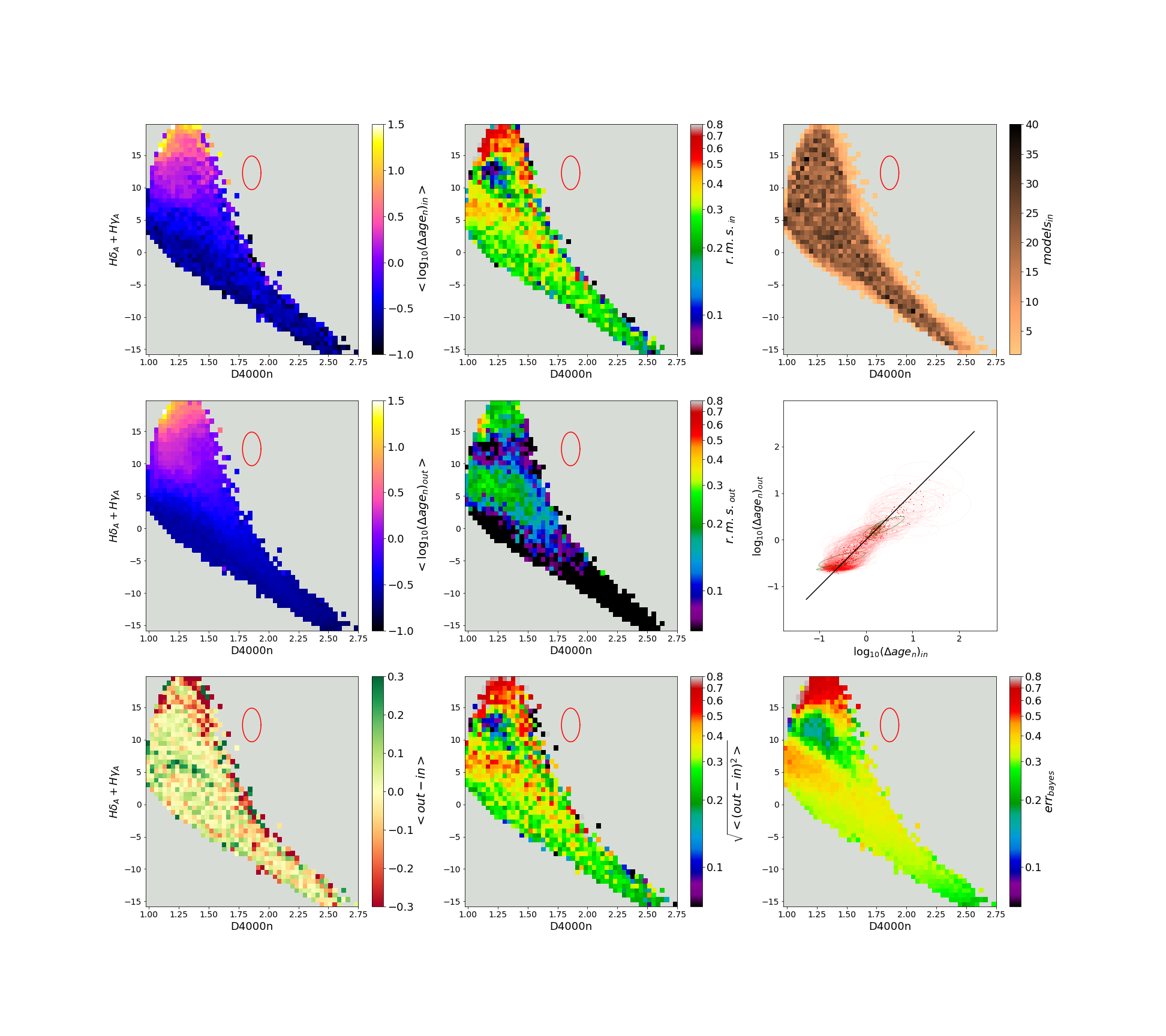}}
    \caption{Same of figure \ref{fig:lwage_snr5}, but for $\Delta age_n$ at SNR=5.}
    \label{fig:d1090n50_snr5}
\end{figure}
\begin{figure}
    \centerline{\includegraphics[width=1.6\textwidth]{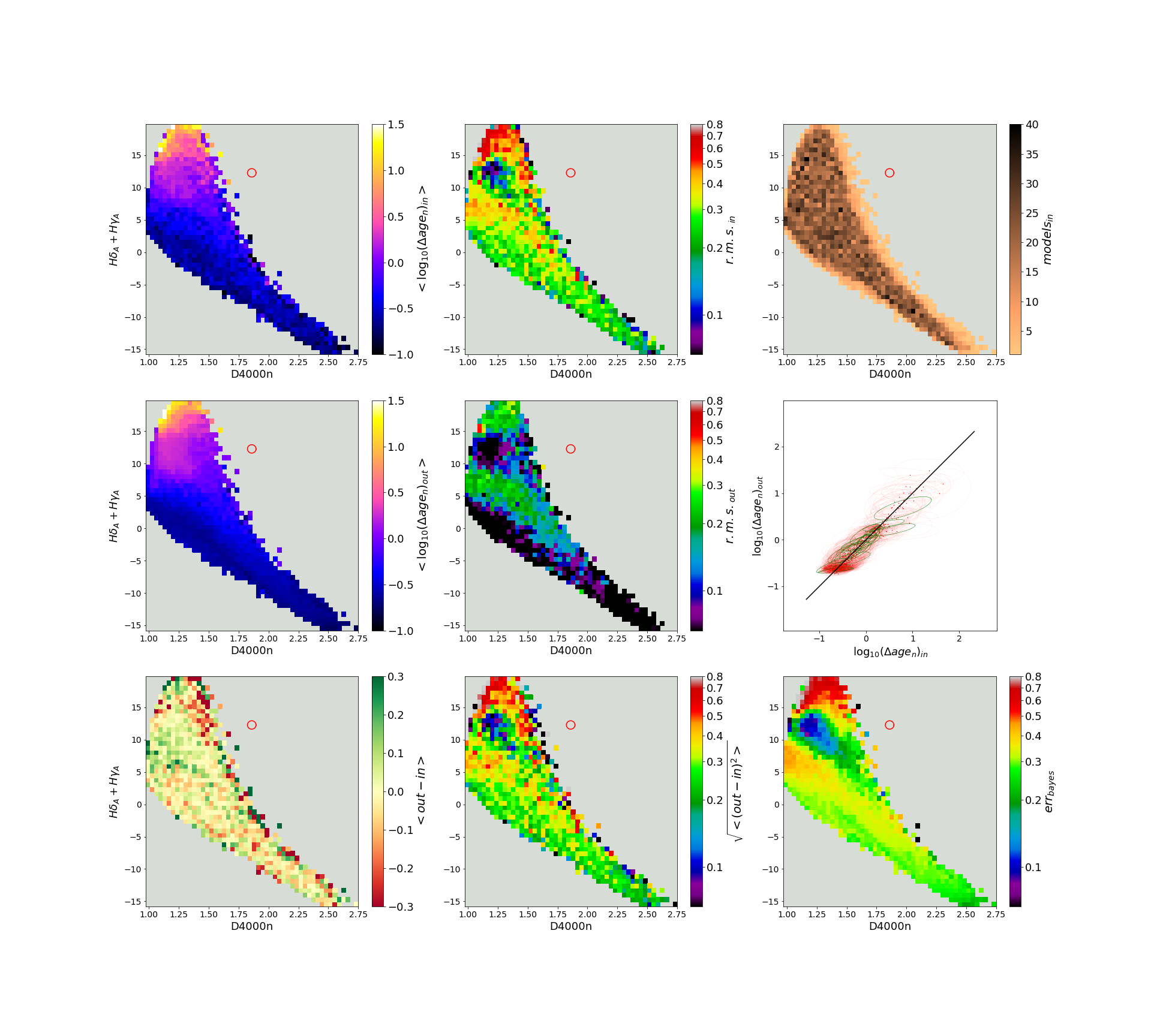}}
    \caption{Same of figure \ref{fig:lwage_snr5}, but for $\Delta age_n$ at SNR=20.}
    \label{fig:d1090n50_snr20}
\end{figure}
\begin{figure}
    \centerline{\includegraphics[width=1.6\textwidth]{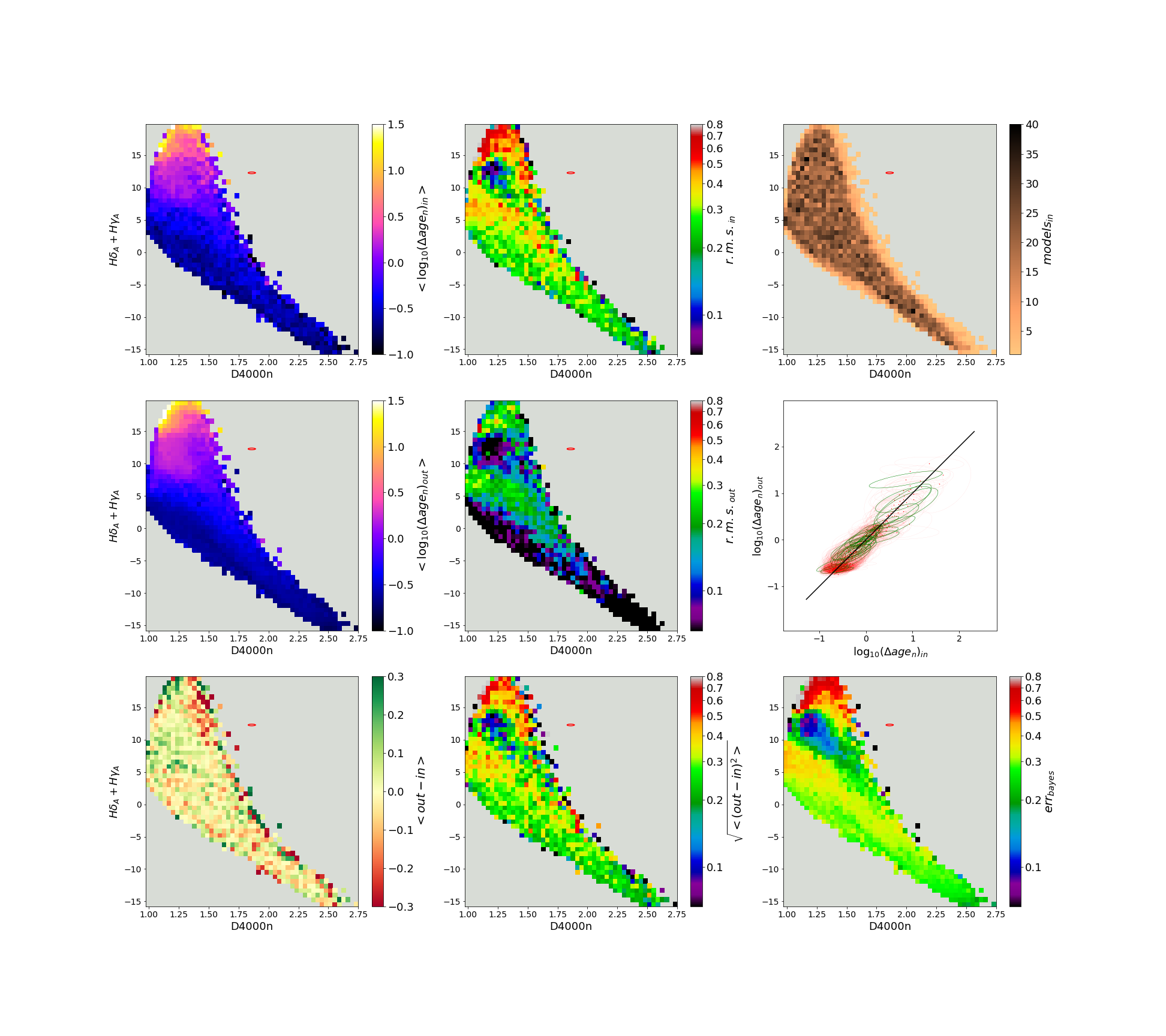}}
    \caption{Same of figure \ref{fig:lwage_snr5}, but for $\Delta age_n$ at SNR=100.}
    \label{fig:d1090n50_snr100}
\end{figure}
The goal of this project is to characterize galaxy SFHs, in particular we aim at finding SFH duration by getting $\Delta age_n$, a new parameter that we have introduced in this work.\\
Looking at the first panel from left of the top row of figure \ref{fig:d1090n50_snr5}, we can observe how this parameter populates the Balmer plane. There is a narrow region in the Balmer peak, in which $\Delta age_n>10$. This is in agreement with what we wrote in the last subsection. In fact, these SFHs are characterized by an old peak of stellar activity and by a young star burst, so that their overall SFH duration is $\sim$10 times their median stellar age. Then, for $10<H\delta_A+H\gamma_A<18$, we have $\Delta age_n\sim 3$, and as values of Balmer lines decrease, $\Delta age_n$ decreases. In particular, only in the lower region of the Balmer plane $\Delta age_n\sim1/3$, while for $D4000n<$2.0 we can reach $\Delta age_n<1/3$.\\

In figure \ref{fig:d1090n50_snr5}, we can observe that the mean output distribution of $\Delta age_n$ (first panel from left of the middle row) reproduces the general trends of the input one, but the small fluctuations observed in the input distribution are lost here. In fact, as the first panel from left of the bottom row shows, there are remarkable fluctuations between input and output mean distributions. In particular, in the region dominated by A-type stars, where there is the peak of the Balmer lines, the bias is of the order of $\sim$-0.3 dex. Even for $D4000n>$1.5, in the upper region of the Balmer plane we observe fluctuations $\sim$0.3 dex in absolute value. However, the mean bias is well centered on zero, and in most of the bins $<out-in>$ is not greater than 0.2 dex in absolute value.\\
Furthermore, the distribution inside the individual bins is not well reproduced, as we can see by comparing the middle panels of the top and middle line, which represent the rms of the input and output values in each bin. The output rms deviation, for the old stellar populations region is lower than 0.05 dex, while for the input in the same region is $\sim0.3$ dex.\\
All these information are summarized in the third panel from left of the middle line. This represents the relation between $log_{10}(\Delta age_n)$ input and output, which can be approximately seen as linear. We have to notice also the strong presence of red ellipses, meaning that data inside the bins are not correlated. So, once we measure $D4000n$ and $H\delta_A+H\gamma_A$ from a galaxy spectrum, the extra information on $\Delta age_n$ that can be provided by the other spectral features is negligible at this low SNR. It is important to notice that we are not able to get output $log_{10}(\Delta age_n)$ $<$-0.5, in agreement with the previous chapter \ref{chap:time_resolution}: such low values would be below the limiting time resolution.\\
As far the uncertainty is concerned, we can observe that the standard deviation (second bottom panel from left) is well represented by the Bayesian error (first bottom panel from right). In particular, in the Balmer peak, there is a Bayesian error $>$0.5 dex, while in most of the plane, the error is between 0.3 dex and 0.45 dex. However, there is a region ($10<H\delta_A+H\gamma_A<15$ and $1.1<D4000n<1.5$), in which the Bayesian error on $log_{10}(\Delta age_n)$ is as low as $\sim$0.2 dex. This region is characterized by an input rms deviation $<$0.05 dex, and this is in agreement with the output rms deviation. So, in this region we get a smaller Bayesian error ($\sim$0.2 dex), because there can only be models with a $log_{10}(\Delta age_)\sim 0.5$.\\

Next, we analyze the same results for SNR equal to 20 (Fig. \ref{fig:d1090n50_snr20}) and then equal to 100 (Fig. \ref{fig:d1090n50_snr100}).\\
At SNR=20, we can observe a better match between the input and output mean distributions than in the previous case (SNR=5), as first bottom panel from left shows ($<out-in>$). There are only few bins in the Balmer peak, in which bias reaches values $\sim$0.3 dex in absolute value. In most of the Balmer plane, the bias is $\sim$0.15 dex in absolute value.\\
However, the $log_{10}(\Delta age_n)$ distribution within each bin is not well reproduced, as we can notice by comparing the second panels from left of the top and middle lines. In most of the bins, the rms of the output values is $<$0.05 dex. This means that we essentially get the median prior value of the corresponding bin. We have to note that in the region with $log_{10}(\Delta age_n)\sim$0.0, the rms of the input values is reproduced in output significantly better than in other regions. In this region the input rms deviation is $\sim$0.4 dex, while the output rms deviation is $\sim$0.3 dex. In fact, as we observe in the first panel from left of the middle row, there is a significant presence of green ellipses for $log_{10}(\Delta age_n)$.\\
Regarding the uncertainty, the Bayesian error is lower than in the previous case (SNR=5), and it is in agreement with the standard deviation values. In particular, we note that in the Balmer peak the Bayesian error is $>$0.5 dex, while in most of the Balmer plane it is between 0.3 dex and 0.4 dex. We can observe a region with $10<H\delta_A+H\gamma_A<15$, in which the error is $\sim$0.1 dex. These bins are characterized by an input rms deviation lower than 0.05 dex. This small variation of $\Delta age_n$ inside these bins means that only models with $log_{10}(\Delta age_n)\sim$0.5 can belong to this region.\\
From this analysis we can conclude that uncertainties manly depends on the degree of degeneracy at a given bin in the Balmer plane, i.e. on the rms of the prior in the bin, as the additional constraints (photometry and metal-sensitive indices) contribute very little even at this SNR.\\

According to figure \ref{fig:tres}, the best time resolution can be reached for signal-to-noise ratio equal to 100. Now, we want to report the results  for SNR=100.\\
In figure \ref{fig:d1090n50_snr100}, we can observe that even in this case in most of the Balmer plane the absolute value of the bias is $\sim$0.15 dex, and only in few bins it reaches values $\sim$0.3 dex.\\
However, the width of the input distribution of $log_{10}(\Delta age_n)$ in each bin is reproduced with similar accuracy as in the case of SNR=20. Even if we have a slightly larger fraction of green ellipses for $log_{10}(\Delta age_n)\sim$0.0, there is a strong presence of red ellipses.\\
The Bayesian error reproduces well the standard deviation distribution and it is comparable to the Bayesian error for SNR=20: error $>$0.5 dex in the Balmer peak; in the most of the Balmer plane 0.3 dex$<err_{bayes}<$0.4 dex; and $err_{bayes}<$0.1 dex in the region with 10$<H\delta+H\gamma<$15.\\

In agreement with the analysis of chapter \ref{chap:time_resolution}, we can observe that at no SNR it is possible to measure $log_{10}(\Delta age_n)<-0.5$, i.e. $\Delta_{10,90}\sim 1/3 age_{50}$. At odds with the limiting time resolution, for which a steady increase of performance is observed for increasing SNR up to 100, in the case of $\Delta age_n$ for complex SFH the accuracy is limited by intrinsic degeneracies already at SNR$\sim$20, and the gain in going at much larger SNR is marginal.\\

\subsubsection{Light-weighted stellar metallicity}
\begin{figure}
    \centerline{\includegraphics[width=1.6\textwidth]{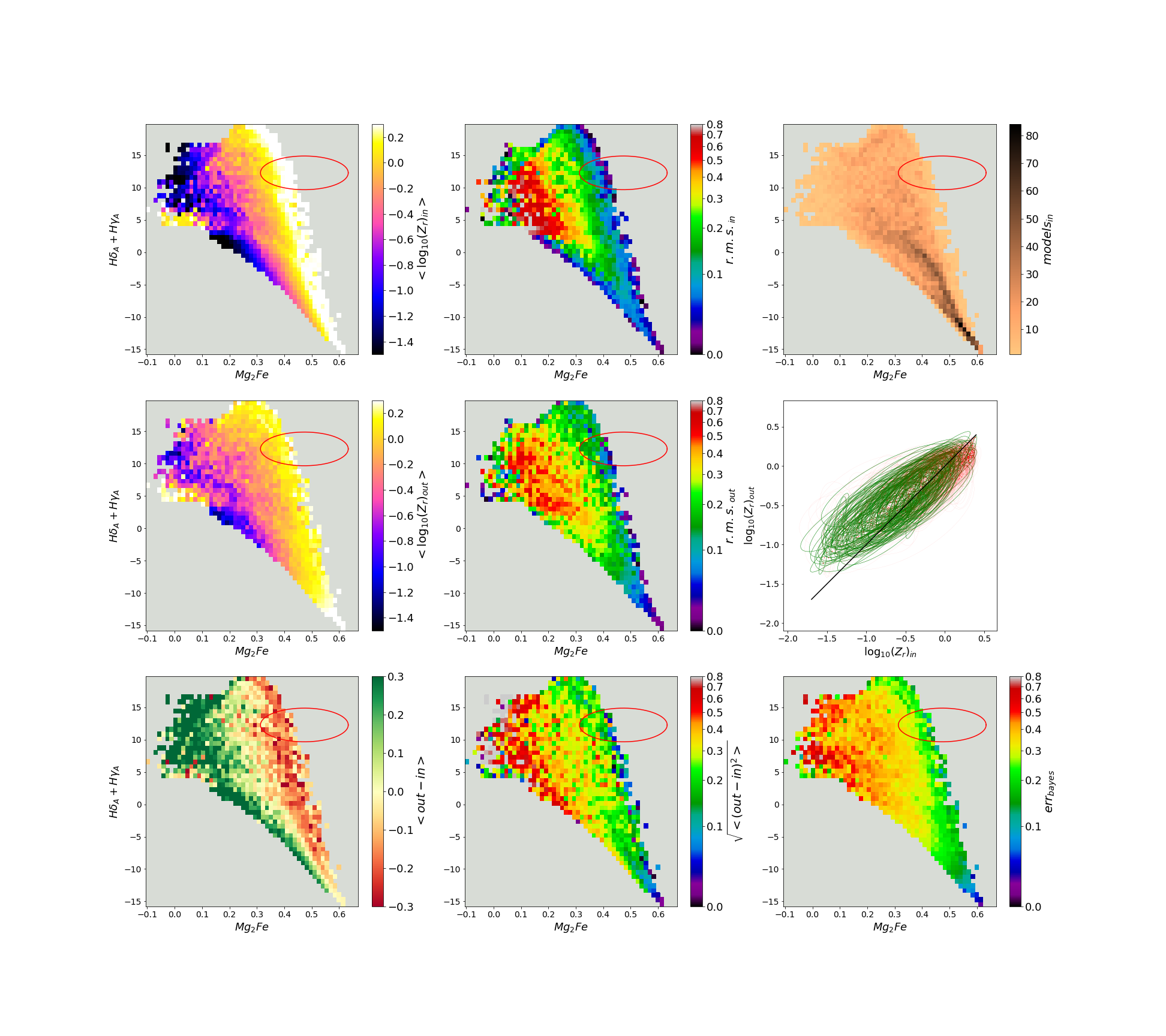}}
    \caption{Same of figure \ref{fig:lwage_snr5}, but for light-weighted metallicity at SNR=5 on ``[$H\delta_A+H\gamma_A$]-[$Mg_{2}Fe$]'' plane.}
    \label{fig:lwmet_snr5}
\end{figure}
\begin{figure}
    \centerline{\includegraphics[width=1.6\textwidth]{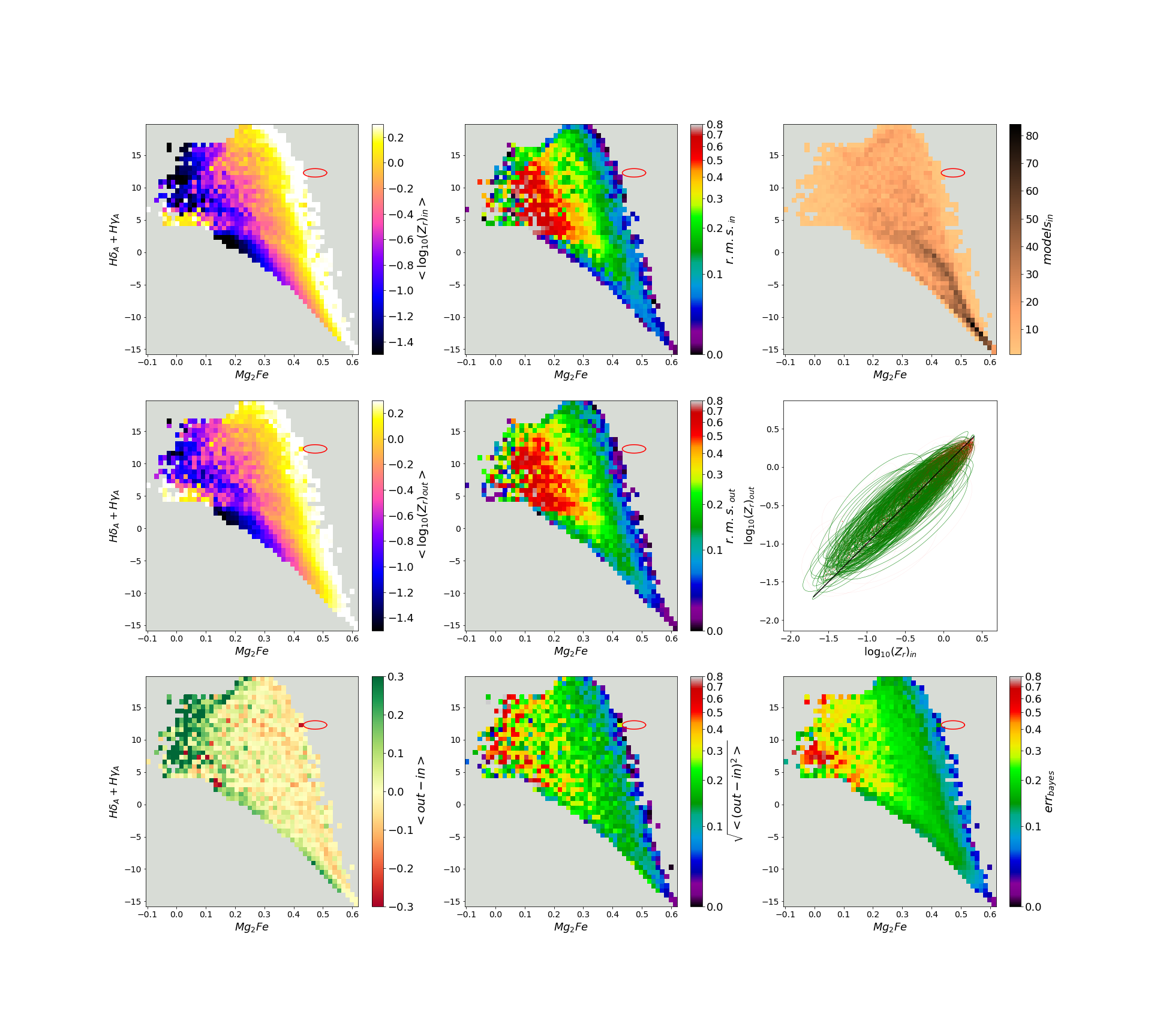}}
    \caption{Same of figure \ref{fig:lwage_snr5}, but for light-weighted metallicity at SNR=20 on ``[$H\delta_A+H\gamma_A$]-[$Mg_{2}Fe$]'' plane.}
    \label{fig:lwmet_snr20}
\end{figure}
\begin{figure}
    \centerline{\includegraphics[width=1.6\textwidth]{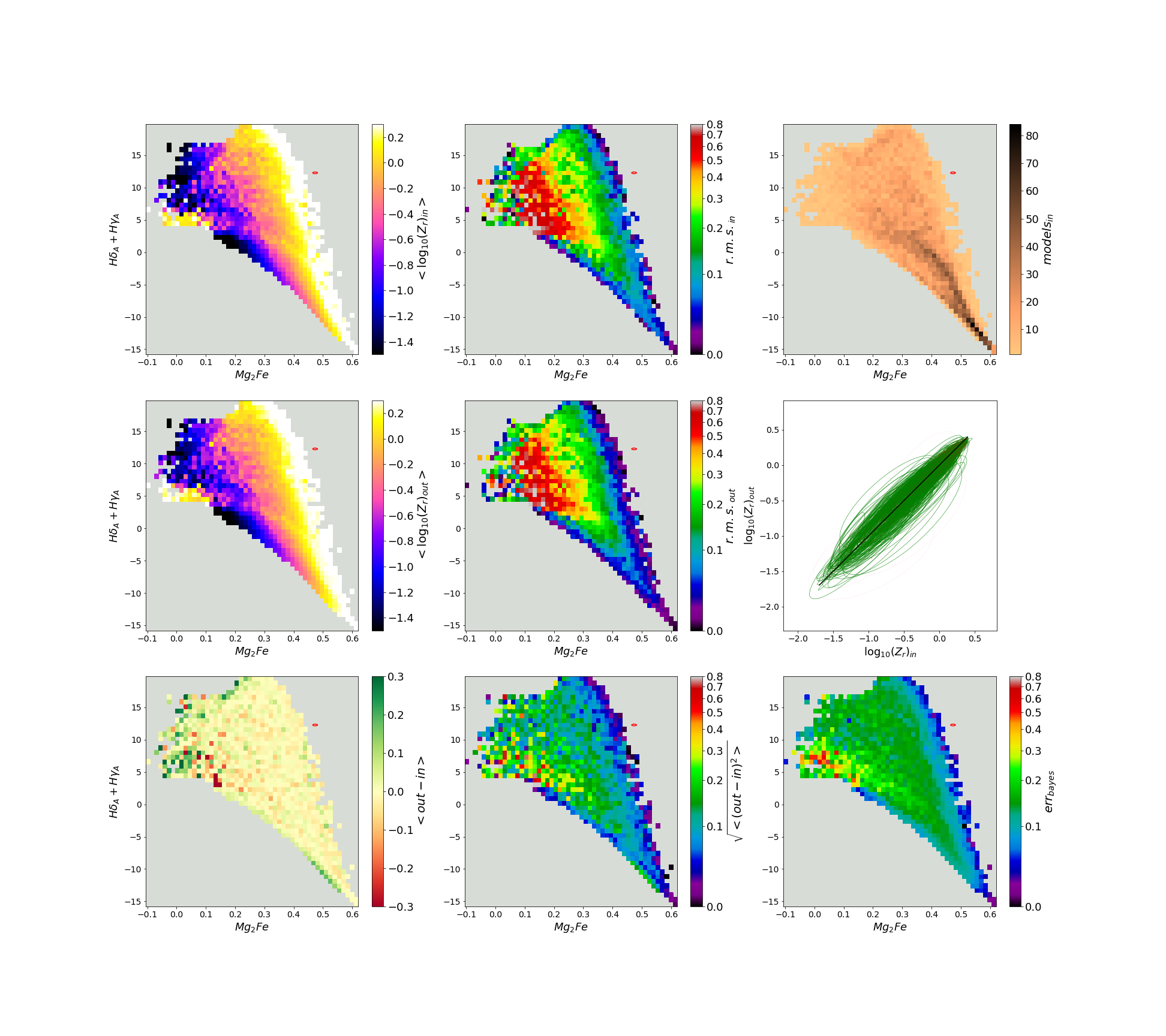}}
    \caption{MSame of figure \ref{fig:lwage_snr5}, but for light-weighted metallicity at SNR=100 on ``[$H\delta_A+H\gamma_A$]-[$Mg_{2}Fe$]'' plane.}
    \label{fig:lwmet_snr100}
\end{figure}
Because of the age-metallicity degeneracy, the accuracy in the characterization of the SFH is intimately related to the accuracy on metallicity estimates. This is explored in this section, in which we consider the ``light-weighted'' metallicity in r-band, $Z_{r}$. Instead of representing these results in the Balmer plane, we represent them in ``[$H\delta_A+H\gamma_A$]-[$Mg_{2}Fe$]'' plane, because the composite magnesium-iron index is mostly metal sensitive.\\

The first panel from left of the top row of figure \ref{fig:lwmet_snr5} displays the $Z_{r}$ distribution on the ``[$H\delta_A+H\gamma_A$]-[$Mg_{2}Fe$]'' plane. The stellar populations on the left side of the Balmer peak\footnote{Similarly at the Balmer plane, it is the region on ``$H\delta_A+H\gamma_A$-$Mg_{2}Fe$'' plane, in which $H\delta_A+H\gamma_A>15$} are characterized by $log_{10}(Z_r)<$-0.7, while on the right side of the Balmer peak, we have $log_{10}(Z_r)>$0.0 for most of the cases.\\
Low [$Mg_2Fe$] values correspond to low metallicity values. In particular, for $[Mg_2Fe]<$0.2 we have $log_{10}(Z_r)<$-0.75. These bins correspond to metal-poor, young stellar populations ($<10^9yr$), whose spectra are dominated by Balmer lines. \\ 

For $Z_{r}$ at SNR=5 (figure \ref{fig:lwmet_snr5}) we have a strong bias (first panel from left of the bottom row), which derives from a poor correlation between the mean output distribution, $<log_{10}(Z_{r})_{out}$, and the input one, $log_{10}(Z_r)_{in}$. In particular, for $[Mg_2Fe]>$0.3 there is a bias on average $\sim$-0.2 dex, while for $[Mg_2Fe]<$0.2 we have a bias $\sim$0.25 dex.\\
However, input and output appear to be well correlated inside each bins, as we can observe by comparing the middle top panel and the middle panel of the second row. This good relation inside the bins can be seen thanks to the notable presence of the green ellipses in the third panel from left of the middle row.\\
We can note also that we have an high Bayesian error for lower metallicities (third panel from left of the bottom row), which is comparable to $\sqrt{<(out-in)^2>}$. The Bayesian error decreases as metallicity increases. Bayesian error reaches values $<$0.25 dex anly for $log_{10}(Z_r)>$0.0. On the left side of the Balmer peak ($[Mg_2Fe]<$0.2), the error on $log_{10}(Z_r)$ that we get from our analysis is greater than 0.45 dex. This is due to the young stellar populations that populate that region, because their spectra have very weak metal absorption, hard to measure reliably.\\

Figure \ref{fig:lwmet_snr20} displays the light-weighted metallicity in r-band for SNR=20.\\
Compared to the previous case (SNR=5), general trends and small fluctuations are much better reproduced, within $\pm0.1$ dex for $[Mg_2Fe]>0.1$. For $[Mg_2Fe]<0.1$ though, there is still a bias $\sim0.3$ dex.\\
The remarkable presence of green ellipses, that overlap the identity line, means that output values are well correlated with the input ones also inside the bins, as we can see by comparing the middle panels of the top and of the middle row.\\
As the uncertain is concerned, at SNR=20 we are able to constrain the light-weighted metallicty to within $\pm0.2$ dex for most of the plane. We reach an error $<$0.15 dex for $[Mg_2Fe]>$0.4. Instead, high values of $err_{bayes}$ ($err_{bayes}>$0.4 dex) are reached for $[Mg_2Fe]<$0.1.\\

If we consider very high SNR spectra, i.e. SNR=100 (fig. \ref{fig:lwmet_snr100}), the mean output distribution, $<log(Z_{r})_{out}>$, reproduces very closely the input data distribution, as we can see by comparing the first panels of the top and of the middle row. In fact, we have a bias well centered on zero, although for $log_{10}(Z_r)<$-1.25 we have a bias $\sim$0.3 dex in absolute value.\\
Furthermore, also inside the single bin we have a good relation of the input and the output data. This implies a good reproduction of the scatter inside each bin, as the comparison between the second panels from left of the top and of the middle row display. Besides, this is underlined by the strong presence of the green ellipses in the first panel from right of the middle row.\\
As the Bayesian error is concerned, in most of the cases we have an error $\sim$0.15 dex on $log_{10}(Z_r)$. For $log_{10}(Z_r)>$0.25 we reach an error$<$0.1 dex. We note also a small region for $Mg_2Fe\sim$0.0 characterized by a Bayesian error $>$0.4 dex and by a bias $>$0.2 dex. Looking at the input distribution, this region corresponds to models with a SFH characterized by young star bursts, which give an important contribute to the light-weighted stellar metallicity. Furthermore, Bayesian error is well reproduced by $\sqrt{<(out-in)^2>}$.\\

In conlusion, high SNR spectra are key to obtain reliable $Z$ estimates. While high SNR is desirable for age and SFH duration as well, in the case of $Z$ SNR$\lesssim 20$ is insufficient to get useful estimates and the accuracy steadily and significantly increases as we move to SNR as high as 100.

\newpage
\section{Conclusions}
Reconstructing the SFH of galaxies is one of the key tasks of modern research on galaxy formation and evolution. The contrast between the ``hierarchical'' growth of DM structure and the ``anti-hierarchical'' scenario for galaxy formation, makes this task even more fundamental. In the past decades several authors have developed methods and codes aimed at characterizing star formation history of galaxies \citep[e.g.][]{MOPED,fernandes2005,ocvirk2006,VESPA, ppxf2012,bagpipes}.\\
In this work, we propose to characterize galaxy SFHs based on two model-independent parameters, namely the mean or median age of the stellar populations and a new parameter that describes the duration of the star formation history relative to the median stellar age. The estimation of these parameters is based on the Bayesian statistical approach described in \cite{kauffmann2003}, and developed by \cite{gallazzi2005}. This method is robust against degeneracy of different SFHs producing very similar spectra. It delivers the posterior probability distribution function (PDF) of the parameter of interest, from which we can compute the associated uncertain and the median value. The PDF of each value is obtained by the multiplication of the prior distribution function, based on initial assumptions, and the likelihood function, which is computed by using ten spectral features: five spectral indices (D4000n, [H$\delta_A$+H$\gamma_A$], H$\beta$, [Mg$_{2}$Fe] and [MgFe]$^\prime$) and five SDSS photometric fluxes in $ugriz$.\\
We defined the SFH duration relative to the median stellar age as $\Delta age_n=(age_{10}-age_{90})/age_{50}$, where $age_{10}$, $age_{50}$, $age_{90}$ are the look-back time at which the 10\%, 50\% and the 90\% of the total formed mass is reached, following the definition in \cite{pacifici2016}.\\

A first result of this project is the limit value of $\Delta age_n$, from which we can distinguish an extended SFH from one of negligible duration, discussed in chapter \ref{chap:time_resolution}. We define the time resolution as the minimum value, $\Delta age_{n,min}$, of $\Delta age_n$ above which our set of spectral features start depending on this parameter. Looking at the median value of each spectral feature as function of $\Delta age_n$, we note a flat trend for $\Delta age_n<-1.0$, which we define as reference value for the spectral feature. In order to take into account all the spectral features in this analysis, we define the quantity $\delta$ (equation \ref{eq:delta}), which describes the difference between each spectral model and a reference spectrum with negligible SFH duration. We aim at finding the limiting time resolution in an ideal case. So, neglecting dust, star bursts and variable metallicity, we create a CSP library composed by 5 millions of models a la Sandage with different $\tau$ and $t_{form}$, splitted in 5 subsamples with different fixed metallicity ($2\cdot 10^{-2}Z_{\odot}, 2\cdot 10^{-1}Z_{\odot}, 4\cdot 10^{-1}Z_{\odot}, 1 Z_{\odot}, 2.5 Z_{\odot} $). After perturbing the spectral features of the models according to different SNR, and after binning the models in 83 logarithmically spaced bins of 0.05 dex in $age_{50}$ from $10^6yr$ up to $10^{10.15}yr$, we find $\Delta age_{n,min}$ for each stellar metallicity and median stellar age.\\
After finding $\Delta age_{n,min}$ for all the $age_{50}$ and $Z$ bins, we find that $\Delta age_n$ varies within a remarkably narrow range over 4 orders of magnitude in median age and over the full metallicity range, as figure \ref{fig:tres} displays. In particular, we found that $log_{10}(\Delta age_n)$ remains between -0.1 and -0.3, for $10^7yr<age_{50}<10^{10.15}yr$ for SNR=20, and it reaches the lowest values for signal-to-noise ratio equal to 100. We have seen that for SNR equal to 5 or 10, the time resolution is very poor and completely driven by photometry, and the analysis of spectral indices does not give any additional information about $\Delta age_n$. The contribution of spectral indices to time resolution becomes relevant above SNR=20, for which we can measure at most $\Delta age_{10,90}\sim1/2\cdot age_{50}$. For SNR$>$50 the accuracy of time resolution begins to saturate, and the difference between SNR=100 and SNR=500 is negligible. \\
Furthermore, we noted a time resolution minimum at $age_{50}\sim10^9yr$, as figure \ref{fig:tres_map} displays. At this median stellar age, we have the maximum rate of variation in the strength of hydrogen lines. This leads to a strong sensitivity to the duration of the SFH, because the Balmer lines go from being dominant in the spectra to being weak shortly after.\\
We can also observe a variation of time resolution for stellar populations with $age_{50}$ lower or greater than $10^9yr$.
In particular, $log_{10}(\Delta age_{n,min})$ reaches $\sim-0.12$ dex for stellar population with $age_{50}<10^9yr$, and $\sim-0.25$ dex for stellar population with $age_{50}>10^9yr$. This is due to the larger fraction of cool stars in older stellar populations, which provide more observational constraints with respect to hot, Balmer-dominated stars that prevail in younger stellar populations. \\

In order to see the accuracy we can obtain on measurements of light-weighted age in r-band ($age_r$), mass formed-weighted age ($age_{mfw}$), light-weighted metallicity in r-band ($Z_r$), median stellar age ($age_{50}$) and $\Delta age_n$, we simulated astronomical observation by perturbing $12\,500$ model spectra of a CSP library of $500\,000$ models characterized by dust, star bursts and variable metallicity. The results were represented on the Balmer plane, except for the light weighted metallicity that was represented on [$H\delta_A+H\gamma_A]-[Mg_2Fe$]. Both stellar metallicity and light-weighted age increase with $D4000$, and using the Bayesian statistical approach, we are able to constrain light-weighted stellar metallicity and light-weighted age to within $\pm$0.15 dex for most of the sample, in agreement with \cite{gallazzi2005}.\\
As the SFH duration is concerned, we observed that its uncertainties does not change a lot between SNR=20 and SNR=100. Considering SNR=20, we found that, for $H\delta_A+H\gamma_A>15$, the Bayesian error is $>0.5$ dex, which correspond to a factor $>3.2$ on $\Delta age_n$. For $10<H\delta_A+H\gamma_A<15$ and $D4000n\sim1.25$, there is a region on Balmer plane in which the error on $log_{10}(\Delta age_n)$ is $\sim0.1$ dex, corresponding to a factor 1.2 on $\Delta age_n$. This is due to a small input rms deviation ($\sim0.1$ dex) at fixed position on the Balmer plane. For $H\delta_A+H\gamma_A<9$, we have an error between 0.4 dex and 0.3 dex. We note that the Bayesian error decreases as the Balmer break increases. In particular, for $D4000n>2.2$ we find an error$\sim0.25$ dex, i.e. a factor $<2$ on $\Delta age_n$. Furthermore, we observe in the first panel from right of the middle row of figure \ref{fig:d1090n50_snr20}, a limit at $\sim$-0.5 dex for $log_{10}(\Delta age_n)_{out}$, below which we are not able to get information on SFH duration, in agreement with our previous results.\\
This results highlights that the main limiting factor in characterizing the (duration of) star formation histories is their largely degenerate effect on spectra, rather than the quality of the observed spectra.\\

This project highlights the limitations of SFH characterization, in particular on the duration of star formation activity. The approximately constant time resolution relative to $age_{50}$, tells us that we can get information only on a minimum time interval of the order of $\sim1/3~age_{50}$. For example, if we observe an elliptical galaxy with a median stellar age $\sim10^{10}yr$, we can not get information about the firsts $3\cdot10^9yr$ of its star formation history. Similarly, if we observe a galaxy with a median stellar age $\sim10^8yr$, we can get information until $\sim3\cdot10^7yr$ after the beginning of star forming processes. However, these limits strictly apply in the idealized case assumed in chapter \ref{chap:time_resolution}. In fact, in chapter \ref{sec:charact_csfh}, we see that, for more realistically complex SFH, we would obtain an error on $log_{10}(\Delta age_n)\sim0.6$ dex, corresponding to a factor $\sim4.0$ on $\Delta age_n$, for stellar population characterized by a median stellar age $<10^9yr$, while for those with a median stellar age $>10^9yr$, we can reach an error on $log_{10}(\Delta age_n)\sim0.3$ dex, which correspond to a factor $\sim2$ on $\Delta age_n$.\\
However, we observed a region on Balmer-plane ($D4000\sim1.25$ and $10<H\delta_A+H\gamma_A<15$), in which there are stellar populations with $age_{50}\sim10^9yr$, and we reach an error $\sim0.1$ dex on $log_{10}(\Delta age_n)$, i.e. a factor $\sim$1.25 on $\Delta age_n$. In this case, we are able to measure a minimum value of $\Delta age_{10,90}$ of the order of $3\cdot 10^8yr$ and we obtain $log_{10}(\Delta age_n)=(0.5\pm 0.1)$dex. This means that at least $\Delta age_{10,90}\sim2.5\cdot 10^9yr$, and at most $\Delta age_{10,90}\sim3.9\cdot 10^9yr$. This high time resolution value is due to the strong dependence of spectral features on the duration of SFH in this region of the parameter space, as we can see in figure \ref{fig:tres}. Physically, it is the moment at which A-type stars (spectra dominated by Balmer lines) get rapidly weaker, and cooler stars start to dominate, with spectra characterized by stronger metal lines and weaker Balmer lines. \\

As mentioned before, in the last two decades a number of methods have been developed to reconstruct galaxy SFH, which mainly compare the full observed spectrum with combination of model spectra. One of these is STARLIGHT \cite{fernandes2005}. This is a code that essentially tries to obtain SFH by choosing the combination of model spectra over a grid of ages and metallicity, so to minimise the $\chi^2$ of the data relative to the model. Any degeneration is neglected with this method, which are, instead, considered in our Bayesian approach. These degeneracies are in fact the main source of uncertainty as we showed before. \\
VESPA (\cite{VESPA}) is another method which aims at recovering star formation and metallicity histories from galactic spectra by using an adaptive parameterization grid. This approach uses a grid with maximum resolution of 16 age bins, logaritmically spaced in look-back time from 0.02 up to 14 Gyr, and characterized by a size $\sim 0.18$ dex. Following our analysis, we can compute for each of the 16 age bins:
\begin{equation}
    \Delta age_n=(age_{i+1}-age_{i-1})/age_{i}
\end{equation}
where $age_{i+1}$ and $age_{i-1}$ are the top and the bottom limit of the $i^{th}$-bin, and $age_i\equiv (age_{i+1}+age_{i-1})/2$. We find for each bin $log_{10}(\Delta age_n)\sim -0.38$, which is in agreement with our $log_{10}(\Delta age_{n,min})$ results. \\

The upcoming new generation of optical and NIR spectrographs will provide information for reconstructing the history of star formation in individual galaxies up to redshifts up to $\sim0.7$. For example, the Stellar Populations at intermediate redshift Survey (StePS), a survey that uses WEAVE (\cite{dalton2012}) spectrograph on the William Herschel Telescope, focuses on $35\,000$ galaxies with redshift$>$0.3, from which high quality spectra (SNR$>$15) are obtained. Another relevant spectroscopic survey is LEGA-C (\cite{lega_c2021}), an ESO/VLT public spectroscopic survey targeting galaxies with 0.6$<z<$1.0, which provides high-SNR spectra for more than 3000 galaxies over their restframe optical range.\\
The method of SFH characterization developed in this thesis will be applied to these datasets, and the results will be compared to other approaches to put in place another piece of the complex puzzle of galaxy formation and evolution.
In doing this, following \cite{costantin2019}, we have to be aware that both theoretical limitations (intrinsic scatter of our models) and technical limitations (limited signal-to-noise ratio that can be obtained in surveys) force us to adopt essential but robust SFH characterizations. This is what we have actually done in this project: 
\begin{itemize}
    \item We proposed to find a parametric characterization of star formation history of galaxies, by defining useful parameter as $\Delta age_n$.
    \item We used a Bayesian statistical approach. It is very robust against the SFH degeneracy on galaxy spectra.
    \item We focused on the theoretical limits of duration of star formation activities, which can be useful to set the time resolution on histories of star formation for new spectroscopic surveys, or can be tested with different stellar population libraries.
\end{itemize}

\clearpage
\appendix
\section{Time resolution maps}
\begin{figure}[h]
    \centering
    \hspace*{-3cm}
    \includegraphics[width=1.2\textwidth]{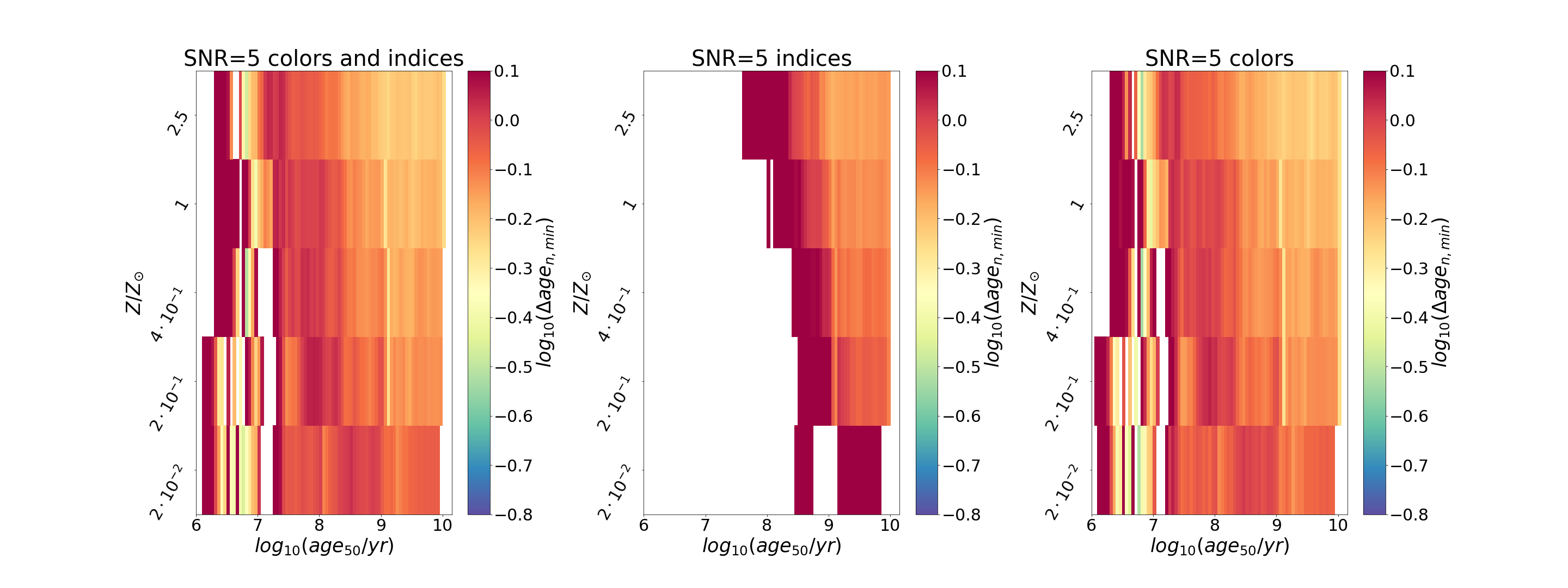}
    \caption{Time resolution map at fixed SNR=5}
    \label{fig:tres_mapsnr5}
\end{figure}
\begin{figure}[h]
    \centering
    \hspace*{-3cm}
    \includegraphics[width=1.2\textwidth]{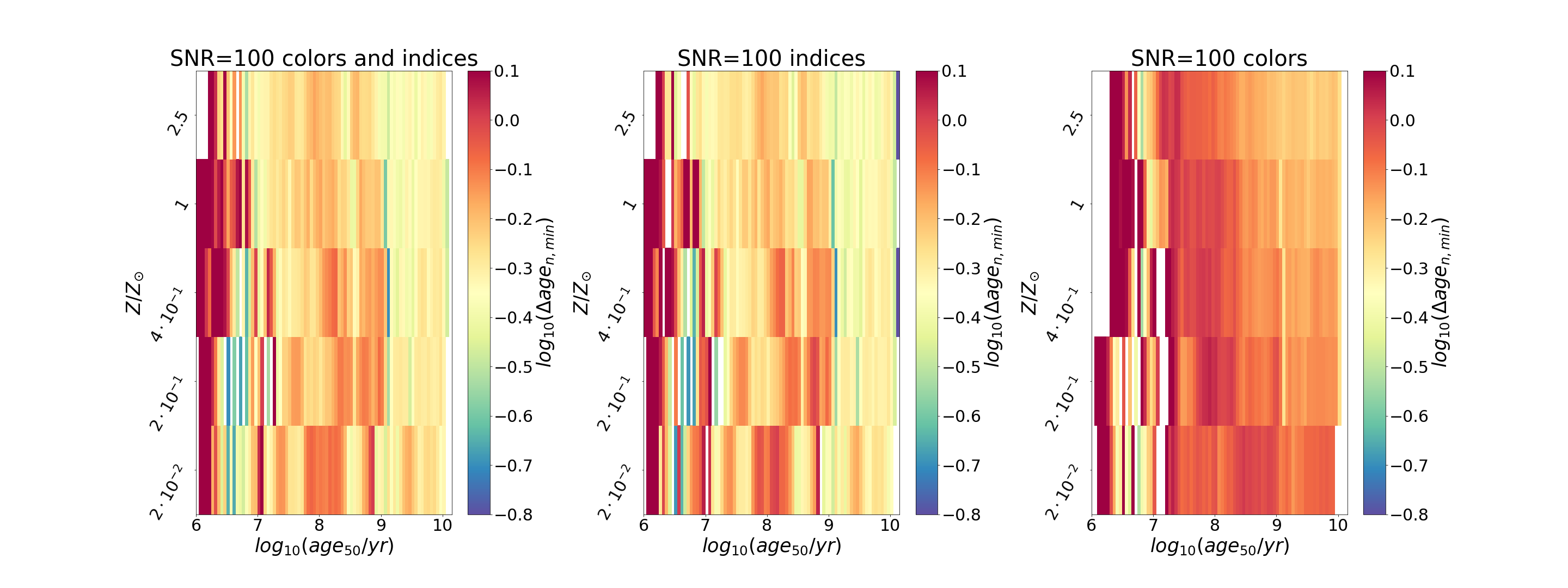}
    \caption{Time resolution map at fixed SNR=100}
    \label{fig:tres_mapsnr100}
\end{figure}

\clearpage

\section{Spectral indices}
\begin{table}[h]
    \centering
    \begin{tabular}{||c c c c c||}
    \hline
  Index name & Index wl range & Blue cont wl range & Red cont wl range & Units \\
  \hline\hline
        $H\beta$ & 4847.875 4876.625 & 4827.875 4847.875 & 4876.625 4891.625 & \AA  \\
  \hline
  $H\delta_A$ & 4083.500 4122.250 & 4041.600 4079.750 & 4128.500 4161.000 & \AA\\
  \hline
  $H\gamma_A$ & 4319.750 4363.500 & 4283.500 4319.750 & 4367.250 4419.750 & \AA\\
  \hline
  $Mg_2$ & 5154.125 5196.625 & 4895.125 4957.625 & 5301.125 5366.125 & mag\\
  \hline
  $Fe4531$ & 4514.250 4559.250 & 4504.250 4514.250 & 4560.500 4579.250 & \AA \\
  \hline
  $Fe5015$ & 4977.750 5054.000 & 4946.500 4977.750 & 5054.000 5065.250 & \AA\\
  \hline
  $Mgb$ & 5160.125 5192.625 & 5142.625 5161.375 & 5191.375 5206.375 & \AA\\
  \hline
  $Fe5270$ & 5245.650 5285.650 & 5233.150 5248.150 & 5285.650 5318.150 & \AA\\
  \hline
  $Fe5335$ & 5312.125 5352.125 & 5304.625 5315.875 & 5353.375 5363.375 & \AA\\
  \hline
    \end{tabular}
    \caption{Lick indices from \cite{worthey94} and \cite{worthey97}}
    \label{tab:indices}
\end{table}

\clearpage
\bibliography{main}

\nocite{*}

\end{document}